\documentclass[12pt]{article}
\usepackage{amsmath,amssymb,amsthm}
\usepackage{latexsym,graphicx,color,bbm}

\def\de{\partial}

\def\6#1{{\underline{#1}}}
\def\m6#1{{\underline{#1}\,}}

\newdimen\Tdim
\def\ispan{{\setbox0=\hbox{i}%
\Tdim\ht0\advance\Tdim\dp0\rule[-\dp0]{0pt}{\Tdim}}}
\def\jspan{{\setbox0=\hbox{j}%
\Tdim\ht0\advance\Tdim\dp0\rule[-\dp0]{0pt}{\Tdim}}}
\def\Tspan#1{{\setbox0=\hbox{#1}%
\Tdim\ht0\advance\Tdim\dp0\advance\Tdim.55ex\rule[-\dp0]{0pt}{\Tdim}\box0}}

\def\be{\begin{eqnarray}}
\def\ben{\begin{eqnarray*}}
\def\ee{\end{eqnarray}}
\def\een{\end{eqnarray*}}
\def\Tr{{\rm Tr}\,}

\def\p{\partial}

\def\D{\mathcal{D}}

\def\=:{=\hspace{-.7em}\raisebox{1.1ex}{.}\hspace{.1em}\raisebox{-0.2ex}{.} }

\newcommand{\NF}{N_{\rm f}}

\newcommand{\NC}{N_{\rm c}}

\newcommand {\Nf}{N_{\rm f}}

\newcommand {\mc}{{\mathbb C}}

\def\ckt{\rangle}
\def\brc{\langle}

\newcommand {\beq}{\begin{eqnarray}}
\newcommand {\eeq}{\end{eqnarray}}

\def\diag{{\rm diag}}

\makeatletter
\@addtoreset{equation}{section}

\renewcommand{\thefootnote}{\fnsymbol{footnote}}
\makeatother

\makeatletter
\newcommand{\thetablename}{Table}
\def\fnum@table{\thetablename\ \thetable}
\makeatother

\begin{document}
\thispagestyle{empty}
\begin{flushright}
RIKEN-TH-154\\
IFUP-TH/2009-10\\
May, 2009 \\
\end{flushright}
\vspace{3mm}

\begin{center}
{\Large \bf Fractional Vortices and Lumps} \\
\vspace{15mm}

Minoru~Eto$^{1,2}$\footnote{
Now at Theoretical Physics Laboratory at RIKEN since April 2009.
}
,
Toshiaki~Fujimori$^3$,
Sven~Bjarke~Gudnason$^{1,2}$,
Kenichi~Konishi$^{1,2}$,\\
Takayuki~Nagashima$^{3}$, 
Muneto~Nitta$^4$,
Keisuke~Ohashi$^5$\footnote{
Now at the department of physics in Kyoto University since April 2009.
}
and Walter~Vinci$^{1,2,5}$

\bigskip\bigskip\bigskip
{\it

$^1$  
~Department of Physics, University of Pisa,
Largo Pontecorvo, 3, Ed. C, 56127 Pisa, Italy
\\
$^2$  INFN, Sezione di Pisa,
Largo Pontecorvo, 3, Ed. C, 56127 Pisa, Italy \\
$^3$ Department of Physics, Tokyo Institute of
Technology, Tokyo 152-8551, Japan\\
$^4$ Department of Physics, Keio University, Hiyoshi,
Yokohama, Kanagawa 223-8521, Japan\\
$^5$ Department of Applied Mathematics and Theoretical Physics,
University of Cambridge, Cambridge, CB3 0WA, UK\\
}

\vspace{12mm}

\abstract{We study what might be called fractional vortices, vortex
  configurations with the minimum winding from the viewpoint of their
  topological stability,  but which are characterized by various
  notable substructures in the transverse energy distribution. The
  fractional vortices occur in diverse Abelian or non-Abelian
  generalizations of the Higgs model. The global and local features
  characterizing these are studied, and we identify the two crucial
  ingredients for their occurrence---the vacuum degeneracy leading to
  non-trivial vacuum moduli ${\cal M}$, and the BPS nature of the
  vortices. Fractional vortices  are further classified into two kinds. 
  The first type of such vortices appear when ${\cal M}$ has orbifold 
  $\mathbb{Z}_n$ singularities; the second type occurs in systems in which 
  the vacuum moduli space  ${\cal M}$ 
  possesses either a deformed geometry or some singularity.
These general features are
  illustrated with several concrete models.  
}

\end{center}

\vfill
\newpage
\setcounter{page}{1}
\setcounter{footnote}{0}
\renewcommand{\thefootnote}{\arabic{footnote}}

\section{Introduction}

Vortices appear in many different areas of physics, from fluid and
plasma dynamics, solid-state physics, particle physics to cosmology.
Usually certain topological properties lie behind their stability in
time and in space. Typically, the energy distribution in the plane
perpendicular to the vortex axis is peaked around the vortex axis,
with a well-defined finite width in the vortex profile.  This is
certainly the case for the single (i.e.~minimum-vorticity) type II
Abrikosov-Nielsen-Olesen (ANO below) vortex
\cite{Abrikosov,NielsenOlesen}, where the origin of the vortex
stability lies in the first homotopy group, $\pi_{1}(U(1))={\mathbbm
  Z}$.  As it turns out, however, when the gauge group and/or
the matter content of the system are of more general kind than the
standard Abelian-Higgs model (Landau-Ginzburg model)---$U(1)$ gauge
group and one charged scalar field---, a variety of interesting
generalized vortex solutions appear.   

The present paper is concerned with a class of vortex-like solutions
which might be called ``fractional vortices''.  They are characterized
by the minimal quantized vorticity (winding or magnetic flux) from the
point of view of topological stability; nevertheless, their transverse
profiles exhibit various non-trivial substructures as if they were
made of smaller vortices, a little like a multi-vortex solution in the
standard type II superconductor.  But in contrast to the latter case
the tension of each of the sub-peak is not quantized, and
their relative weights,  distances  and shapes depend on the details of the system,
such as the coupling constants, the scalar VEVs (vacuum expectation
values) and the symmetry breaking pattern, etc.  In all cases, a sub-peak cannot be removed by sending its position to
infinity while keeping others in a finite region. \footnote{An analogous
  phenomenon  occurs in a simple extension of the Abelian Higgs
  model---the Landau--Ginzburg model---with two coupled scalar fields.
This can be realized in a certain unconventional superconductor  \cite{Babaev}. 
On a broader prospect, our fractional vortices
  share also some features with the fractional instantons---the torons 
  \cite{Torons} or the calorons \cite{Caloron} 
which exist when the base space has a period such as a torus.
Fractional vortices in a torus or a cylinder have also been studied 
\cite{Tong:2002hi,Eto:2004rz}, see also \cite{Bruckmann:2007zh}.
} 

The aim of this work is to show that such a phenomenon occurs very
generally, and to make a preliminary study of these solutions, trying
to find what characterizes the occurrence and substructures of these
fractional vortices. 
As the arena of our study, we consider various Abelian and non-Abelian
extensions of the Higgs  model. The degrees of freedom will be a set
of complex scalar fields with various charges and gauge fields.  For
definiteness and for simplicity, we take the models whose Lagrangians
have the form of the bosonic sector of ${\cal N}=2 $ supersymmetric
gauge theories. They are natural generalizations of the Abelian-Higgs
model in the Bogomol'nyi-Prasad-Sommerfield(BPS) limit. 
The constant which characterizes the VEVs and
which forces the system into the Higgs phase, arises as the
Fayet-Iliopoulos (FI) term in the supersymmetric setting. Although
most of our results are independent of supersymmetry, we shall
mention also results more specific to the supersymmetric versions of
the models, when appropriate. 

It will be shown that the fractional vortices can be further classified into two different classes. 
The first type exists when the vacuum moduli ${\cal M}$ has 
an orbifold  ${\mathbb Z}_n$ singularity.   The second type occurs  when the vacuum moduli ${\cal M}$ has a 2-cycle 
with a deformed geometry. 

This paper is organized as follows.
In Sec.~\ref{sec2} we review the semi-local vortices 
in the extended Abelian-Higgs model  (EAH).
In Sec.~\ref{Sec:Structures} general properties
of vortices in degenerate vacua are discussed. 
In Sec.~\ref{sec4} we present various concrete models 
admitting fractional vortices based on 
a  ${\mathbb C}P^1$ target space.  The first two models, in particular,  provides the simplest examples 
of the fractional vortices of the first and second types, respectively.
In Sec.~\ref{sec5} we discuss an $SO(N)\times U(1)$ gauge theory,  which exhibits fractional vortices 
of the second type.   
Some useful technical details are collected in the Appendices.

\section{Semi-local vortex in the Extended Abelian-Higgs model}\label{sec2}

Something quite non-trivial occurs already in the Abelian-Higgs model,
if the number of charged fields is greater than one
\cite{EAH,Achucarro:1992hs,Hindmarsh}.    The model is  
\beq \mathcal{L} =  -\frac{1}{4e^2}F_{\mu\nu}F^{\mu \nu} 
+ \D_\mu q \left(\D^\mu q\right)^\dag 
- \frac{\lambda}{2}\left(q q^\dag - \xi \right)^2 \ , 
\label{AbelianSL}\eeq
where $\D_{\mu }= \de_{\mu} - i A_{\mu}$ is the standard covariant
derivative, $q = (q_{1}, q_{2}, \ldots, q_{\Nf})$ represents a set of
complex scalar matter fields of the equal charge.  This model is
sometimes called the {\it semi-local} model since not all global
symmetries, i.e.~$G=U(\Nf)$ here, are gauged. Even if we restrict to
the minimum vorticity, the vortex profile turns out to depend on the
particular solution considered. 

In order to have a finite-energy (the energy per unit length---the
tension) configuration, the complex scalar fields must asymptotically  
approach a vacuum configuration,
\beq  M \equiv  \{ q_{i} \},    \qquad  
\sum_{i=1}^{N_{f}} \left|q_{i}\right|^{2}  =  \xi \ ,  \label{vacuumM}
\eeq
far from the vortex center. 
By the  $SU(N_{F})$ global and $U(1)$ local symmetry they can be
chosen to be  
\beq  \brc q  \ckt  = (q_{0},  0, \ldots, 0), \qquad    q_{0} = \sqrt{\xi},  
\label{vacuum}   \eeq
breaking the global symmetry to $SU(\Nf-1)\times U(1)$ 
In other words, the vacuum moduli space is 
\beq  {\cal M} = \mc P^{\Nf-1} = SU(\Nf)/[SU(\Nf-1)\times
  U(1)] \ .   \label{symbr}\eeq    
The vacuum configurations $M$ represent a non-trivial $U(1)$ fibration
over the vacuum moduli space $\mc P^{\Nf-1}$.
As the $U(1)\subset U(N_{f})$ part is gauged, its breaking
does not lead to any further vacuum degeneracy. 
Since the first homotopy group of the vacuum configurations $M$
(\ref{vacuumM}) is trivial,  
vortex solutions may not necessarily be stable. 

Indeed,  for $\beta \equiv \lambda/e^{2}> 1$ (i.e.~type II
superconductors), an ANO vortex solution embedded in the first flavor
is found to be unstable against fluctuations of the extra fields
($i=2,3, \ldots, N_{f}$) which increase its size; the vortex flux
spreads out all over the transverse space 
\cite{Achucarro:1992hs,Hindmarsh}.   

For $\beta < 1$ (i.e.~type I superconductors), instead, an ANO vortex
\cite{Abrikosov,NielsenOlesen} embedded in one of the flavors is
found to be stable. The origin of the stability of such a vortex can
be traced to the fact that the asymptotic scalar field must be
actually the vacuum configuration~(\ref{vacuum}) {\it modulo} gauge
transformations. 
The vortex winds a non-trivial $U(1)$ fiber 
over the vacuum moduli
$\mc P^{\Nf-1}$:
the relevant homotopy is 
\beq \pi_{1}(U(1))  = {\mathbb Z} \ , \label{firsthmtp} \ee
just as in the case of the ANO vortex (which indeed it is). 

In the interesting special (BPS) case, $\beta=1$, we find a family of
degenerate vortex solutions with the same tension, $T=2\pi\xi$.
Except for the special point of the vortex moduli space (i.e.~the space of
solutions),  which represents the ANO vortex (sometimes called a
``local vortex''), the vortex has a power-like tail in the profile
function, and the width of the vortex (thickness of the string) can be
of an arbitrary size\footnote{This type of vortex solutions has been
  termed ``semi-local vortices''. Again although this is not an entirely
  adequate terminology we shall stick to it as it is
  commonly used in the literature.}.
Far from the vortex center, 
the vortex configuration essentially 
reduces to the  $\mathbb{C}P^{\NF-1}$ sigma-model lump (or
two-dimensional Skyrmion), characterized by  
\beq \pi_{2}\left(\mathbb{C} P^{\NF-1}\right) = {\mathbb Z} \ .
\label{secondhmtp}  \eeq 
In terms of an effective potential as a function of the vortex radius,
the $k=1$ (minimum-winding) sector of the system has the minimum at
the origin for  $\beta < 1$; at infinity for $\beta > 1$ (a
``run-away vacuum'' behavior);  and has no potential---a flat
direction---in the BPS case.   

\section{Vortices in degenerate vacua \label{Sec:Structures}}

\begin{figure}
\begin{center}
\includegraphics[width=4in]{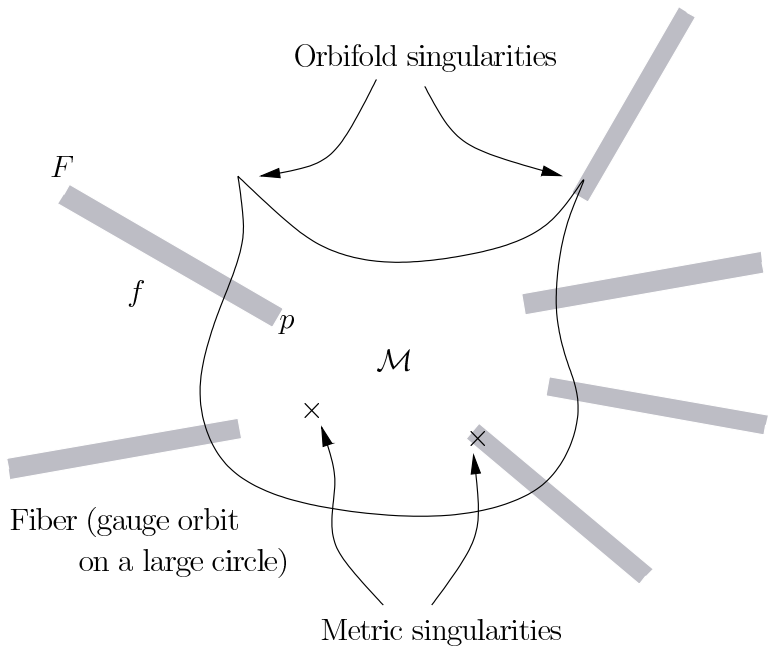}
\caption{\footnotesize Vacuum moduli ${\cal M}$,  fiber $F$ over it,
  and possible singularities} 
\label{Gatto}
\end{center}
\end{figure}

There are two crucial ingredients which lead to the interesting
varieties of degenerate vortex solutions in systems such as the
extended Abelian-Higgs model (with $\beta=1$) just considered: the
{\it vacuum degeneracy} and the BPS saturated nature of the vortices.
The first means that, in contrast to the cases in which the vacuum
moduli is trivial (a point)\footnote{This is the case for the standard
  Abelian-Higgs model, of course, but so it is in the case of the
  $U(N)$ gauge theory with $\NF=N$ squarks in the fundamental
  representation. The latter model and its generalizations, after the
  discovery of the non-Abelian vortices in the color-flavor locked
  vacuum \cite{HT,ABEKY}, has attracted considerable attention
  \cite{ABEK}-\cite{DKO}.}, 
we must consider in general all vortex solutions in all possible
points of the vacuum moduli simultaneously. See Fig.~\ref{Gatto}.
Even if we restrict ourselves to the minimally winding vortex
solutions only---we shall do so in this article for definiteness---the
vortices represent non-trivial fiber bundles over the vacuum moduli
${\cal M}$.  The BPS saturated nature of the vortices implies that the
vortex equations linearize, and in turn half of the equations (the
matter equations of motion) being solved by the moduli-matrix Ansatz, 
as is well known \cite{Isozumi:2004vg,Eto:2005yh,Eto:2006pg}, see
Eqs.~(\ref{Ansatze}), (\ref{complexif}).  The other equations (the
gauge field equations) reduce, in the strong coupling limit or
anyway sufficiently far from the vortex center, to the vacuum
equations for the scalar fields.  In other words, the vortex
solutions tend to sigma model lumps. 
   
\subsection{Structures of the vacuum moduli \label{generalst}} 

Let us first consider what we regard as the global aspect of our
vortices. More precisely, our first concern is the {\it vacuum moduli}
${\cal M}$ on each point of which the vortex solutions are defined.  
Let the symmetry group of the underlying system be 
\beq  K = L \otimes G_{F} \ , \eeq
where $L$ is the local gauge group, while $G_{F}$ is  the global
symmetry group. Let $M$ be the manifold of the minima of the scalar
potential, the vacuum configuration $M=\{q_{i} \ | \ q^\dag T^I
q = \xi^I\}$.   The vacuum
moduli ${\cal M}$ is given by the points  
\beq  p \in  {\cal M}=  M / F \ , \eeq
where the fiber $F$ is the sum of the gauge orbits of a point in $M$
\beq  f  \in  F = \{q^{g}\, |  \, q^{g}=   g q  \} \ , \qquad   
g \in  L  / L_{0} \ , \eeq
where we have taken into account the possibility that a given vacuum
configuration might leave a subgroup $L_{0} \subset L$ unbroken: 
\beq L_{0}^{\{q\}} = \{ \ell_{0}\in L \, |  
\, \ell_{0} q = q \}\ .  \eeq
In other words the vacuum moduli are made of the points of $M$ in which
gauge-equivalent points are identified. 

A subgroup of the global group
\be {\tilde G}^{\{q\}} \subset  G_{F} \ , \quad
{\textrm{such that}}\ 
{\tilde G}^{\{q\}} = \{ g_{f} \in G_{F} \, |  \, g_{f} q = \ell q \}, \qquad 
\ell  \in L \ , \label{symme}  \eeq
represents the unbroken global symmetry group of the
system\footnote{In other words, ${\tilde G}^{\{q\}}$ is the
  subgroup of $G_{F}$ i.e.~transformations which can be ``undone''
  by---or equivalent to---a local gauge transformation.}.

\begin{figure}
\begin{center}
\includegraphics[width=3in]{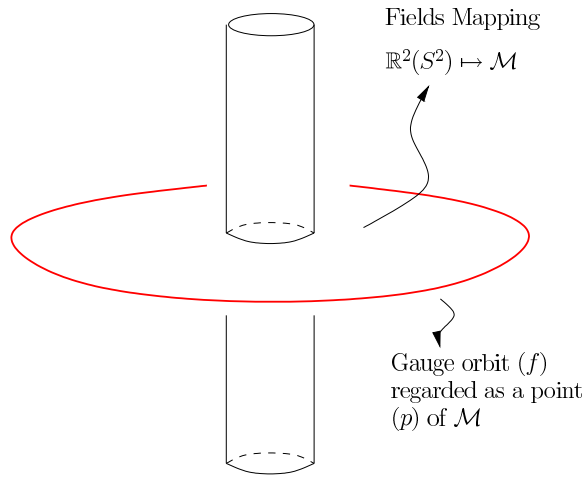}
\caption{Far from the vortex center the fields trace, along a circle,
  the gauge orbits ($f$) regarded as a single point of ${\cal M}$.
  Inside the circle, the fields provide a map from $R^2 (S^2)$  to
  ${\cal M}$. }
\label{VortexFig}
\end{center}
\end{figure}

A vortex solution is defined on each point of ${\cal M}$, in the sense
that the scalar configuration along a sufficiently large circle
($S^{1}$) surrounding it traces a non-trivial orbit in $F$ (hence a
point in ${\cal M}$).  The existence of a vortex solution 
at a point $f \in F$ requires that 
\beq \pi_{1}\left(F, f\right) \neq \mathbbm{1} \ ; \eeq
a vortex corresponds to a non-trivial element of $\pi_{1}(F, f)$.  The
field configuration on a disk $D^{2}$ encircled by $S^{1}$ traces
${\cal M}$, apart from points at finite radius where it goes off $M$
(hence from ${\cal M}$).  In other words it represents an 
element of $\pi_{2}({\cal M}, p)$ \footnote{This notion can be made more precise by considering the strong-coupling limit 
of our systems where the gauge field equations reduce to the vacuum condition, see Eq.~(\ref{vaccond}) below.  In the presence of an ANO like sub-peaks, however, this correspondence leads to 
a singular lump, as will be seen in several examples below.   
}, where $p$ is the gauge orbit
containing $f$, or 
\beq p = \pi(f) \ : \eeq
 $\pi$ is the projection of the fiber onto a point of the basis
${\cal M}$.   The exact sequence of homotopy groups for the fiber bundle reads 
\beq \cdots \to 
\pi_{2}\left(M, f\right) \to 
\pi_{2}\left({\cal M}, p\right) \to
\pi_{1}\left(F, f\right) \to 
\pi_{1}\left(M, f\right) \to 
\pi_{1}\left({\cal M}, p\right) \to
\cdots   \label{homotseq} \eeq 
where $\pi_{2}(M/F, f) \sim\pi_{2}({\cal M}, p)$.  
Note that in our application of such a sequence to the physical,
vortex solutions, the reference point $f$ or $p$ appearing in the
definition of the homotopy groups, corresponds to the field
configurations along the large circle $S^{1}$ encircling the given
solution, see Fig.~\ref{VortexFig}. 

Given the points $f, p$ and the space ${\cal M}$, the vortex solution
is still not unique.  Any exact symmetry of the system  (internal
symmetry  ${\tilde  G}^{\{q\}}$ as well as spacetime symmetries
such as Poincar\'e invariance) broken by an individual vortex
solution gives rise to vortex zero modes (moduli), $\mathcal{V}$.
The vortex-center position moduli $\mathcal{V} \sim {\mathbb C}$, for
instance,  arise as a result of the breaking of the translation
invariance in ${\mathbb R}^{2}$.  The breaking of the internal symmetry
${\tilde  G}^{\{q\}}$ (Eq.~(\ref{symme})) by the individual vortex
solution gives rise to orientational zero-modes in the $U(N)$
models extensively studied in last several years. See
\cite{FGK,General,grande} for more recent results on this issue.  

Our main interest here, however, is the vortex moduli which arises
from the non-trivial vacuum moduli ${\cal M}$ itself.  Due to the BPS
nature of our vortices, the gauge field equation (see  
Eq.~(\ref{left}))\footnote{The index $I$ denotes generally all the generators of the gauge group 
considered. A non-vanishing (FI) parameter  $\xi$ is assumed only for $U(1)$ factor(s).}
\beq F_{12}^{I} = g_{I}^{2}\left(q^{\dag} T^{I} q - \xi^I\right) \ , \label{vaccond} \eeq
reduces,  in the strong-coupling limit (or in any case, sufficiently
far from the vortex center),  to the vacuum equation
defining $M$. 
This means that a vortex configuration can be approximately seen  as a
non-linear $\sigma$-model (NL$\sigma$M) lump with target space ${\cal
  M}$ (for non-trivial element of $\pi_{2}(\cal M)$).  Various
distinct maps 
\beq S^{2}  \mapsto {\cal M} \ , \eeq  
of the same homotopy class correspond to physically inequivalent  
solutions;  each of  these corresponds to a vortex  with the equal
tension  
\beq T_{\rm min}=  -\xi^{I} \int  d^{2}x \, F_{12}^{I} > 0 \ , \eeq
because of their BPS nature. 
They thus represent non-trivial {\it vortex moduli}.

The semi-local vortices of the extended-Abelian Higgs (EAH) model
reviewed in the previous section arise precisely this way.   
In the EAH model with $N$ flavors of (scalar) electrons,  
\beq M = S^{2N-1} \ , \qquad  
F = S^{1} \ , \qquad 
\mathcal{M} = S^{2N-1}/S = \mathbb{C}P^{N-1} \ , \eeq
and the homotopy sequence reads 
\beq 
\cdots \to 
\mathop{\pi_{2}\left(S^{2N-1}\right)}_{\begin{array}{c}\parallel\\\mathbbm{1}\end{array}} \to 
\mathop{\pi_{2}\left(\mathbb{C}P^{N-1}\right)}_{\begin{array}{c}\parallel\\\mathbb{Z}\end{array}} \to
\mathop{\pi_{1}\left(S^{1}\right)}_{\begin{array}{c}\parallel\\\mathbb{Z}\end{array}} \to 
\mathop{\pi_{1}\left(S^{2N-1}\right)}_{\begin{array}{c}\parallel\\\mathbbm{1}\end{array}} \to 
\cdots \label{SeqEAH} \eeq
The usual argument tells us then that $\pi_{2}(\mathbb{C}P^{N-1})$ and
$\pi_{1}(S^{1}) $ are isomorphic: each (i.e.~minimum) vortex solution
corresponds to a minimal $\sigma$-model lump solution.  
As in this model the vacuum moduli ${\cal M}$ is a (smooth) manifold,
the above relations do not depend on the reference point $f$ (or $p$). 

In most cases discussed below, however, the base space ${\cal M}$
will be various kinds of {\it singular manifolds}: a manifold with
singularities.  
The nature of the singularity depends on the system and on the
particular point(s) of ${\cal M}$.  
Some of them are simple conic (orbifold) singularities, due to the
fact that some discrete (e.g.~${\mathbb Z}_{N}$) symmetry is
restored at that point. The fiber (the gauge orbits) is smaller by
some discrete quotient, with respect to $F$ at neighbouring points.
Other singularities at isolated points, or along some submanifold,
reflect an even more drastic change of $F$ such as a different
unbroken gauge group at those points, as compared to that in
surrounding regular (or less singular) points of ${\cal M}$. The
fiber itself goes through a discrete change in its dimension and in
the type, at or along the singularity(ies).

\subsection{Classification of fractional vortices \label{causes}}

There are basically two distinct causes or mechanisms leading to the appearance of  multiple peaks in the
energy density even if the vortex under consideration has a minimum
vorticity  required by the regularity and the topological stability.   
The first type 
is related to the presence of  orbifold singularities in ${\cal M}$. 
For example, let us consider a ${\mathbbm Z}_{2}$ point $p_{0}$ such as the one appearing in a simple $U(1)$ model with two scalars (Section~\ref{sec:21model}).   At this singularity, both elements of $\pi_{2}\left({\cal M}, p\right) $ and $\pi_{1}\left(F, f\right) $  make a discontinuous change.   The minimum element of $\pi_{1}\left(F_{0}, f_{0} \right) $ is half  of that of $\pi_{1}\left(F, f \right)$ defined off the singularity, and similarly for 
$ \pi_{2}\left({\cal M}, p_{0} \right) $  with respect to $ \pi_{2}\left({\cal M}, p \right) $, $p \ne p_{0}$. 
Even though the exact sequence such as Eq.~(\ref{homotseq}) continues to hold  on and off the orbifold point, the vortex defined near such a point will look like a doubly-wound vortex, with two centers (if the vortex moduli parameters are chosen appropriately).  

\begin{figure}[ht]
\begin{center}
\includegraphics[height=5.5cm]{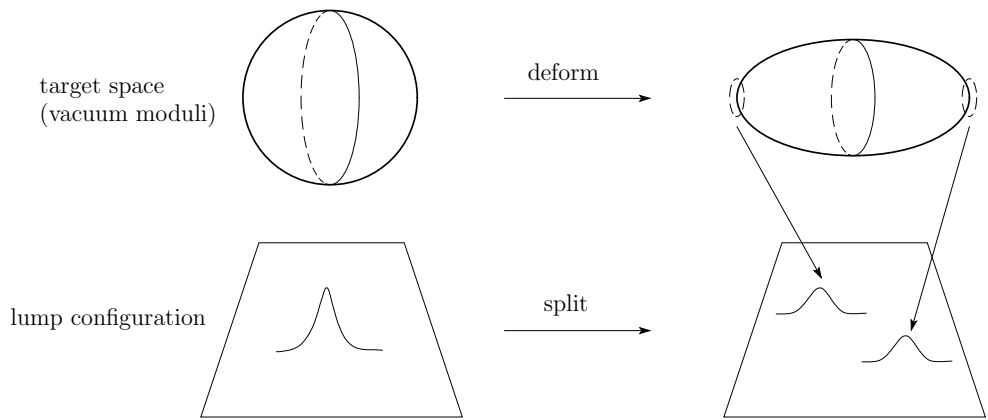}
\caption{\footnotesize{A sketch of fractional lumps  {\it of the second type}  in a sigma model limit.}}
\label{fig:type_b}
\end{center}
\end{figure}

Another cause for the appearance of fractional peaks, 
which we call the second type,   is best understood by considering the strong coupling limit where the vortex reduces to a sigma-model lump, as already noted.  Even if the base point $p$ is a perfectly generic, regular point of ${\cal M}$, not close to any singularity, 
the field configurations in the transverse plane ($S^{2}$) trace 
a 2-cycle in the vacuum moduli space ${\cal M}$.    The energy  distribution reflects the nontrivial structure of $\mathcal{M}$ as
the volume of the target space is mapped into the transverse plane,
$\mathbb{C}$
\beq E = 2\int_{\mathbb{C}} \frac{\p^2 K}{\p\phi^I\p\phi^{\dag\bar{J}}}
\p\phi^I\bar{\p}\phi^{\dag\bar{J}} = 2\int_{\mathbb{C}} \bar{\p}\p K
\ .\eeq
Let us consider the case in which a 2-cycle in the vacuum moduli space ${\cal M}$ 
(the target space of the non-linear sigma model) 
is endowed with a deformed geometry,    with 
regions of relatively larger scalar curvature, and 
possibly with some singularities.   When the field configuration sweeps such regions,  
the energy density will show 
sub-peaks as illustrated in Fig.~\ref{fig:type_b}\footnote{The
  existence of the directions in the target space, which are not
  related to any isometry, is necessary for the fractional lumps of
  the second type. Such directions are parametrized by
  so-called quasi-Nambu-Goldstone modes in the context of
  supersymmetric theories \cite{Higashi} while the directions of
  isometries correspond to Nambu-Goldstone modes.}. 

The field configuration may also simply 
hit one of the singularities (conic or not), 
which could represent a sick point in the non-linear sigma model limit.  
Even at finite coupling, the vortex tension density will exhibit a similar substructure. 
The existence of the singularity is, however,  not essential for the occurrence of 
fractional vortices of the second type, in contrast to the first type.

\section{Models based on  $\mathbb{C}P^{1}$} \label{sec4}

Several concrete models will be studied below. The fractional vortices
appearing in these systems are caused by one or the other  of the above
mechanisms,  or by collaboration of the two.   The actual manifestation
of these singularities could sometimes look quite complicated.   We
first discuss in this section models where the base space (vacuum
moduli) is a $\mathbb{C}P^{1}$ with one or two
singularities, 
or a smooth but deformed $\mathbb{C}P^{1}$.

\subsection{Abelian Higgs model with two fields of different charges\\
-- fractional vortex of the first type --
\label{sec:21model}}

The first model is a simple extension of the Abelian-Higgs model
with $\NF=2$ flavors $H = (A, B)$  but with unequal charges.
We assign the $U(1)$ charges $(\{m,n\}) $  to the fields  $A$ and $B$, respectively. The gauge transformations take the form,
\beq
H = (A,B) \to (e^{im\alpha(x)} A, e^{in\alpha(x)} B) \ .
\label{fiber}
\eeq
For simplicity, we assume that the charges are relatively prime, i.e.,   ${\rm g.c.d}(\{m,n\}) = 1$.
The vacuum manifold ($D$-flatness condition) is topologically
equivalent to $S^3$ and the vacuum moduli are topologically 
the same as $\mathbb{C}P^1$ but with some conical singularities
\beq
&& M = \{A,B\;|\; m|A|^2 + n|B|^2 = \xi\}\ ,\quad \\
&& {\cal M} = M/U(1) \simeq 
   W{\mathbb C}P^1_{(m,n)}\simeq \mathbb{C}P^1/(\mathbb{Z}_m \times
\mathbb{Z}_n) \ .
\label{eq:mn_vac}
\eeq
The vacuum moduli can be also described by the following quotient
\beq
(A,B) \sim (\lambda^m A, \lambda^n B)\ ,\quad
\lambda \in \mathbb{C}^* \ .
\eeq
Clearly, $A=0$ is a $\mathbb{Z}_n$ fixed point and $B=0$ is a
$\mathbb{Z}_m$ fixed point. The $U(1)$ gauge symmetry is broken at
every point of the vacuum moduli, thus topologically stable 
vortices can appear. 

Such a vortex solution is characterized by the broken $U(1)$-winding
number $\nu$ given in Eq.~(\ref{eq:tension}). The BPS energy density
and mass are  
\beq
{\cal E} &=& - \xi F_{12} + \p_i^2 J\ ,\quad
J \equiv \frac{1}{2} HH^\dagger\ ,
\label{eq:energy_form}\\
T &=& \int dx^2\, {\cal E} =  2\pi \xi \nu \ .
\label{eq:energy_form2}
\eeq
By using the moduli matrix method  in Appendix, it can be
expressed as
\beq
&&\nu = - \frac{1}{2\pi} \int dx^2\, F_{12} = \frac{1}{\pi} \int
dx^2\, \p\bar\p \log |s|^2 \ ,\\
&& \bar W = - i\bar\p\log s  \ ,\\
&&H = (A,B) = \left(s^{-m} A_0(z)\ ,\ s^{-n} B_0(z)\right)\ ,
\eeq
where $\nu$ is a positive number, $s$ is an everywhere non-zero function 
and {\it the moduli matrices} 
$A_0(z)$ and $B_0(z)$ are polynomial functions of $z$.
The first equation determines the asymptotic behavior of $s$ as
\beq
|s|^2 \to |z|^{2\nu}\qquad\text{as}\quad |z| \to \infty \ .
\eeq
We choose the boundary condition
\beq
(A,B) \to (A_{\rm vev} e^{im\nu \theta},B_{\rm vev} e^{in\nu\theta })
\in M\qquad\text{as}\quad |z| \to \infty \ .
\eeq
The BPS equations (see Appendix) lead  to the  master equation
\begin{align}
\bar{\partial}\partial\log\omega = -\frac{e^2}{4}
\left[m\,\omega^{-m}|A_0|^2 + n\, \omega^{-n}|B_0|^2 - \xi\right]
\ ,
\label{model21_mastereq}
\end{align}
where $\omega\equiv s s^\dag$.

Before going into details, let us make a comment. If we fix $A \equiv
0$ ($B\equiv0$) everywhere, we can think of the system as just the
Abelian-Higgs model with one complex scalar field $B$ ($A$) whose
$U(1)$ charge is $n$ ($m$). 
The vortices there are the normal ANO solutions, though the $k$-vortex
solutions will have the $U(1)$-winding number $k/n$ ($k/m$) with
tension $T_B = 2\pi \xi k/n$ ($T_A = 2\pi \xi k/m$).
Indeed, when only one field is active while  the other is inert,
 the $U(1)$ gauge coupling constant and the FI term can be rescaled 
such that the system looks exactly as the standard Abelian-Higgs model with unit  $U(1)$
charge.   What we are trying to study in this section is an intermediate situation
between two kinds of vortices where both fields contribute non-trivially. 
Such intermediate states should have the energy  $T \equiv m T_A = n T_B$, and  we shall see configurations which have
$m$ peaks in one limit and $n$ peaks in another limit.

First we choose a generic point such as $A_{\rm vev} \neq 0$ and
$B_{\rm vev} \neq 0$. 
The moduli matrices behave asymptotically as follows
\beq
A_0(z) = s^m A \to |z|^{m\nu} e^{im\nu\theta} A_{\rm vev}\ ,\quad
B_0(z) = s^n B \to |z|^{n\nu} e^{in\nu\theta} B_{\rm
  vev}\ ,\qquad\text{as}\quad |z| \to \infty \ .
\eeq
Holomorphy of $A_0,B_0$ requires 
$m\nu \in \mathbb{Z}_+$ and $n\nu \in \mathbb{Z}_+$.
 As we have chosen $m$ and $n$ to be relatively prime, this is satisfied by
$\nu \equiv k \in \mathbb{Z}_+$. 
Thus we have obtained the non-trivial condition for $A_0,B_0$
\beq
\nu = k:\quad
A_0(z) = A_{\rm vev} z^{mk} + {\cal O}(z^{mk-1})\ ,\quad
B_0(z) = B_{\rm vev} z^{nk} + {\cal O}(z^{nk-1})\ .
\eeq
Note that $k$ vortices have $(m+n)k$ moduli parameters with the
boundary vacuum modulus. 
They may correspond to positions and sizes of the fractional vortices. 

When we choose the special point $A_{\rm vev} = 0$ ($\mathbb{Z}_m$
fixed point) or $B_{\rm vev} = 0$ ($\mathbb{Z}_n$ fixed point) as a
boundary condition, the conditions for the moduli matrix drastically
change. 
Say $|A_{\rm vev}| = \sqrt{\xi/m}$ and  $B_{\rm vev} = 0$.
Immediately we get $\nu = k/m$ and the conditions
\beq
\nu = \frac{k}{m}:\quad
A_0 = \sqrt{\frac{\xi}{m}}\, z^k + \cdots,\quad
B_0 = b z^{\beta} + \cdots,
\eeq
where $\beta$ is a semi-positive definite integer less than 
$n\nu = \frac{n}{m}k$.
If we set $B_0 = 0$, the solution is identical to the ANO vortex as we
mentioned before. 
When $B_0$ is not zero, the solutions significantly differ from the
ANO solution and also from the semi-local vortices in EAH model.
Similarly, if we choose $|B_{\rm vev}| = \sqrt{\xi/n}$ and 
$A_{\rm vev} = 0$, the $U(1)$ winding number
becomes $\nu = k/n$ and the conditions change as 
\beq
\nu = \frac{k}{n}:\quad
A_0 = a z^\alpha + \cdots,\quad
B_0 = \sqrt{\frac{\xi}{n}}\, z^k + \cdots,
\eeq
where $\alpha$ is a semi-positive definite integer less than $m\nu =
\frac{m}{n}k$. 
Note that the $U(1)$ charge $\nu$ is fractionally quantized at the
conical singularities. 
The present model thus nicely illustrates the first mechanism for the fractional vortices
discussed in the previous section.

\begin{figure}[ht]
\begin{center}
\includegraphics[width=2.5in]{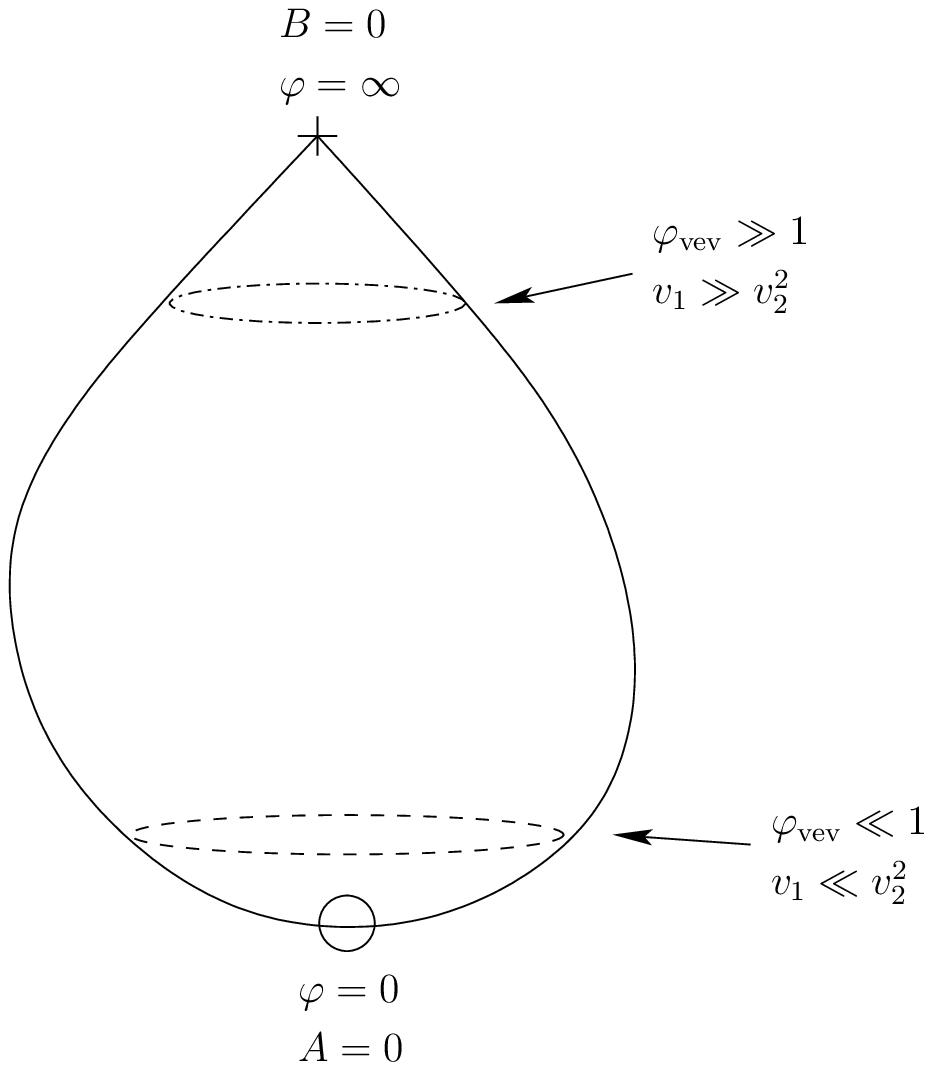}
\caption{\footnotesize{A sketch of the $\mathbb{C}P^1/\mathbb{Z}_2$ which is
   a sphere with a conical singularity at a  (i.e., north)
    pole. }}
\label{fig:cp1dz2}
\end{center}
\end{figure}

From this point on,  we shall concentrate on the special  concrete case $m=2$ and
$n=1$,  in order to illustrate in detail the properties of a
fractional vortex.  The vacuum moduli 
${\cal M} = W\mathbb{C}P^1_{(2,1)} \simeq \mathbb{C}P^1/\mathbb{Z}_2$
has a $\mathbb{Z}_2$ conical singularity
at $B_{\rm vev} = 0$ (north pole), see Fig.~\ref{fig:cp1dz2}. 
We consider the minimal-energy vortex  configuration ($k=1$).
When we choose a generic point ($B_{\rm vev} \neq 0$) as the boundary condition, 
the minimal configuration has $U(1)$ winding $\nu=1$ whose energy is
\beq
T_{\rm min} = 2\pi\xi \ ,\quad(\nu=1) \ .
\label{eq:winding_u1}
\eeq
The corresponding moduli matrix is given by
\beq
A_0(z) = A_{\rm vev} z^2 + a_1 z + a_2\ ,\quad 
B_0(z) = B_{\rm vev} z + b_1 \ ,\qquad 
a_1,a_2,b_1 \in \mathbb{C}^3\ ,
\label{eq:mn_gene}
\eeq
with $2 |A_{\rm vev}|^2 + |B_{\rm vev}|^2 = \xi$.
Although this is the minimal-energy configuration, 
we have three complex moduli parameters $a_1,a_2,b_1$.
Remember that $A$ ($B$) is zero at a point where $A_0$ ($B_0$) is zero.
Note that $A_0$ has two zeros and $B_0$ has one zero because $A$ winds
twice and $B$ winds once when we go around the boundary, $S^1$ at
spatial infinity.  
An important observation is that the $U(1)$ gauge symmetry is not
generally recovered at the zeros. 
Only when $A$ and $B$ vanish
simultaneously, the $U(1)$ gauge symmetry 
is recovered (this would happen  if  some of the zeros of $A_0$ and $B_0$ are
coincident).

Consider now the vortex at the special point of the vacuum moduli,   $B_{\rm vev} = 0$.
The minimal configuration $k=1$ corresponds to $\nu=1/2$ and has a
tension 
\beq
T_{\rm min}^{\rm special} = \pi \xi \ ,\quad(\nu = 1/2) \ .
\eeq
The moduli matrix takes the form (for $k=1$) 
\beq 
A_0 = A_{\rm vev} z + a \ ,\quad
B_0 = b \ ,\qquad A_{\rm vev} = \sqrt{\xi/2} \ ,
\label{Vortex1bis}
\eeq
where $a,b$ are the moduli parameters. 
Comparing this with Eq.~(\ref{eq:mn_gene}) with $B_{\rm vev} = 0$,
one immediately sees that the latter is not a minimal-energy solution. 

\begin{figure}[h!tp]
\begin{center}
\begin{tabular}{ccl}
\includegraphics[width=9.5cm]{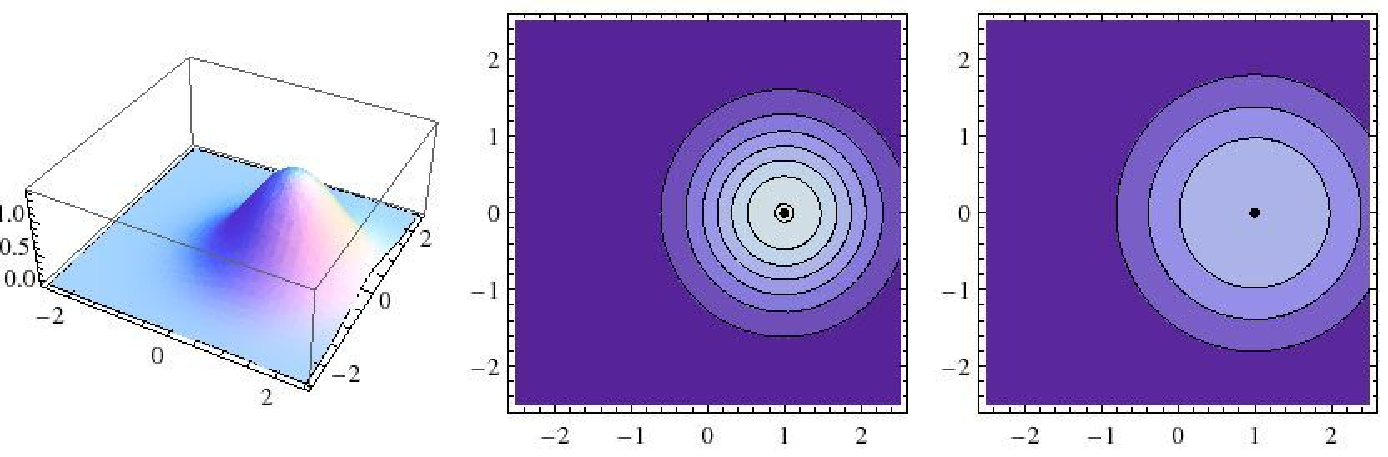} && \includegraphics[height=2.8cm]{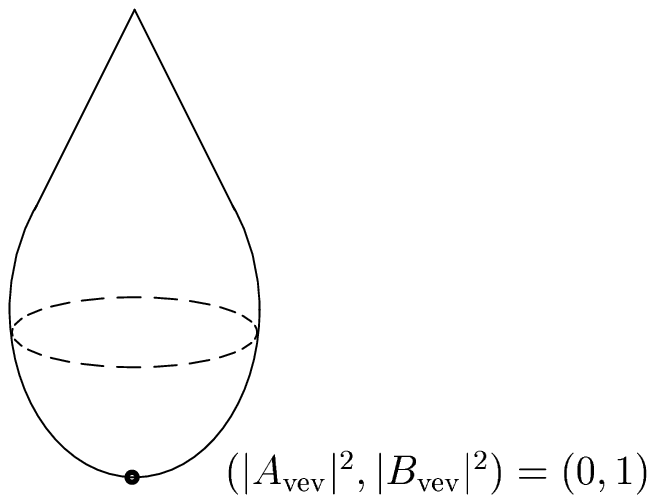}\\
\includegraphics[width=9.5cm]{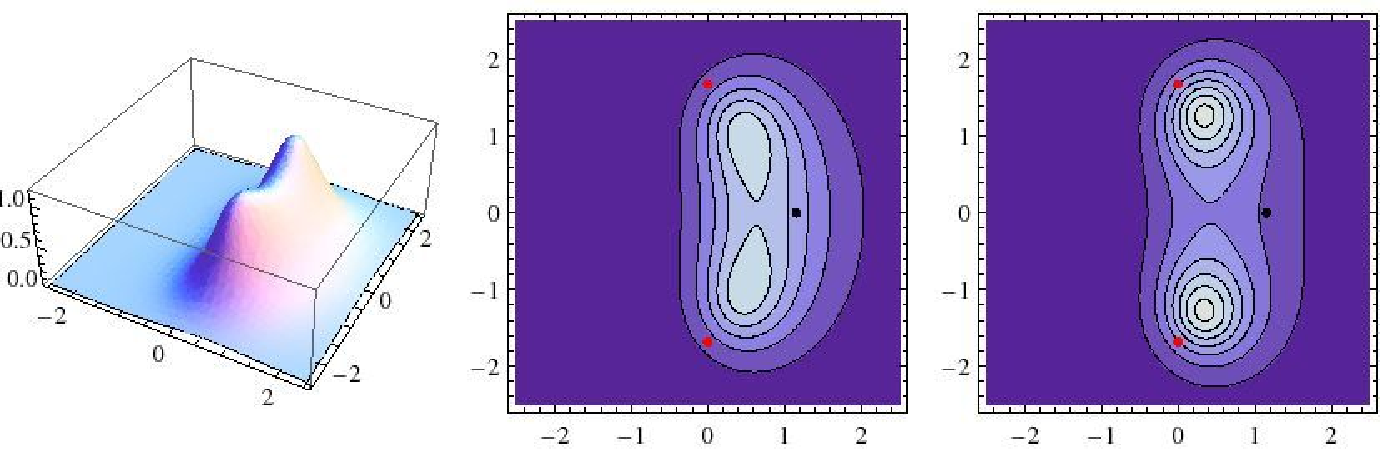} && \includegraphics[height=2.8cm]{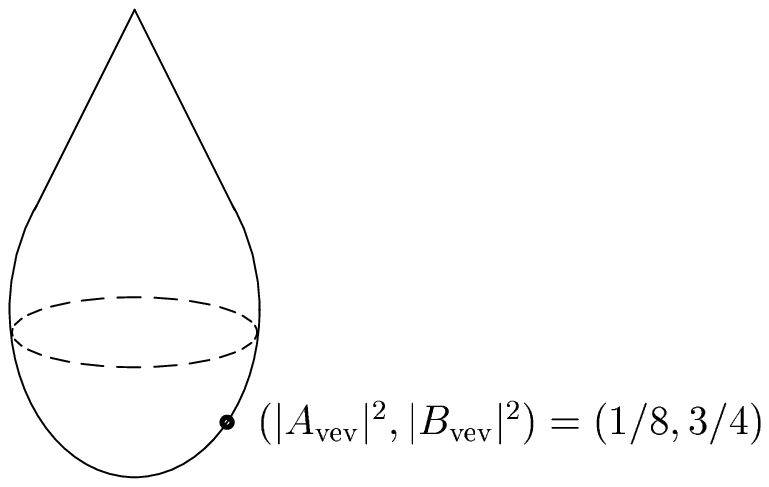}\\
\includegraphics[width=9.5cm]{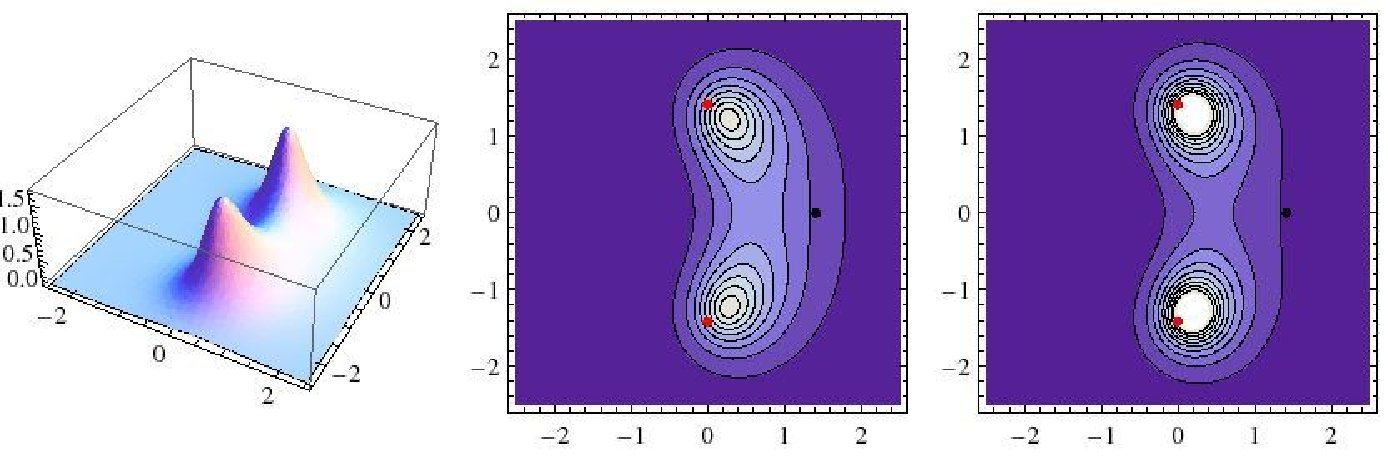} && \includegraphics[height=2.8cm]{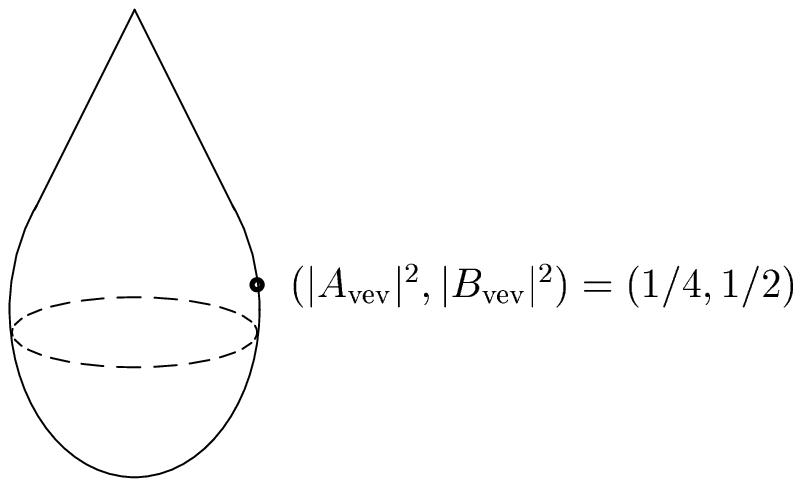}\\
\includegraphics[width=9.5cm]{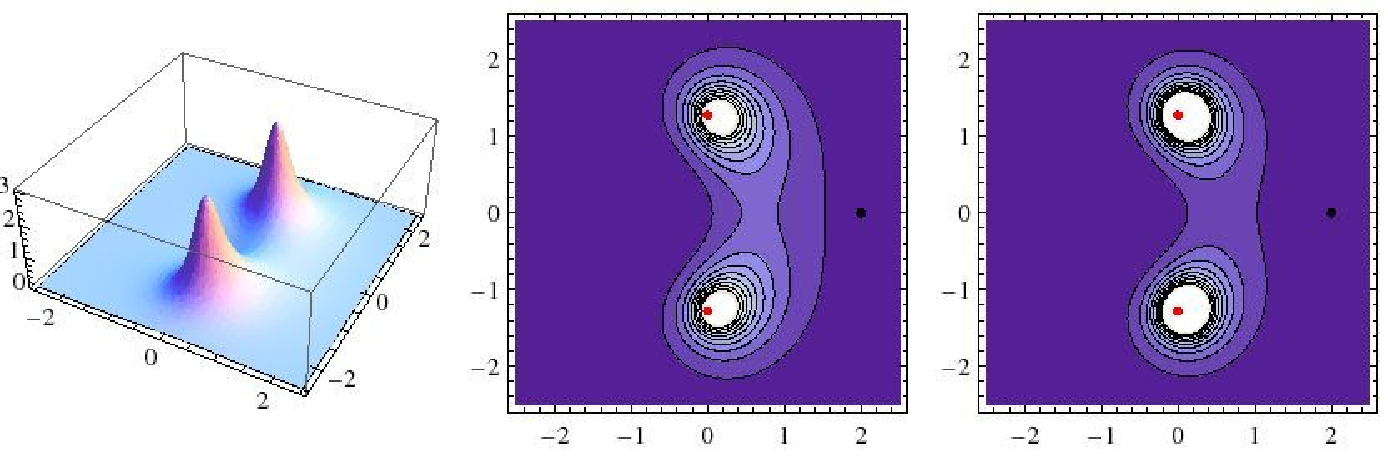} && \includegraphics[height=2.8cm]{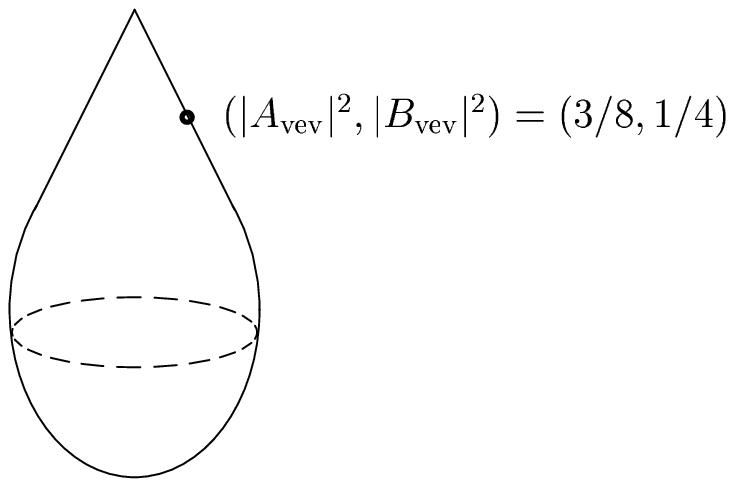}\\
\includegraphics[width=9.5cm]{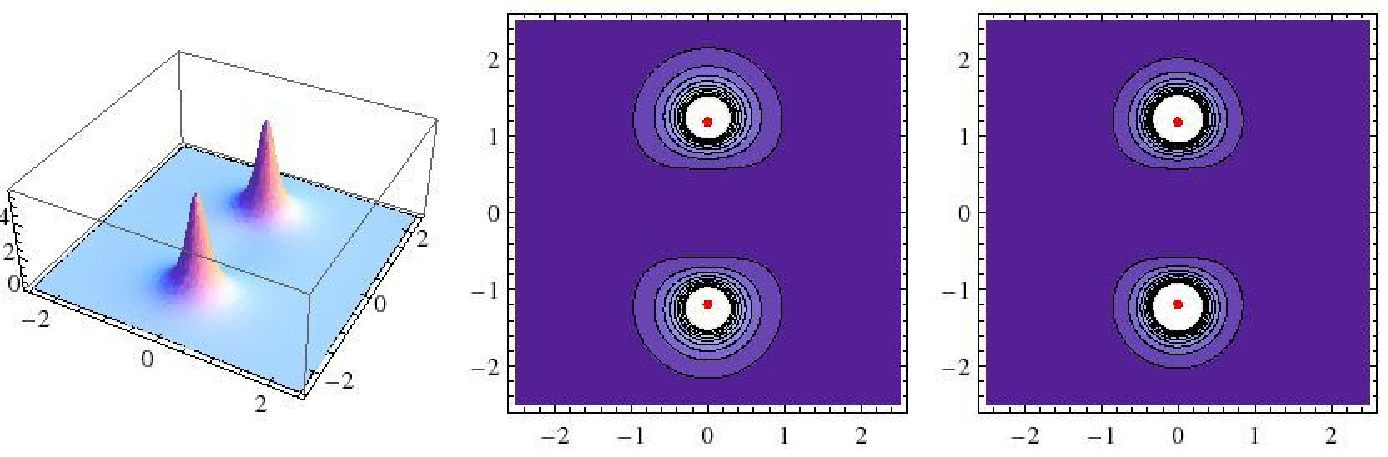} && \includegraphics[height=2.8cm]{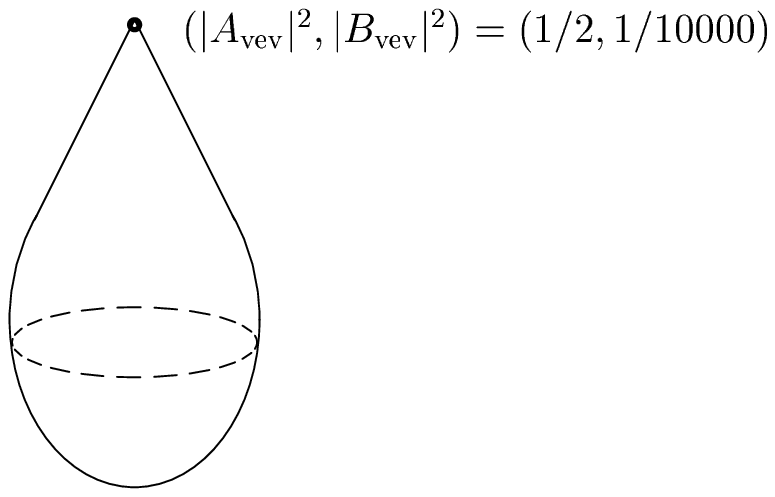} \\
\includegraphics[width=9.5cm]{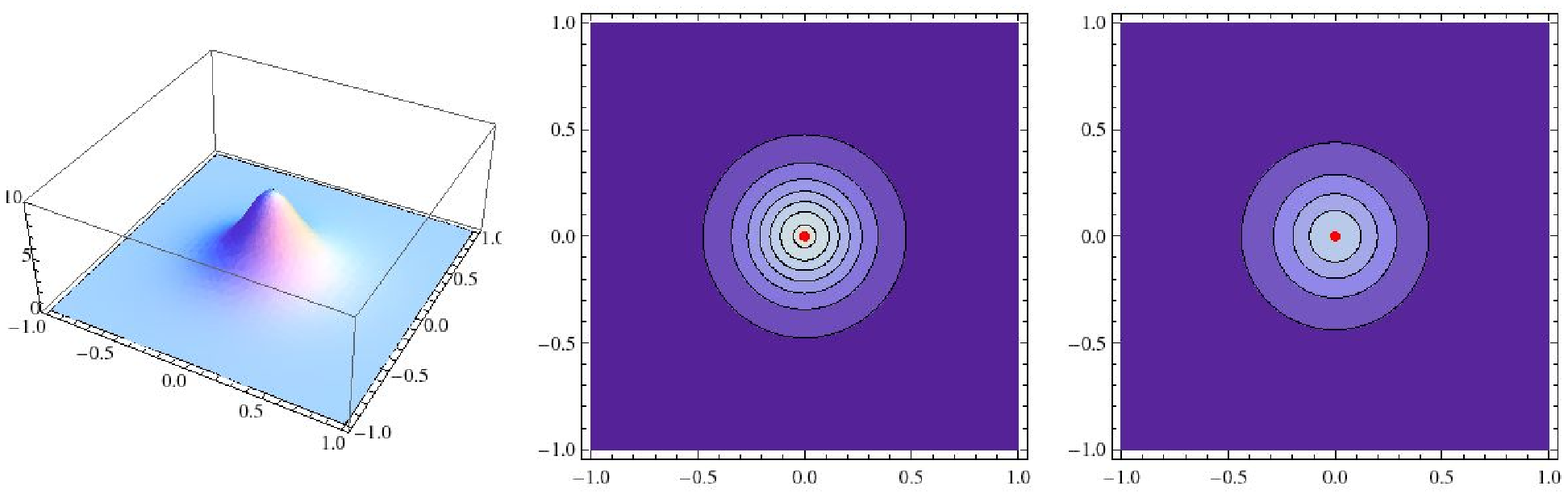} && \includegraphics[height=2.8cm]{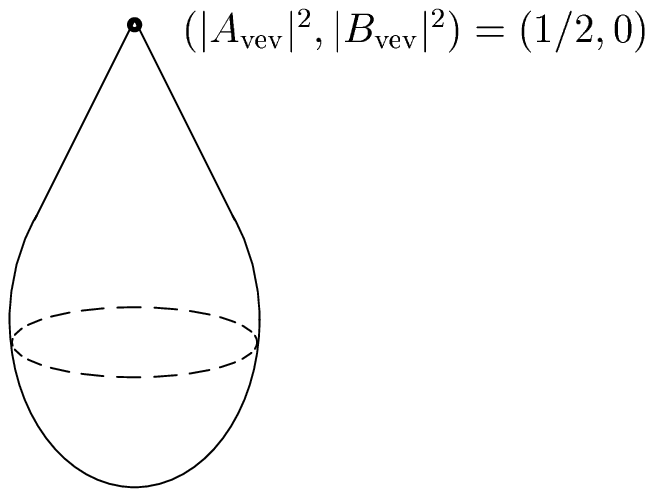}
\end{tabular}
\caption{\footnotesize{
  The energy (the left-most and the 2nd left panels) and
 the  magnetic flux  (the 2nd right panels) density are shown, together with
  the boundary values $(A,B)$ (the right-most panels) for the
  minimal lump of the first type in the strong gauge coupling limit.   
  The moduli parameters are fixed as $a_1=0,a_2=1,b_1=-1$ in
  Eq.~(\ref{eq:mn_gene}). The red dots are zeros of $A$ and the
  black one is the zero of 
  $B$.   $\xi=1$.  The last figures illustrates the minimum lump defined at exactly the orbifold point
 (see Eq.~(\ref{Vortex1bis}))  with $A_{vev}= 1/\sqrt{2}$,  and with $b=0.8$. 
  }}
\label{FigMod21}
\end{center}
\end{figure}

The vortex (energy, and magnetic flux) profiles can be approximately 
determined from the strong-coupling limit consideration.
The gauge theory reduces to the non-linear sigma model whose target
space is the vacuum moduli ${\cal M}$ in Eq.~(\ref{eq:mn_vac}). The
K\"ahler potential is given, in the supersymmetric version of our model,  by 
\beq
K = |A|^2 e^{-2V} + |B|^2e^{-V} + \xi V \ .
\label{eq:kahler_21}
\eeq
Integrating out the $U(1)$ vector multiplet $V$, we get the following
K\"ahler potential in terms of an inhomogeneous coordinate:
$\varphi = 2\sqrt{\xi}A/B^2$ 
\beq
K = \xi \log f(\varphi,\bar\varphi) + \xi f^{-1}(\varphi,\bar\varphi)
\ ,\quad
f(\varphi,\bar\varphi) \equiv 1 + \sqrt{1 + 2|\varphi|^2} \ .
\label{eq:kahler_cp1}
\eeq
Note that the first term is due to the magnetic flux $F_{12}$ and the
second term corresponds to the surface term $\p_i^2J$ in
Eq.~(\ref{eq:energy_form}). 
All the regular BPS solutions are analytically solved by
\beq
2 \xi |s|^2 
= |B_0|^2 +\sqrt{|B_0|^4 + 8 \xi |A_0|^2}
= |B_0|^2 f(\varphi,\bar\varphi) \ ,\quad
\varphi = \varphi(z) = \frac{2\sqrt{\xi} A_0(z)}{B_0(z)^2} \ .
\eeq
Only the solutions which have points where $A$ and $B$ simultaneously
vanish cannot be seen in this limit, because the $U(1)$ gauge symmetry
would remain unbroken there. Such solutions contain small lump
singularities and we should go back to the original gauge theory in
order to observe the configurations correctly. 
A numerical result is shown in Fig.~\ref{FigMod21}. 

As we move in the vacuum moduli space  ${\cal M}$ by varying the VEVs
$A_{\rm vev}, B_{\rm vev}$ 
(or $\varphi_{\rm vev}\equiv 2 \sqrt{\xi} A_{\rm vev}/B_{\rm vev}^2$)
 and  change the vortex moduli parameters  the
tension density  profile shows varying substructures. 
Since the zeros of the fields do not imply necessarily the restoration of a $U(1)$ gauge
symmetry, the positions of the peaks do not always coincide with the zeros of
$A,B$.    Although it is very complicated to specify the positions of peaks
analytically, it is easy to visualize it numerically. In Fig.~\ref{FigMod21},
we have shown the zeros of $A,B$ and the peaks.
We observe that there are no direct relations between the zeros of
fields and the positions of the peaks, except at   
the two poles, $A_{\rm vev}=0$ (south pole) and
$B_{\rm vev}=0$ (north pole), of the space $\cal M$.

An axially symmetric peak appears at the zero $z=z^{\rm S}$ of $B_0(z)$  
in the limit $A_{\rm vev} \to 0$;  as 
 $A_{\rm vev}$ departs from $0$, it decomposes into two sub-peaks. 
We cannot remove  one of the two sub-peaks pushing its position to
infinity. This feature can be easily observed 
for large $|\varphi_{\rm vev}|\gg 1$
where positions of the two peaks are naturally approximated by 
the zeros $z=z_i^{\rm S}(i=1,2)$ of $A_0(z)$.
(Here $A_{\rm vev}(z_1^{\rm S}+z_2^{\rm S})=-a_1, 
A_{\rm vev}z_1^{\rm S}z_2^{\rm S}=a_2$, $B_{\rm vev}z^{\rm N}=-b_1$.)
The energy density $E=2\partial \bar \partial K$ at those points
is given by
\begin{eqnarray}
 E|_{z=z^{\rm S}_i}
&=&2\xi |\varphi_{\rm vev}|^2
\frac{|z^{\rm S}_1-z^{\rm S}_2|^2}{|z^{\rm S}_i-z^{\rm N}|^4}.
\end{eqnarray}
For instance, if the three zeros get separated by large distances,
then we see that the sub-peaks are diluted.
If, instead,  only one of the zeros, $z=z_2^{\rm N}$ is pushed toward 
infinity, that is $|z_2^{\rm N}-z^{\rm S}|, |z_1^{\rm N}-z_2^{\rm N}|
\gg|z_1^{\rm N}-z^{\rm S}|$, 
the peaks at $z=z_1^{\rm N}$ becomes singular.
In either case,  the isolated one peak is not allowed as 
a vortex(lump) solution.
This solution consisting the two sub-peaks 
is one of typical examples of fractional vortices.
Only when $B_{\rm vev} = 0$, they become independent. Such a limiting configuration is  no longer a minimal
energy configuration,  however.   The minimal configuration at exactly $B_{\rm vev} = 0$   (with 
(\ref{Vortex1bis}))  has only one peak.    Its tension is
half of the minimal configuration for 
$B_{\rm vev} \neq 0$. 

The reason why the minimum vortex at  $B_{\rm vev} \neq 0$  must have twice the energy
with respect to the minimal object  at $B_{\rm vev} = 0$  is as
follows.  Our vacuum moduli has a $\mathbb{Z}_2$ singularity at 
$|B_{\rm vev}| = 0$. If  the vacuum is chosen at  $B_{\rm vev} \neq 0$   the solution touches the singularity at a
finite point in the $z$-plane and would get singular there.  To remove such a singularity, the solution must 
wrap twice around the vacuum moduli.  On the other hand, 
if   one is at exactly  the $\mathbb{Z}_2$ point  the solution never touches it and  a regular
solution can be constructed with just a single winding. 

As discussed in Section~\ref{Sec:Structures}, these characteristics of
the vortex-energy profile are thus  deeply rooted in the property of the
vacuum moduli ${\cal M}$ itself and to its singularity structure.  
In understanding the qualitative features of the vortex tension
distribution and their dependence on 
$p = \varphi_{\rm vev}= 2\sqrt{\xi}A_{\rm vev}/B_{\rm vev}^2$  
just described, 
the crucial fact is that the elements of the homotopy groups 
corresponding to regular configurations make a discontinuous jump at 
$p=\varphi=\infty$ ($B_{\rm vev} = 0$). 
In fact
\beq \frac{\pi_{2}\left({\cal M}, p\right)}
{\pi_{2}\left({\cal M}, \infty\right)} ={\mathbb Z}_{2} \ , \qquad 
\frac{\pi_{1}\left(F, f\right)}{\pi_{1}\left(F, f_{0}\right)}=
{\mathbb Z}_{2}\ , \qquad  p \ne \infty \ . \eeq
The $S^1$ fiber itself reduces to half at the orbifold singularity
\beq f = \pi^{-1}(p) = S^{1}\ , \qquad  
f_{0}= \pi^{-1}(\infty) = S^{1}/{\mathbb Z}_{2}\ . \eeq
Thus even though $\pi_{2}(M)=\mathbbm{1}$ and 
$\pi_{1}(M)=\mathbbm{1}$, just as in the case of the EAH model
Eq.~(\ref{SeqEAH}),  each minimum  element of $\pi_{2} ({\cal M}, p) $
is  a double cover of the minimum element of $\pi_{2}({\cal M},
\infty)$, just as the fiber at the generic $p$ ($S^{1}$:  $\alpha=0
\to 2 \pi$ in Eq.~(\ref{fiber})) is a double cover of the 
fiber at  $B=0$  ($\alpha=0 \to \pi$ in Eq.~(\ref{fiber})). 
This is the (global) reason for the double peaks observed
in Fig.~\ref{FigMod21}.

The argument here can be easily extended to more general cases with
the multiple flavors  
$H = (A,B,C,D,\cdots)$ with generic $U(1)$ charges $Q = (m,n,o,q,\cdots)$, which are all relatively prime.
The moduli manifold is then 
${\cal M} = \mathbb{C}P^{\NF-1}_{(m,n,\cdots)} \simeq \mathbb{C}P^{\NF-1}/(\mathbb{Z}_m \times \mathbb{Z}_n \times \cdots)$.
Near a $\mathbb{Z}_m$ singular point,
$(|A_{\rm vev}|,|B_{\rm vev}|,|C_{\rm vev}|,\cdots) = (\sqrt{\xi/m},0,0,\cdots,)$   $m$ peaks appear  in the energy
distribution. 

\subsection{An Abelian  $U(1) \times U(1)$ Higgs model\\
-- fractional vortex of the second type --
\label{sec:simplemodel}
}

The next system we consider,  which has the same target space as  in
the previous model,   is a $U(1)_1\times U(1)_2$ gauge theory  with three flavors of scalar
electrons $H = (A,B,C)$ with charges $Q_1 = (2,1,1)$ for 
$U(1)_1$ and $Q_2 = (0,1,-1)$ for $U(1)_2$. The gauge transformations act as
\beq
(A,B,C) \to \left(e^{i 2 \alpha(x)} A, e^{i\alpha(x) + i\beta(x)} B,
e^{i\alpha(x)-i\beta(x)} C\right) \ .
\eeq
Note that the transformation $(\alpha,\beta) = (\pi , \pm \pi)$ leaves the fields invariant:  the true gauge group is
$$ [U(1)_1 \times U(1)_2]/\mathbb{Z}_2. $$
Some details about this model are given in the Appendix.
The vacuum manifold and vacuum moduli space are 
\beq
M &=& \{A,B,C \quad |\quad 2|A|^2 + |B|^2 + |C|^2 = \xi_1,\ |B|^2 -
|C|^2 = \xi_2 \}\ ,\\
{\cal M} &=& M/ \left[\left( U(1)_1 \times
  U(1)_2\right)/\mathbb{Z}_2\right] \ .
\eeq
Here $\xi_1$ is the FI term for the first $U(1)_1$ and $\xi_2$ is that
for the second $U(1)_2$. Note that 
the existence of the supersymmetric vacuum requires $\xi_1\ge |\xi_2|$.
Below we shall mainly be interested in the case of $\xi_2 = 0$ 
(in the Appendix  the case with a non-vanishing  $\xi_2$ is also discussed).

\begin{figure}[ht]
\begin{center}
\includegraphics[height=5cm]{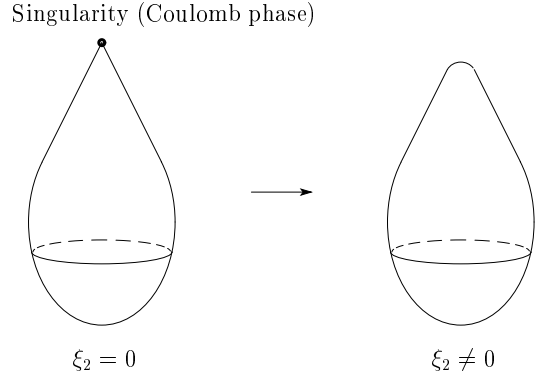}
\caption{\footnotesize{A sketch of the vacuum moduli space.}}
\label{fig:drop2}
\end{center}
\end{figure}

When $\xi_2=0$, the vacuum manifold is the same as one in
Section \ref{sec:21model}.  In fact, the K\"ahler potential of the vacuum manifold is the same as
Eq.~(\ref{eq:kahler_cp1}) 
with replacements $\xi \to \xi_1$ and 
$\varphi = 2\sqrt{2\xi}A/B^2 \to 2 \sqrt{2\xi_1}A/(BC)$.
Although the vacuum moduli manifolds are the same, there is an
important difference.  The singular point in the previous section was
a $\mathbb{Z}_2$ conical (orbifold) singularity whereas 
the singular point $|B| = |C| = 0$ here represents a theory with a restored
 $U(1)_2$ gauge symmetry. i.e.,  in a Coulomb phase. 
Since the $\mathbb{Z}_2$ action has been modded out from the
beginning, the singular point is not a $\mathbb{Z}_2$ fixed point
and therefore can be smeared by the introduction of  non-vanishing
$\xi_2$ as illustrated in Fig.\ref{fig:drop2}.  (See however the discussions below.)

BPS vortex solutions can be treated by the moduli matrix formalism.
The tension is determined by the $U(1)_1$ winding number $\nu_1$
(and the $U(1)_2$ winding number $\nu_2$ if $\xi_2 \neq 0$) 
as
\begin{align}
{\cal E} &= - \xi_1 F_{12}^{(1)} - \xi_2 F_{12}^{(2)} + \p_i^2 J\ ,
\qquad 
J = \frac{1}{2}HH^\dagger\ ,
\label{tensionmod02} \\
T &= \int dx^2\, {\cal E} = 2\pi \xi_1 \nu_1 + 2\pi \xi_2 \nu_2\ ,
\label{tensionmod2} 
\end{align}
In the moduli-matrix formalism, the winding number can be expressed as
\begin{align}
\nu_I &= - \frac{1}{2\pi} \int dx^2\, F_{12}^{(I)} = \frac{1}{\pi}\int
dx^2\, \p\bar\p\log|s_I|^2\ ,\quad
\bar W_I = - i\bar\p\log s_I\ ,\\
H &= (A,B,C) = \left( s_1^{-2} A_0(z),\ s_1^{-1}s_2^{-1}
B_0(z),\ s_1^{-1}s_2 C_0(z) \right)\ ,\qquad
\end{align}
where $s_1$ and $s_2$ are everywhere non-zero functions and the holomorphic functions $A_0,B_0,C_0$ are the elements of the
moduli matrix. 
The BPS equation for the gauge fields reduces to the 
master equations
\begin{align}
\bar{\partial}\partial\log\omega_1 &= -\frac{e^2}{4}
\left[\omega_1^{-1}\left(2\omega_1^{-1}|A_0|^2 + \omega_2^{-1}|B_0|^2
  + \omega_2|C_0|^2\right)-\xi_1\right] \ , \\
\bar{\partial}\partial\log\omega_2 &= -\frac{g^2}{4}
\left[\omega_1^{-1}\left(\omega_2^{-1}|B_0|^2 - \omega_2|C_0|^2\right)
  -\xi_2\right] \ ,
\label{u1u1model_mastereqs}
\end{align}
where $\omega_i\equiv s_is_i^\dag$,  $i=1,2$.

To specify a solution, we need to choose  the winding numbers $\nu_1$
and $\nu_2$. 
The condition for $\nu_1$ and $\nu_2$ depends on the boundary condition 
$(A,B,C) \to (A_{\rm vev},B_{\rm vev},C_{\rm vev}) \in M$ as $|z| \to \infty$.
So we first need to determine $(A_{\rm vev},B_{\rm vev},C_{\rm vev})$,
(see the Appendix). 
It should be noted that, when $\xi_2 =0$,  although the magnetic flux $F_{12}^{(2)}$ does not contribute to the tension  it contributes to the tension {\it density} through the surface
term, $J$ of Eq.~(\ref{tensionmod02}). 
In other words  $\nu_2$  as well as   $\nu_1$  is needed to determine a solution. 
 $F_{12}^{(2)}$ is non-trivial even when $\xi_2 = 0$. 

Let us concentrate on the minimal energy configuration viz.~$\xi_2 = 0$
in the following. 
We first choose a generic point ($|B_{\rm vev}|=|C_{\rm vev}| \neq 0$)
as the boundary condition. 
The minimal configuration is given by two different choices of the
winding numbers 
(the generic tension formula is given in
Eq.~(\ref{eq:tenxion_2U(1)_1})) 
\beq
T_{\rm min} = \pi \xi_1\ ,\quad{\rm with}\quad
(\nu_1,\nu_2) = \left(1/2,\pm 1/2\right)\ .
\label{eq:winding_u1u1}
\eeq
The corresponding moduli matrix is given by
\begin{align}
A_0(z) &= A_{\rm vev} z + a\ ,\quad
B_0(z) = B_{\rm vev} z + b\ ,\quad
C_0(z) = C_{\rm vev}\ ,\quad \text{for}\quad 
(\nu_1,\nu_2) = \left(1/2,1/2\right)\ ,
\label{eq:mm_u1u1_mod}\\
A_0(z) &= A_{\rm vev} z + a\ ,\quad
B_0(z) = B_{\rm vev} \ ,\quad
C_0(z) = C_{\rm vev}z + c\ ,\quad \text{for}\quad 
(\nu_1,\nu_2) = \left(1/2,-1/2\right)\ .\nonumber
\end{align}
The two complex parameters ($a,b$) represent the moduli parameters.

At exactly  $B_{\rm vev} = C_{\rm vev} = 0$ the $U(1)_2$ gauge symmetry
is restored at infinity: 
the system is in a Coulomb phase.  
We shall not discuss this case:  it is beyond the reach of the moduli-matrix formalism.
On the contrary, at an orbifold singularity such as those considered in the previous subsection the system remains in a
Higgs phase even though with a different property from the surrounding vacua.

\begin{figure}[!th]
\begin{center}
\begin{tabular}{ccl}
\includegraphics[width=9cm]{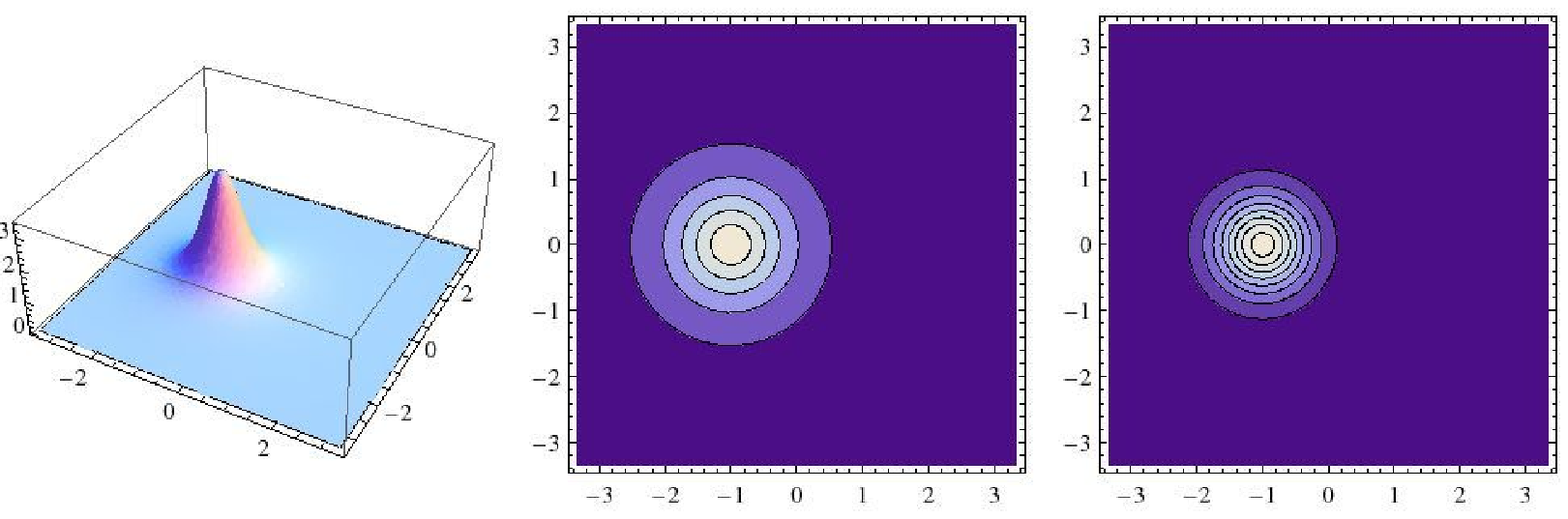} && \includegraphics[height=3cm]{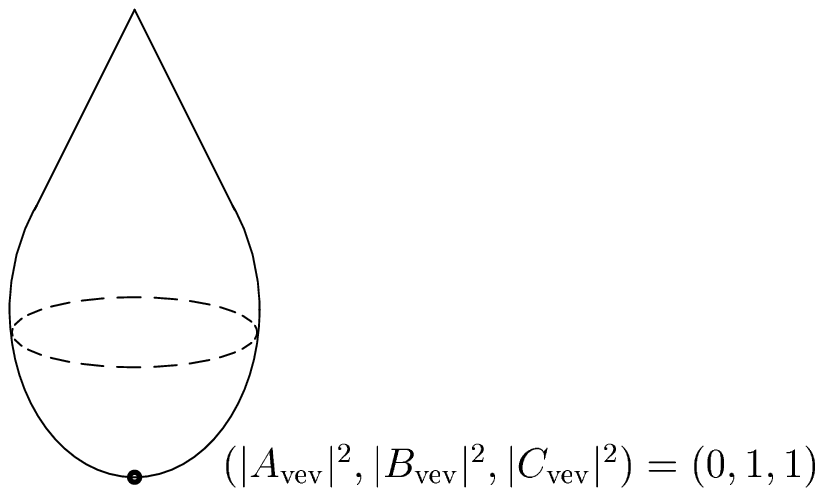}\\
\includegraphics[width=9cm]{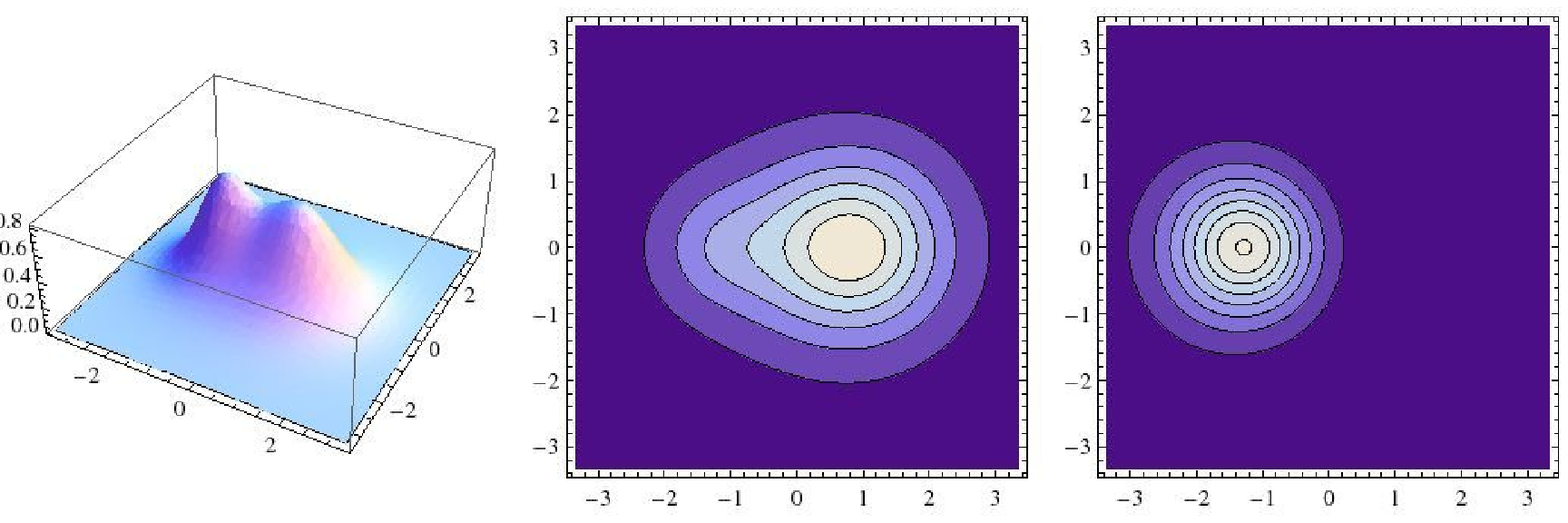} && \includegraphics[height=3cm]{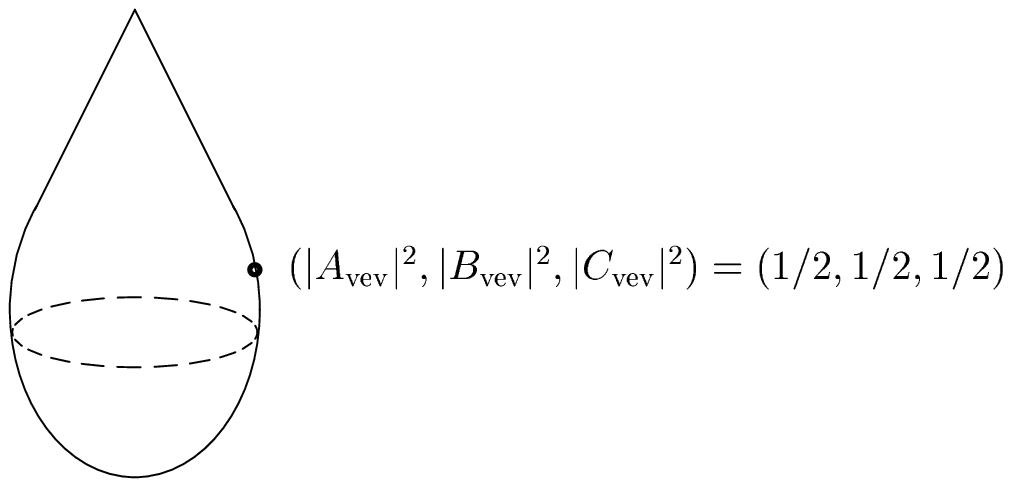}\\
\includegraphics[width=9cm]{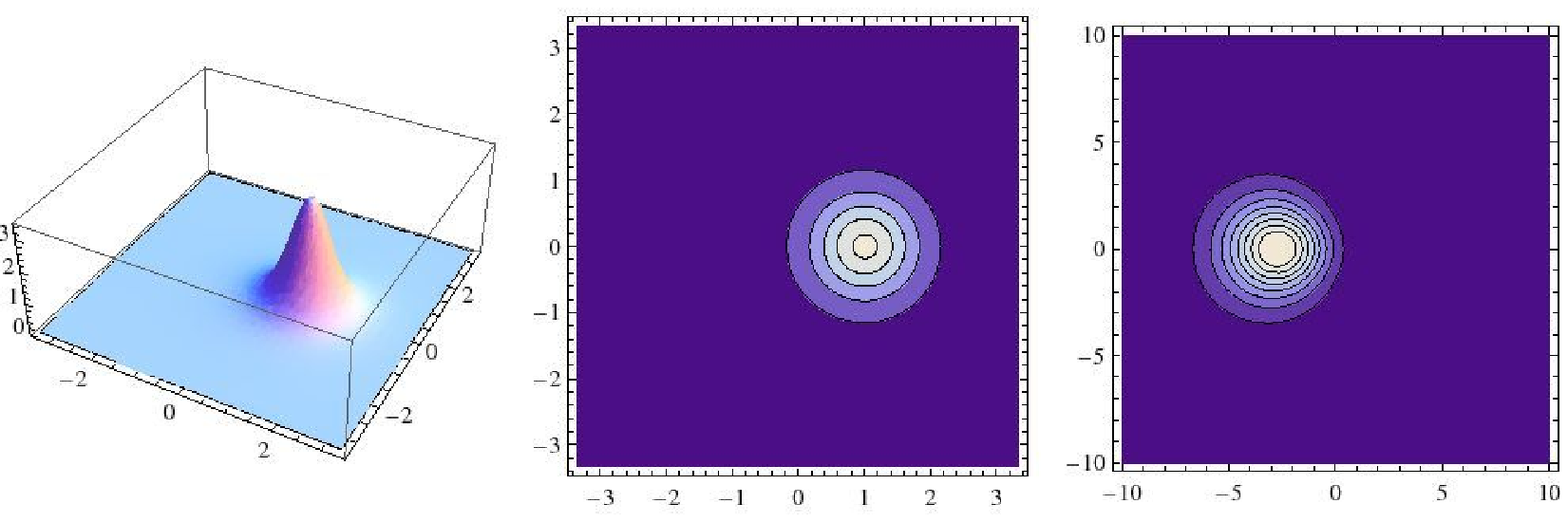} && \includegraphics[height=3cm]{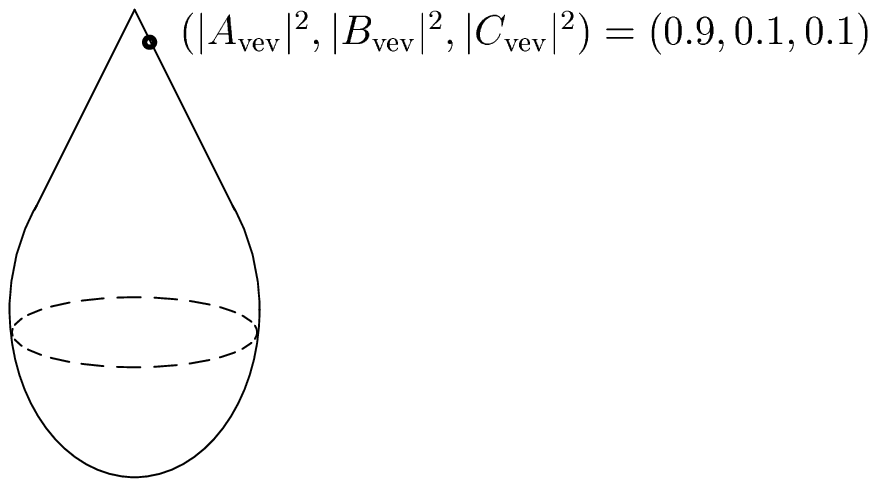}
\end{tabular}
\caption{\footnotesize{The energy density (left-most) and the magnetic
    flux density $F_{12}^{(1)}$ (2nd from the left), $F_{12}^{(2)}$
    (2nd from the right) and the boundary condition (right-most) are
    shown for
    the vortex of the second type. 
    We have chosen $\xi_1=2$, $\xi_2=0$, $e_1 = 1$, $e_2=2$ and
    $a=-1,b=1$ in Eq.~(\ref{eq:mm_u1u1_mod}). }}
\label{fig:u1u1_finite}
\end{center}
\end{figure}

Numerical solutions with various choices of 
$A_{\rm vev}$, $B_{\rm vev}$ and $C_{\rm vev}$ are 
shown in Fig.~\ref{fig:u1u1_finite}.
When a point $|B_{\rm vev}| = |C_{\rm  vev}| \simeq 0$ is chosen as
the boundary condition, one peak is observed near $z_a=-a/A_{\rm vev}$
where $A(z_a) = 0$. 
On the other hand, another single peak appears near 
$z_b = - b/ B_{\rm vev}$ where $B(z_b) = 0$,
when we choose $|A_{\rm vev}| = 0$ as the boundary condition.
These two peaks are smeared in the intermediate values of 
$(A_{\rm vev},B_{\rm vev},C_{\rm vev})$.
Although the vacuum manifold is the same as in Section~\ref{sec:21model},
the vortex solutions are significantly different. The difference can
be clearly seen in the winding numbers and tensions 
in Eqs.~(\ref{eq:winding_u1}) and (\ref{eq:winding_u1u1}).
In the model of Section~\ref{sec:21model}, the $\mathbb{Z}_2$ quotient
requires us to choose $\nu=1$ 
except for at the $\mathbb{Z}_2$ fixed point. However, exactly in that
point there are no such discrete symmetries, so there exist solutions
with $\nu_1 = 1/2$. On the contrary, in the model of this section, we
cannot choose the singular point (in the Coulomb phase) as the
boundary condition.

\begin{figure}[!th]
\begin{center}
\begin{tabular}{ccl}
\includegraphics[width=9cm]{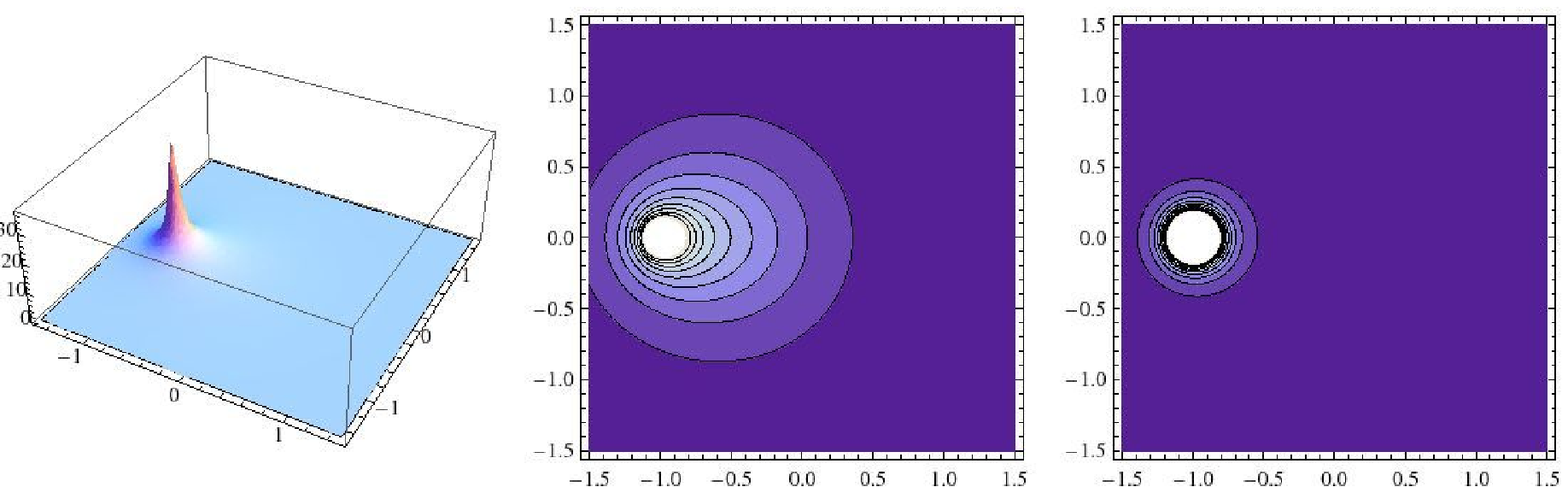} && \includegraphics[height=3cm]{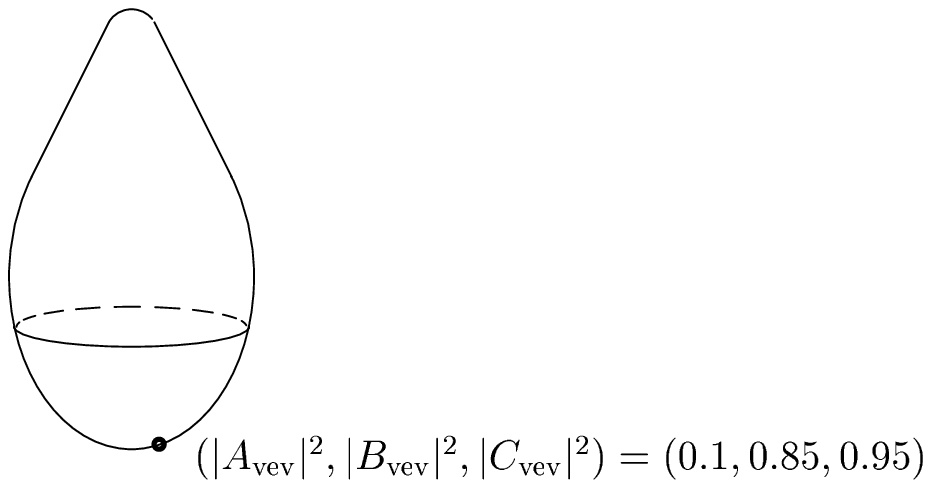}\\
\includegraphics[width=9cm]{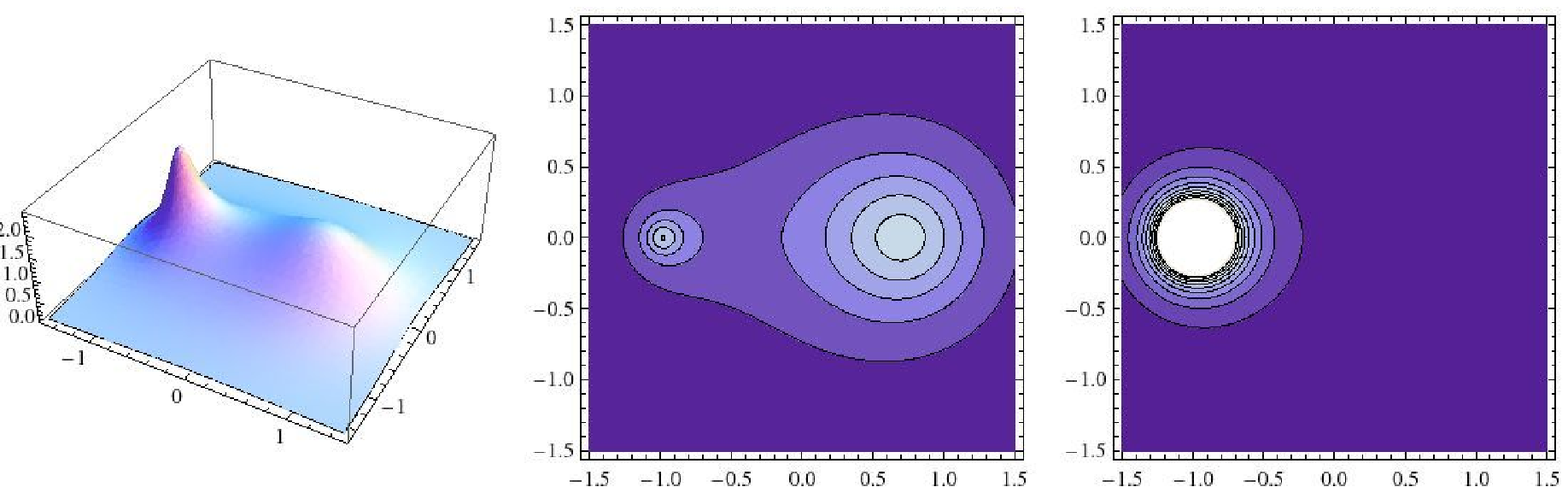} && \includegraphics[height=3cm]{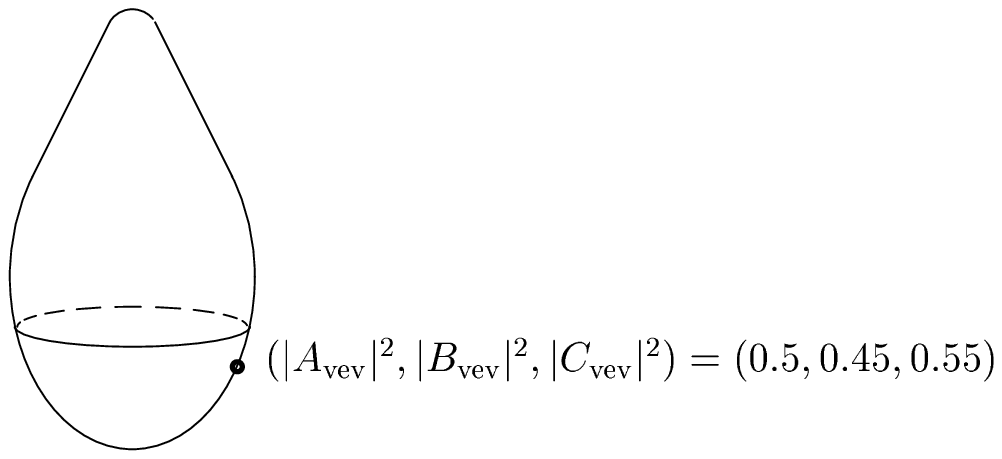}\\
\includegraphics[width=9cm]{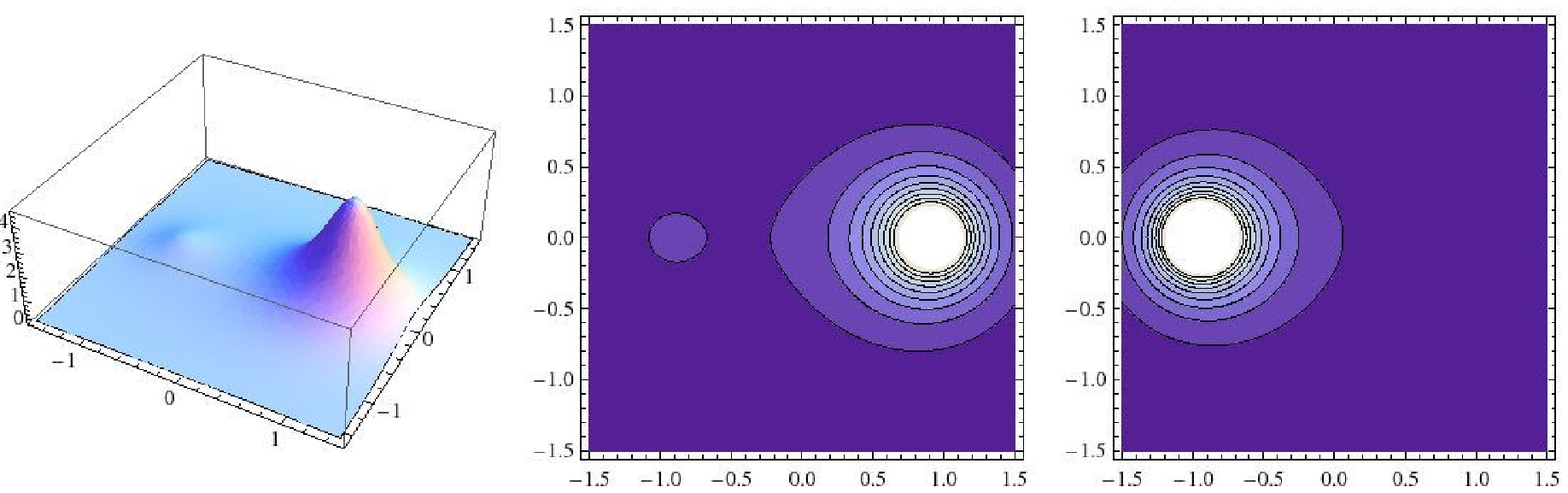} && \includegraphics[height=3cm]{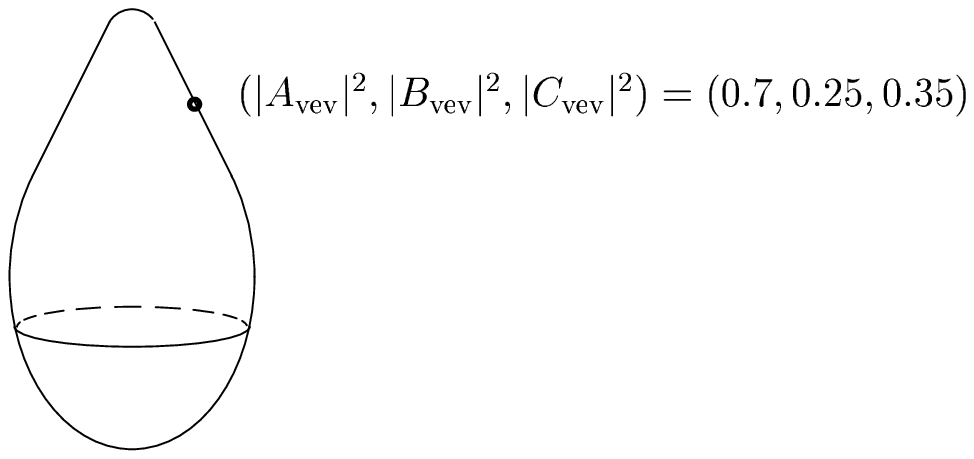}\\
\includegraphics[width=9cm]{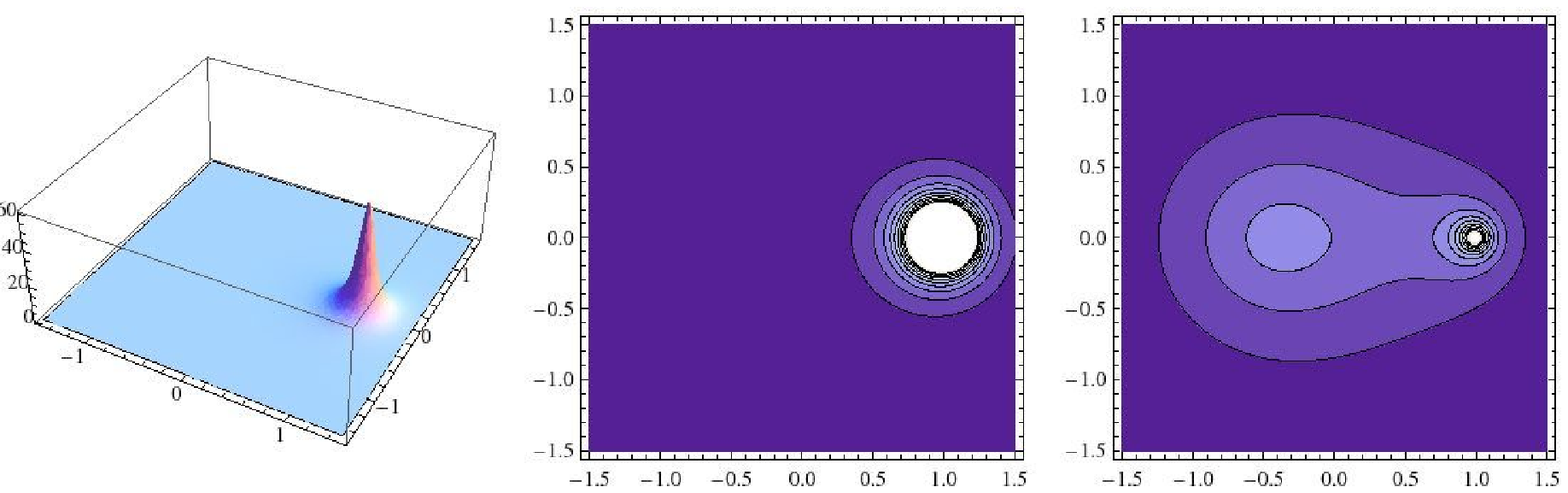} && \includegraphics[height=3cm]{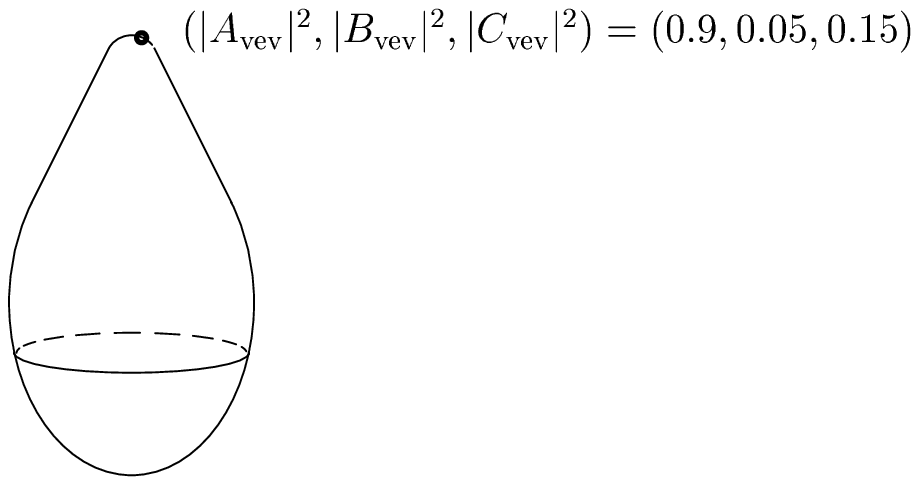}
\end{tabular}
\caption{\footnotesize{
  The energy density (left-most) and the magnetic flux density 
  $F_{12}^{(1)}$ (2nd from the left), $F_{12}^{(2)}$ (2nd from the
  right) and the boundary condition (right-most) for the lump of the
  second type in the strong gauge 
  coupling limit with $a=-1$ and $b=1$ in
  Eq.~(\ref{eq:mm_u1u1_mod}). $\xi_1=2$ and $\xi_2 = -0.1$.}} 
\label{fig:frac1}
\end{center}
\end{figure}

In order to understand better the characteristics of the vortex energy
profile in this model, we study the underlying sigma model,
of which the K\"ahler potential with non-vanishing $\xi_2$ 
is given by
\begin{eqnarray}
 K=\xi_1 f_\lambda(\varphi,\bar \varphi)^{-1}
+\xi_1 \log f_\lambda(\varphi,\bar \varphi)
-\frac{\xi_1-\xi_2}2 \log(1-\lambda f_\lambda(\varphi,\bar \varphi))
\ ,
\end{eqnarray}
where $f_\lambda(\varphi,\bar \varphi)$ is a function of an
inhomogeneous coordinate  
$\varphi=\sqrt{\xi_1}A/(BC)$ as
\begin{eqnarray}
 f_\lambda(\varphi,\bar \varphi)\equiv \frac{1-\lambda|\varphi|^2
+\sqrt{1+2|\varphi|^2+\lambda^2|\varphi|^4}}{1+\lambda},\quad 
\lambda\equiv \frac{\xi_2}{\xi_1} \ .
\end{eqnarray}
By setting $\xi_2=0$ $(\lambda=0)$, we find the same K\"ahler potential
as in Eq.~(\ref{eq:kahler_cp1}). 
Inside the disk of an arbitrary large radius  
the vortices reduce to sigma-model lumps,
\[\varphi=\varphi(z)\equiv
\frac{\sqrt{\xi_1}A_0(z)}{B_0(z)C_0(z)} \ , \]
characterized by $\pi_2({\cal M})$. 
There the minimal lump solution  has a tension 
$T_{\rm lump}=\pi (\xi_1-|\xi_2|)$.   
In the strong coupling limit giving rise to the above non-linear sigma
model, solutions for $s_1$ and $s_2$ are explicitly given by 
\begin{eqnarray}
|s_1|^2 = 
\frac{f_\lambda(\varphi,\bar \varphi)}
{\sqrt{1-\lambda f_\lambda(\varphi,\bar \varphi)}}
\ \frac{|B_0C_0|}{\xi_1}\ ,\qquad 
|s_2|^2 =  
\sqrt{1-\lambda f_\lambda(\varphi,\bar \varphi)}\,\left|
\frac{B_0}{C_0} \right| \ . 
\end{eqnarray}
Since the vortex-lump solutions in the sigma-model limit,
 suffer from the Coulomb singularity, we introduce a small $\xi_2$ to regularize such a singularity.
For instance, the analytic solution $|s_2|^2 = |B_0|/|C_0|$ in the
sigma model limit ($\xi_2=0$) leads to the singular magnetic flux for
$U(1)_2$ since 
$F_{12}^{(2)} = -2 \p\bar\p\log|s_2|^2 = - \p\bar\p\log|z|^2 = - \pi
\delta^{(2)} (z)$.
The price we have to pay is that  the two degenerate
configurations (Eq.~(\ref{eq:mm_u1u1_mod}))  are now split.
Suppose $-\xi_1 \ll \xi_2 < 0$, 
then the energy changes as
\beq
T_{(1/2,1/2)} = \pi (\xi_1 + \xi_2)=T_{\rm lump} \quad<\quad T_{(1/2,-1/2)} = \pi
(\xi_1 - \xi_2) \ , 
\label{eq:ene_u1u1}
\eeq
although the lump solution cannot distinguish the two.
The difference $2\pi |\xi_2|$ 
between the two is caused by the existence 
of an ANO-vortex-like singular peak attached on a tip of the lump
solution for $(\nu_1,\nu_2)=(1/2,-1/2)$. 
Actually, the solution 
for $|s_2|^2$ with $\xi_2<0$ is still singular at the zero of $C_0(z)$
since no configuration with $C=0$ 
can satisfy the $D$ term condition for $\xi_2<0$.
Therefore, we consider only the vortex solution with
$(\nu_1,\nu_2)=(1/2,1/2)$ in Eq.~(\ref{eq:mm_u1u1_mod}) as a smooth
lump solution.
The vortex configurations  in a model with $\xi_{2}$ in
Fig.~\ref{fig:frac1} (strong gauge coupling)
are quite similar to the
ones in the model with $\xi_{2}=0$, Fig.~\ref{fig:u1u1_finite} (finite
gauge coupling).
With $\varphi_{\rm vev}\not=0,\infty$,
we again observe two peaks.
The positions of the two are estimated by zeros of $A_0(z)$ and
$B_0(z)$, $z_a=-a/A_{\rm vev},z_b=-b/B_{\rm vev}$,  
which are mapped onto the two different poles of the vacuum moduli space
$\cal M$.  Note that the two peaks for the fractional vortices 
in the previous subsection are mapped onto 
the same pole (the south pole $\varphi_{\rm vev}=0$). 
The energy density $E=2\partial \bar \partial K$ at those zeros is 
\begin{eqnarray}
 E|_{z=z_a}=\frac{\xi_1(1-\lambda^2)}2
\frac{|\varphi_{\rm vev}|^2}{|z_a-z_b|^2},
\quad E|_{z=z_b}=\xi_1\frac{1-|\lambda|}{|\lambda||\varphi_{\rm vev}|^2}
\frac{1}{|z_a-z_b|^2}.
\end{eqnarray}
From this observation we see that the energy density for the
fractional vortices gets diluted as $|z_a-z_b|$ is increased.

In spite of the similar structure of ${\cal M}$,  
the difference in the gauge group, matter content and in the fiber,
manifest themselves  in distinct ways that the vacuum manifold is
covered by the vortex-sigma model lump solution  here,  as compared to
the previous model. 
A comparison of Eqs.~(\ref{eq:winding_u1}) and (\ref{eq:winding_u1u1}) 
shows that the minimum  lump solution in a generic point of ${\cal M}$
here clearly is seen to be half of the corresponding lump element 
$\pi_{2}({\cal M})$ in Section~\ref{sec:21model}
(on the $\mathbb{Z}_2$ singularity of the latter the solutions are similar).
In the present model there is no jump  in the homotopy-group elements
at the singularity of  ${\cal M}$.  
A sub-peak appears simply  because the target space ${\cal M}$ which
is a distorted sphere is warped.  If the target space were  the
standard $\mathbb{C}P^1$ sigma model  (a perfect sphere) only one peak
would have appeared.

Just as the model considered in Section~\ref{sec:21model} showed a
good example of the first mechanism for the  
vortex substructures discussed in Section~\ref{causes}, the present model nicely illustrates the second mechanism for the 
fractional vortex. 

\subsection{A model with $U(1)\times SU(2)$}
\label{sec:u1u2}

The third and last example of a model with the target space 
of the droplet type is a gauge theory with a $U(1)\times
SU(2)$ gauge group with Higgs fields $H = (A,{\bf B})$ 
\begin{center}
\begin{tabular}{c||cc}
& $U(1)$ & $SU(2)$ \\
\hline
\hline
$A$ & $2$ & $\mathbbm{1}$ \\
${\bf B}$ & $1$ & $\square$ 
\end{tabular}
\end{center}
namely, a complex scalar filed $A$ of $U(1)$ charge $2$, and two
complex scalars ${\bf B} = (\vec B_1,\vec B_2)$ 
in the fundamental representation of 
the $SU(2)$ group, and with the Abelian charge $1$, while $A$ is a
singlet. 
The latter is conveniently denoted by a color-flavor mixed $2 \times
2$ matrix, ${\bf B}$. There is furthermore a global symmetry 
$SU(2)_{\rm f}$. The gauge group acts on the fields as
\beq
(A,{\bf B}) \to (e^{2i\alpha} A, e^{i\alpha} g {\bf B})\ ,\quad
e^{i\alpha} \in U(1)\ ,\quad g\in SU(2) \ .
\eeq
Note that the transformation $(\alpha,g) = (\pm \pi, -{\bf 1}_2)$ is a
symmetry, thus the  gauge group is really
$[U(1)\times SU(2)]/\mathbb{Z}_2 \simeq U(2)$.

The vacuum manifold is given by the $D$-flatness condition
\beq
M = \left\{(A,{\bf B})\quad | \quad 2|A|^2 {\bf 1}_2 + 2 {\bf B}{\bf
  B}^\dagger = \xi {\bf 1}_2\right\}\ .
\eeq
By using $SU(2)$ gauge and flavor  symmetries, the vacuum configuration for ${\bf B}$ fields can be taken in the form
${\bf B} = {\cal B}{\bf 1}_2$,  showing that the
vacuum manifold $M$  is an  $S^3$ defined by $|A|^2 + |{\cal B}|^2 = \xi/2$.  
The vacuum moduli space ${\cal M}$    is a $U(1)$ quotient of this $S^3$ with the
weighted charges given above, topologically the same as $\mathbb{C}P^1$.
At a generic point $({\bf B}\neq0$) of the vacuum moduli, the flavor
symmetry $SU(2)_{\rm f}$ is broken but the color-flavor diagonal 
 symmetry $SU(2)_{\rm c+f}$ is  preserved. At the
special point ${\bf B}=0$, neither the $SU(2)$ gauge nor the flavor
symmetry is broken.
To be precise, we can compute the K\"ahler metric on the vacuum moduli
space by eliminating all the gauge multiplets from the K\"ahler
potential 
\beq
K = |A|^2 e^{-2V} + \Tr \left({\bf B}{\bf B}^\dagger e^{-V-V'}\right)
+ \xi V \ ,
\eeq
with $V$ being a $U(1)$ vector multiplet and $V'$ being an $SU(2)$
vector multiplet.  
Once we eliminate $V'$ from the Lagrangian, we obtain
\beq
K = |A|^2 e^{-2V} + \left|\left(2\det{\bf
  B}\right)^{\frac{1}{2}}\right|^2  e^{-V} + \xi V \ .
\eeq
Comparing this with Eq.~(\ref{eq:kahler_21}), it is easy to see that
the two spaces have the same metric, 
if we replace the inhomogeneous coordinate $\varphi = 2A/B^2 \to
A/\det{\bf B}$. 
Thus the vacuum moduli space is of the droplet form as in
Fig.~\ref{fig:cp1dz2}, but there is no conical
singularity. Instead, the $SU(2)$ gauge symmetry is unbroken at the 
tip of the droplet (${\bf B}=0$).

Construction of the BPS vortex solutions in this model is summarized
in the Appendix in detail. 
The energy formulae are the same as Eqs.~(\ref{eq:energy_form}) and
(\ref{eq:energy_form2}) with 
$J = |A|^2/2 + \Tr ({\bf B}{\bf B}^\dagger) /2$ and the solutions
are determined by $U(1)$ winding number $\nu$
\begin{align}
\nu &= - \frac{1}{2\pi}\int dx^2\, F_{12} = \frac{1}{\pi}\int
dx^2\,\p\bar\p\log |s|^2 \ ,\\
W_{U(1)} &= - i\bar\p\log s \ ,\quad 
W_{SU(2)} = - i {S'}^{-1}\bar\p S'\ ,\\
H &= (A,{\bf B}) = \left(s^{-2} A_0(z), s^{-1}{S'}^{-1} 
      {\bf B}_0(z)\right) \ , \label{eq:mm_u1xu2}
\end{align}
where $s \in \mathbb{C}^*$ and $S' \in SL(2,\mathbb{C})$ and $A_0$ and
all the elements of ${\bf B}_0(z)$ must be holomorphic functions of
$z$. The BPS equations for the gauge the gauge fields lead to the
master equations
\begin{align}
\bar{\partial}\partial\log\omega &= 
-\frac{e^2}{8}\left(\frac{1}{\omega}
\Tr\left(\mathbf{B}_0\mathbf{B}_0^\dag{\Omega'}^{-1}\right) 
+\frac{2}{\omega^2}|A_0|^2 - \xi\right) \ , \nonumber\\
\bar{\partial}\left(\Omega'\partial{\Omega'}^{-1}\right) &=
\frac{g^2}{4\omega}\left(\mathbf{B}_0\mathbf{B}_0^\dag{\Omega'}^{-1} 
-\frac{\mathbf{1}_2}{2}
\Tr\left(\mathbf{B}_0\mathbf{B}_0^\dag{\Omega'}^{-1}\right)\right) \ , 
\end{align}
where $\omega\equiv ss^\dag$ and $\Omega'\equiv S'{S'}^\dag$.

The first equation determines the 
asymptotic behavior of $|s|^2 \sim |z|^{2\nu}$ as $|z| \to \infty$.
The choice of $\nu$ must be consistent with the given boundary condition 
$(A,{\bf B}) \to (A_{\rm vev},{\cal B}_{\rm vev}{\bf 1}_2)$ satisfying 
$|A_{\rm vev}|^2 + |{\cal B}_{\rm vev}|^2 = \xi/2$.
To find consistent solutions, it is useful to consider the
following $SU(2)$ gauge invariant  
$I = \det {\bf B} = s^{-2} \det {\bf B}_0(z)$.
When a generic point $({\bf B}_{\rm vev} \neq 0)$ is chosen,
holomorphy of $A_0(z)$ and $\det {\bf B}_0(z)$ requires 
$2\nu \in \mathbb{Z}_+$ because of the asymptotic behavior
\beq
A_0 \sim A_{\rm vev} z^{2\nu} + \cdots, \quad
\det {\bf B}_0 \sim {\cal B}_{\rm vev}^2 z^{2\nu} + \cdots,\quad
{\rm as}\quad|z| \to \infty \ .
\eeq
In this way the vortices in this system are characterized by the half
quantized $U(1)$ winding number $\nu \in \mathbb{Z}_+/2$. 

The minimal configuration with tension $T = \pi \xi$ is described by
the moduli matrices 
\beq
A_0 = A_{\rm vev} z + a \ ,\quad
\det {\bf B}_0 = {\cal B}_{\rm vev}^2 (z + b) \ .
\eeq
Note that the matrix ${\bf B}_0(z)$ is not uniquely determined by
these conditions. 
The simplest one is 
${\bf B}_0 = {\cal B}_{\rm vev} {\rm diag}(z+b,1)$. This matrix breaks 
the color-flavor symmetry $SU(2)_{\rm c+f}$ of the vacuum into
$U(1)_{\rm c+f}$. Henceforth, the generic 
configurations are generated by $SU(2)_{\rm c+f}$, so that the vortex
has an internal orientation 
${\mathbb C}P^1 \simeq SU(2)_{\rm c+f}/U(1)_{\rm c+f}$
(Nambu-Goldstone mode). The moduli space of the single vortex is 
\beq 
\mathcal{V} = \mathbb{C} \times {\cal V}^{U(2)}_{k=1}\ , \qquad 
{\cal V}^{U(2)}_{k=1} = \mathbb{C}\times\mathbb{C}P^{1}\ , \eeq
which is a product of the orientational zero-modes   ($b$)  and a
center of mass and a ``size'' parameter.

When we choose $A_{\rm vev} = 0$ as the boundary condition, the vortex 
is a semi-local extension of the so-called $U(2)$ local vortex. In
fact, if we turn off the size moduli, viz.~$a=0$, it is precisely 
the $U(2)$ local vortex. The transverse width of the vortex becomes
large as $|a|$ increases while the orientational moduli are still
localized near the vortex core. This is completely different from the 
so-called non-Abelian semi-local vortex in $U(\NC)$ gauge theories
with $\NF > \NC$ Higgs fields in the fundamental representation.

We see that the global features of the vortices of this model
are similar to that of the $U(1)\times U(1)$ model (except, of course,
for the internal, orientational modes which are present here).  
The key fact is that the non-Abelian sector of this model (with the
matter fields ${\bf B}$) is actually a 
$U(2)\sim SU(2)\times U(1)/{\mathbb Z}_{2}$ theory. Taking into
account the $U(1)$ charge of $A$, we see that the fiber is generated
by $\pi_{1}(S^{1}/{\mathbb Z}_{2})$,  on and off the orbifold
singularities (${\bf B}=0$).

The system at  ${\cal B}_{\rm vev} = 0$ is in a (non-Abelian) Coulomb phase classically and  
no vortex solutions exist. In the  $U(1)^{2}$  model of  Section~\ref{sec:simplemodel} we avoided 
 the problem by turning on the   FI parameter for the second
$U(1)$ factor.  The same cannot be done here,  as no
FI-like term exists for a non-Abelian gauge group.  Therefore there exist neither  a smooth non-linear sigma-model limit  nor  any regular lump solutions.   Such solutions would necessarily be singular.

\subsection{An alternative  $U(1)^2$ model: the lemon space}
\label{sec:u1u1simplified}

\begin{figure}[ht]
\begin{center}
\includegraphics[height=6cm]{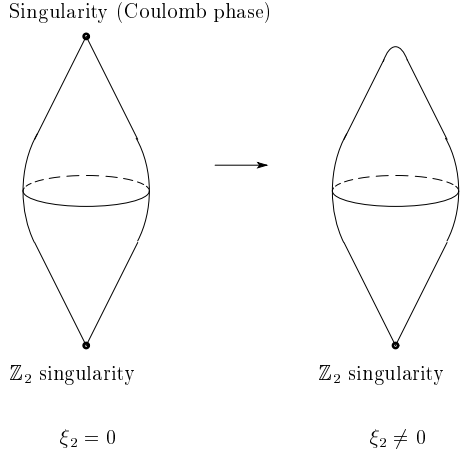}
\caption{\footnotesize{The lemon space.}}
\label{fig:lemon}
\end{center}
\end{figure}

The model of Section~\ref{sec:simplemodel} was chosen to have a
minimal fiber $F$ with the fewest possible fields. However, the
energy density (in the lump limit) and the K\"ahler metric turned out
to be rather elaborate. 
In this subsection, we analyze a related model, with a slightly
different charge assignment, 
$Q_1 = (1,1,1)$ for $U(1)_1$ and $Q_2 = (0,1,-1)$ for $U(1)_2$ as
\beq
(A,B,C) \to \left( e^{i\alpha(x)}A,\, e^{i\alpha(x)+i\beta(x)}B,\,
e^{i\alpha(x) - i\beta(x)}C\right) \ .
\eeq
An important difference from Section~\ref{sec:simplemodel} is that the
 gauge symmetry is  now  $U(1)_1 \times U(1)_2$,  without a $\mathbb{Z}_2$ division.
The points 
$(A,B,C) $ and  $(-A,B,C)$ related by 
$(\alpha,\beta) = (\pi,\pm\pi) \in \mathbb{Z}_2$, are distinct points. 
The vacuum manifold and vacuum moduli space are given by
\beq
M &=& \left\{A,B,C\quad |\quad |A|^2 + |B|^2 + |C|^2=\xi_1,\ |B|^2 -
|C|^2 = \xi_2\right\}\ ,\\ 
{\cal M} &=& M/ \left(U(1)_1 \times U(1)_2\right)\ .
\eeq
We see that  $A=0$ is a  $\mathbb{Z}_2$ orbifold point, whereas  
the point  $B=C=0$ represents a system in Coulomb phase (which can be Higgsed and regularized by  
$\xi_2 \ne  0$).  See
Fig.~\ref{fig:lemon}.   
Clearly this model shares  aspects both of the simple $U(1)$ model of Section~\ref{sec:21model} and  of the
   $U(1)\times U(1)$ model of  Section~\ref{sec:simplemodel}.  
In the following we shall consider mainly the case of $\xi_2=0$,  except when we consider the sigma-model limit, which is well defined only for 
a non-vanishing $\xi_2$. 

 The vortex Ansatz   is 
\beq (A,B,C) = \left( s_1^{-1} A_0(z),\ s_1^{-1}s_2^{-1}
B_0(z),\ s_1^{-1}s_2 C_0(z) \right)\ , \eeq
with the gauge field equations
\begin{align}
\bar{\partial}\partial\log\omega_1 &= -\frac{e^2}{4}
\left[\omega_1^{-1}\left(|A_0|^2 + \omega_2^{-1}|B_0|^2
  + \omega_2|C_0|^2\right)-\xi_1\right] \ , \\
\bar{\partial}\partial\log\omega_2 &= -\frac{g^2}{4}
\left[\omega_1^{-1}\left(\omega_2^{-1}|B_0|^2 - \omega_2|C_0|^2\right)
  -\xi_2\right] \ ,
\label{u1u1model_simpl_mastereqs}
\end{align}
where $\omega_i\equiv s_is_i^\dag$, for $i=1,2$. In order to avoid
repetition, all the details are summarized in the Appendix. 
As in the case of Section~\ref{sec:simplemodel}, the
winding numbers are   $\nu_1$ and $\nu_2$ for $U(1)_1$ and $U(1)_2$,
respectively.  The tension depends  only on  $\nu_1$ 
for  $\xi_2 = 0$:
\beq
T = 2\pi\xi_1\nu_1 \ ,\qquad \nu_1 \in \mathbb{Z}_+ \ .
\eeq
The minimal-energy solutions with the generic boundary condition
($0<|A_{\rm vev}|^2 <\xi_1$) have $T=2\pi\xi_1$ and are obtained by
the following three different moduli matrices
\begin{alignat}{4}
&A_0 = A_{\rm vev} z + a\ ,\quad
B_0 = B_{\rm vev} \ ,\quad
C_0 = C_{\rm vev} z^2 + c_1 z + c_2\ ,\qquad
&(\nu_1,\nu_2) = (1,-1)\ ,\\
&A_0 = A_{\rm vev} z + a\ ,\quad
B_0 = B_{\rm vev}z + b\ ,\quad
C_0 = C_{\rm vev} z + c \ ,\qquad
&(\nu_1,\nu_2) = (1,0)\ ,\label{eq:1101}\\
&A_0 = A_{\rm vev} z + a\ ,\quad
B_0 = B_{\rm vev} z^2 + b_1 z + b_2\ ,\quad
C_0 = C_{\rm vev}\ ,\qquad
&(\nu_1,\nu_2) = (1,1)\ .\label{eq:1102}
\end{alignat}
As they obey different boundary
condition for $\nu_2$, they belong to different topological
sectors. Each configuration has three moduli parameters.

\begin{figure}[h!tp]
\begin{center}
\begin{tabular}{ccl}
\includegraphics[width=10cm]{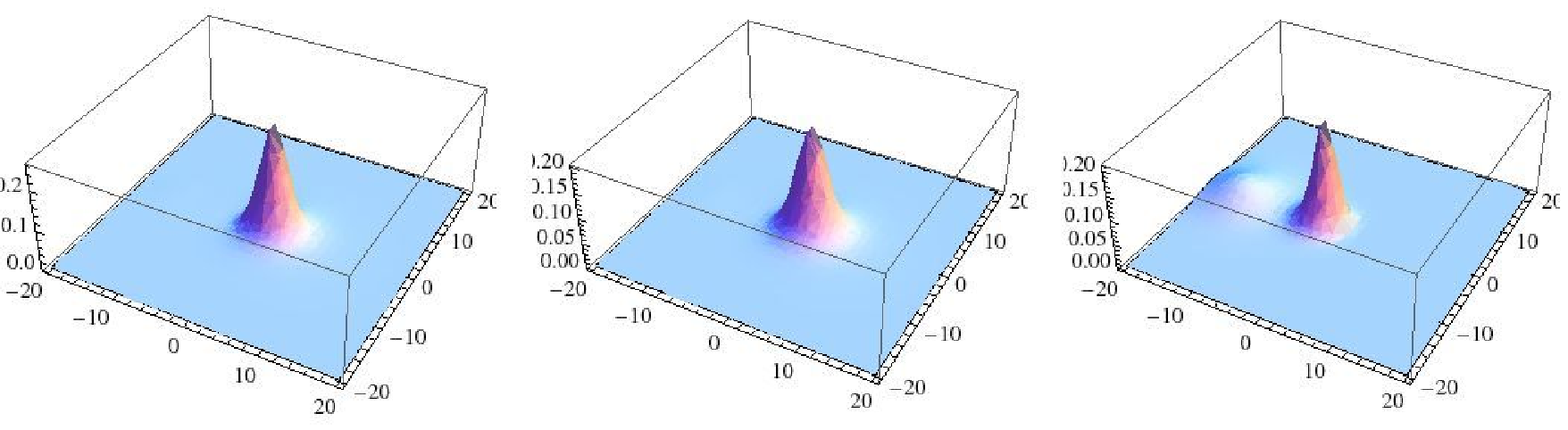} && \includegraphics[height=4cm]{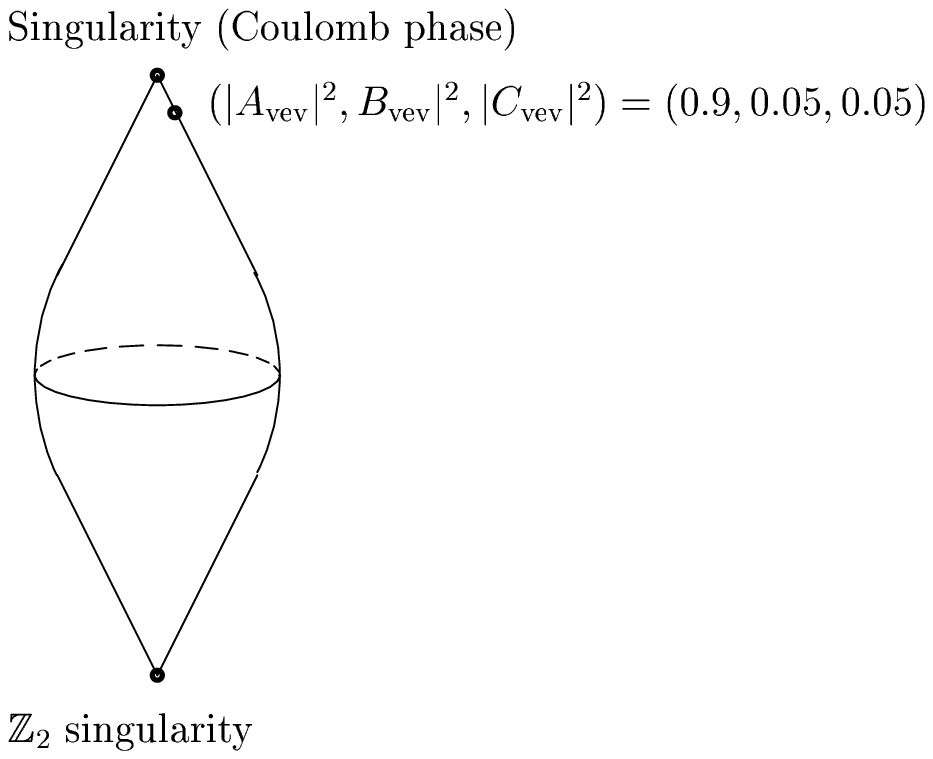}\\
\includegraphics[width=10cm]{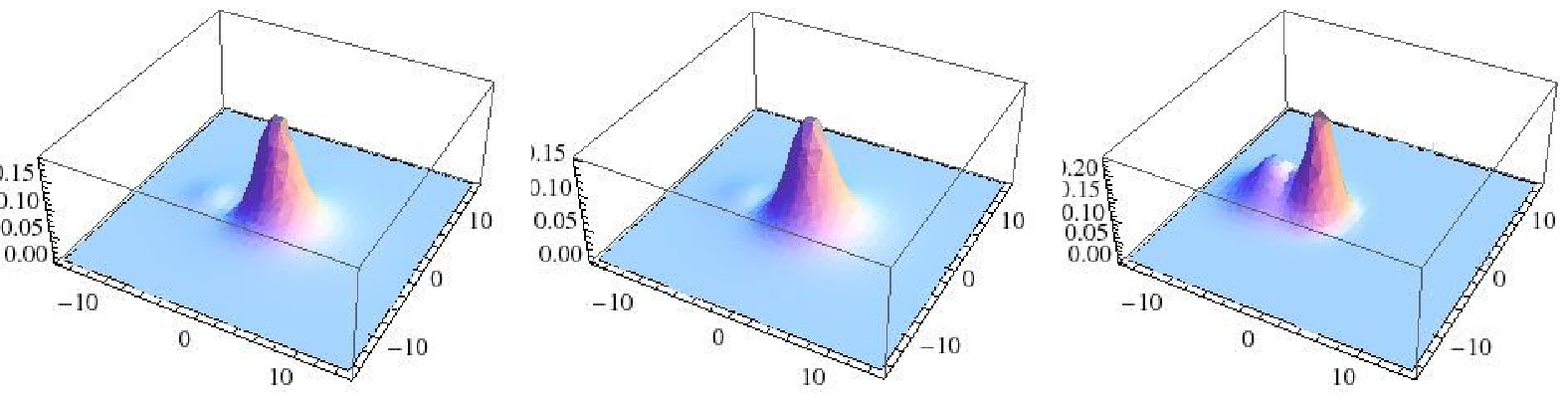} && \includegraphics[height=4cm]{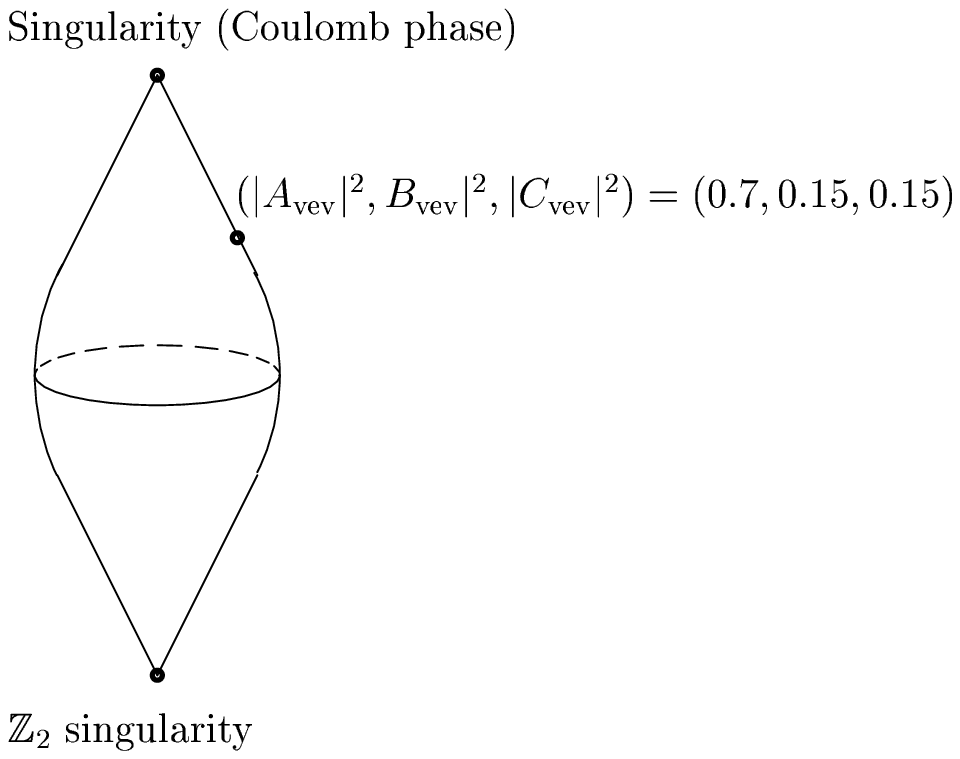}\\
\includegraphics[width=10cm]{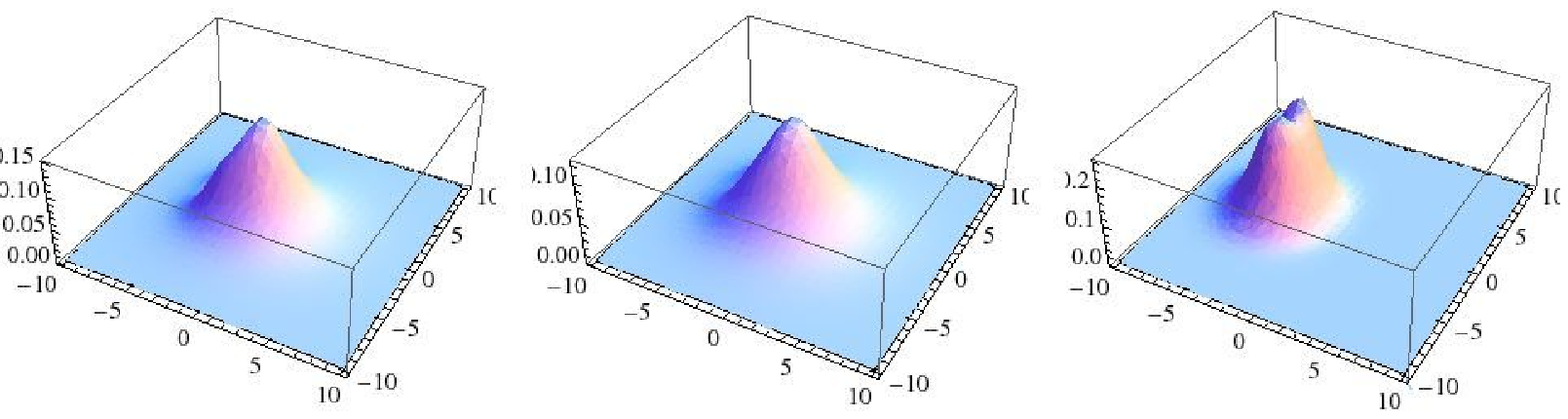} && \includegraphics[height=4cm]{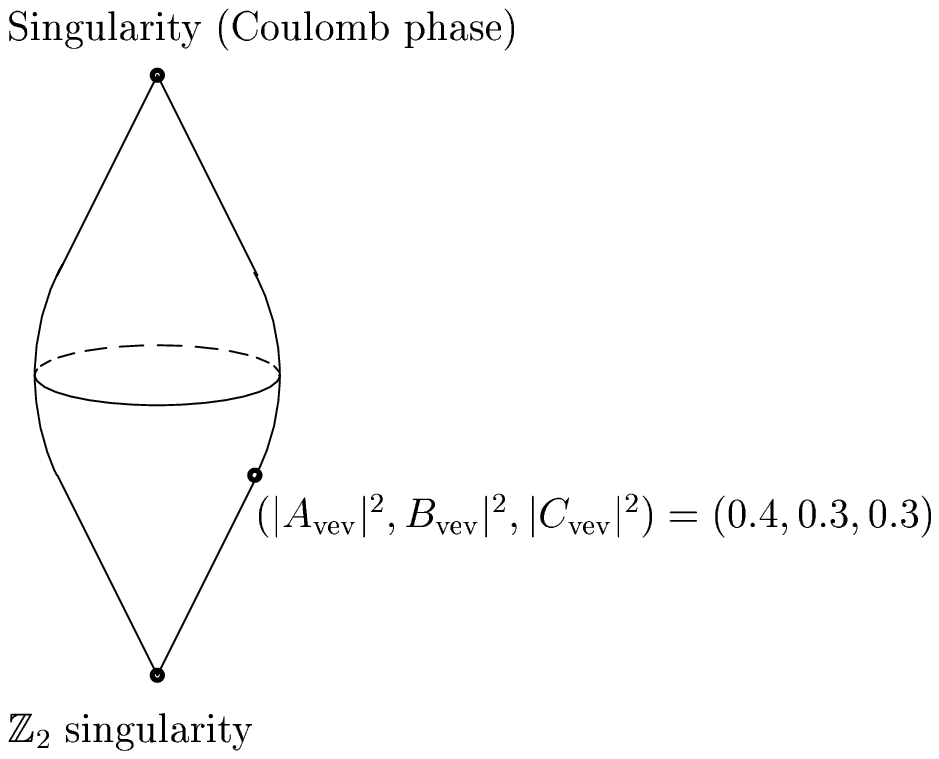}\\
\includegraphics[width=10cm]{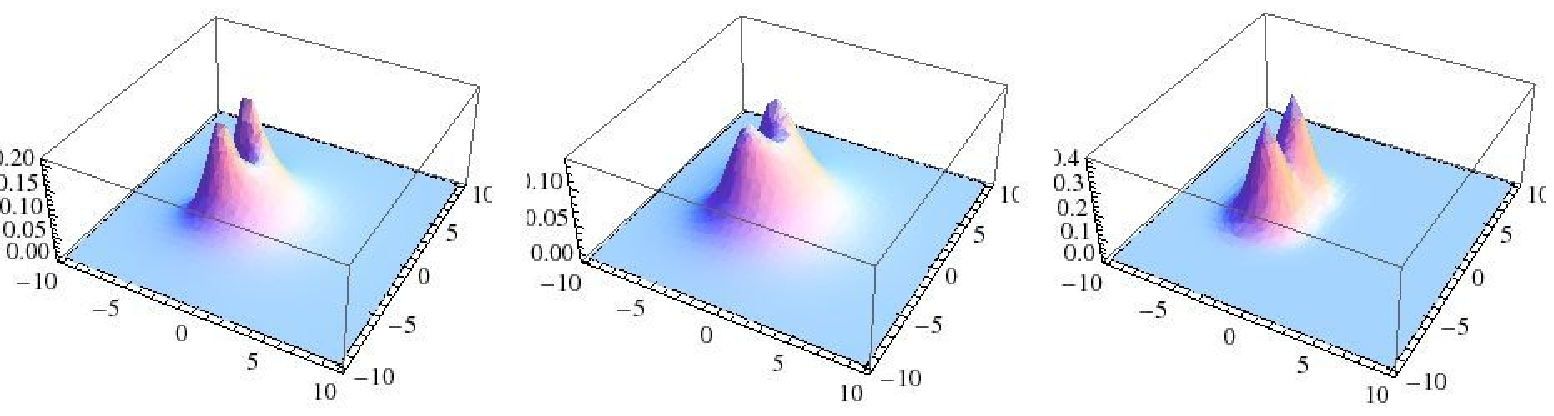} && \includegraphics[height=4cm]{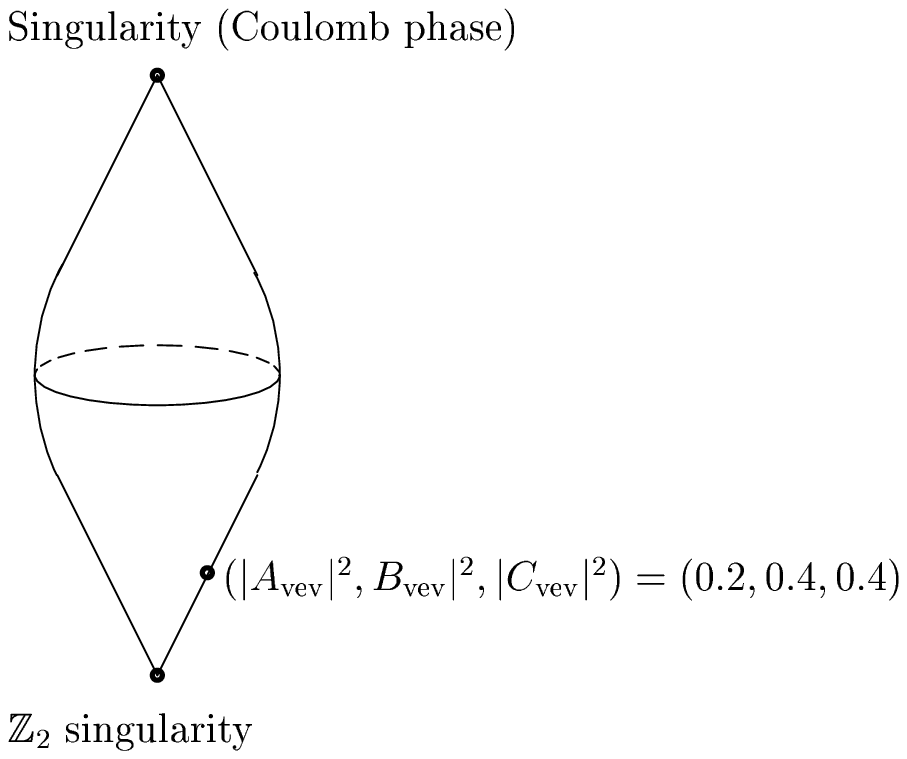}\\
\includegraphics[width=10cm]{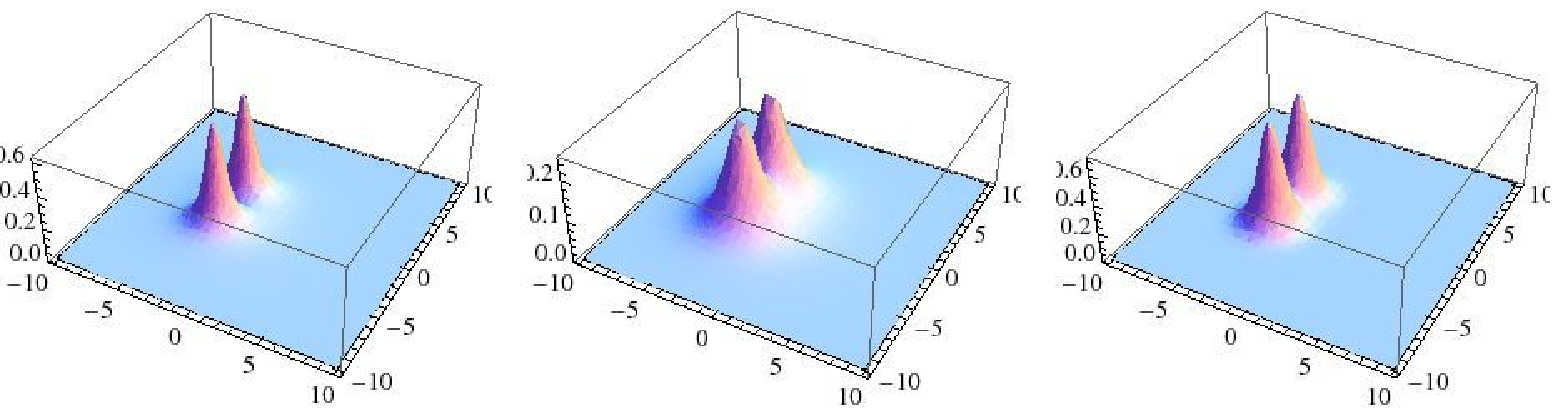} && \includegraphics[height=4cm]{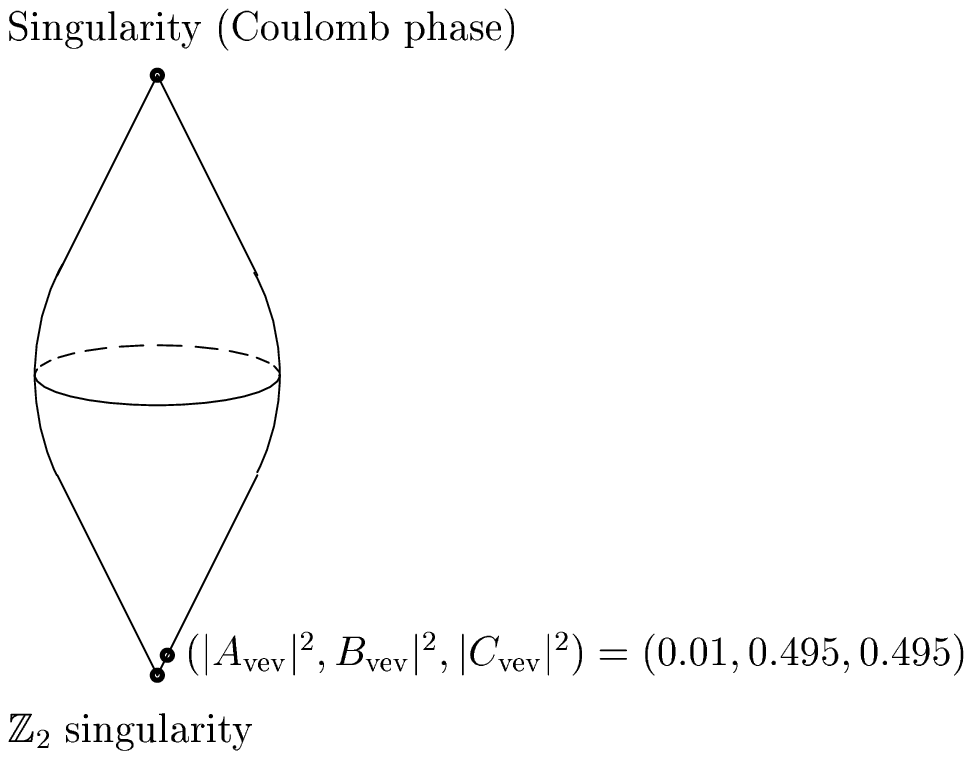}
\end{tabular}
\caption{\footnotesize{
  The energy density (left-most) and the magnetic flux density 
  $F_{12}^{(1)}$ (2nd from the left), $F_{12}^{(2)}$ (2nd from the
  right) and the boundary condition (right-most) for
  Eq.~(\ref{eq:1102}) with $\xi_1=1$ and $\xi_2 = 0$ and $e_1=1$,
  $e_2=2$.  }}
\label{fig:frac1b1}
\end{center}
\end{figure}

\begin{figure}[!th]
\begin{center}
\begin{tabular}{ccl}
\includegraphics[width=10cm]{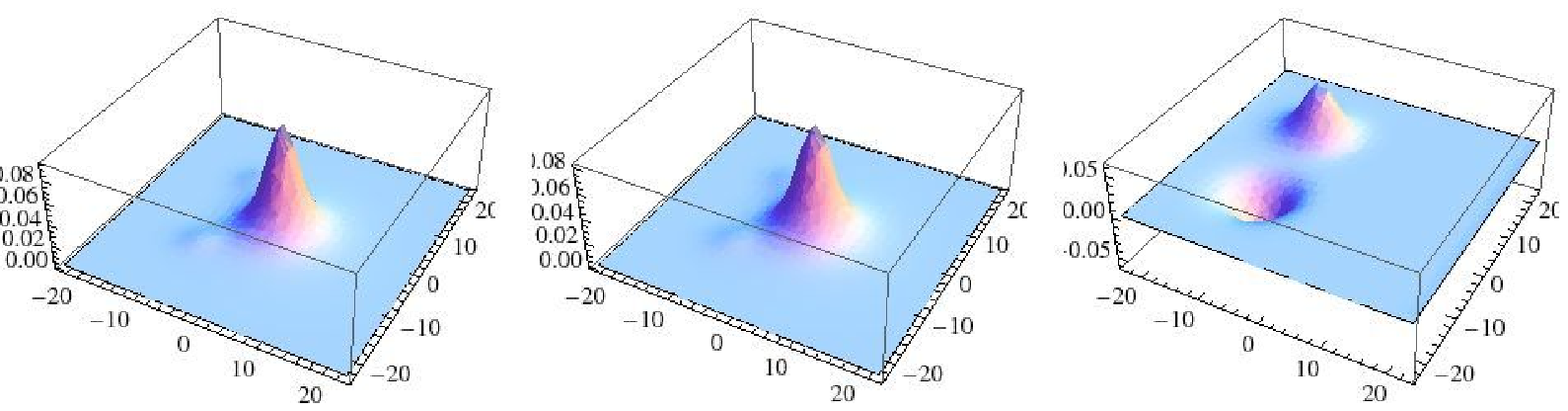} && \includegraphics[height=4cm]{u1u1_11_2_01a}\\
\includegraphics[width=10cm]{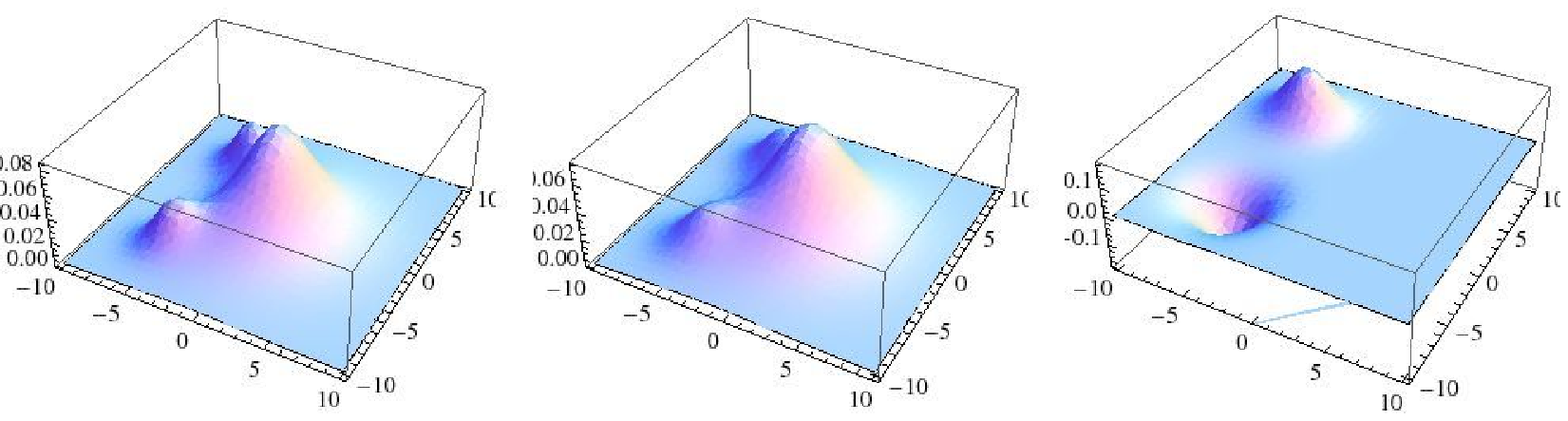} && \includegraphics[height=4cm]{u1u1_11_2_02a}\\
\includegraphics[width=10cm]{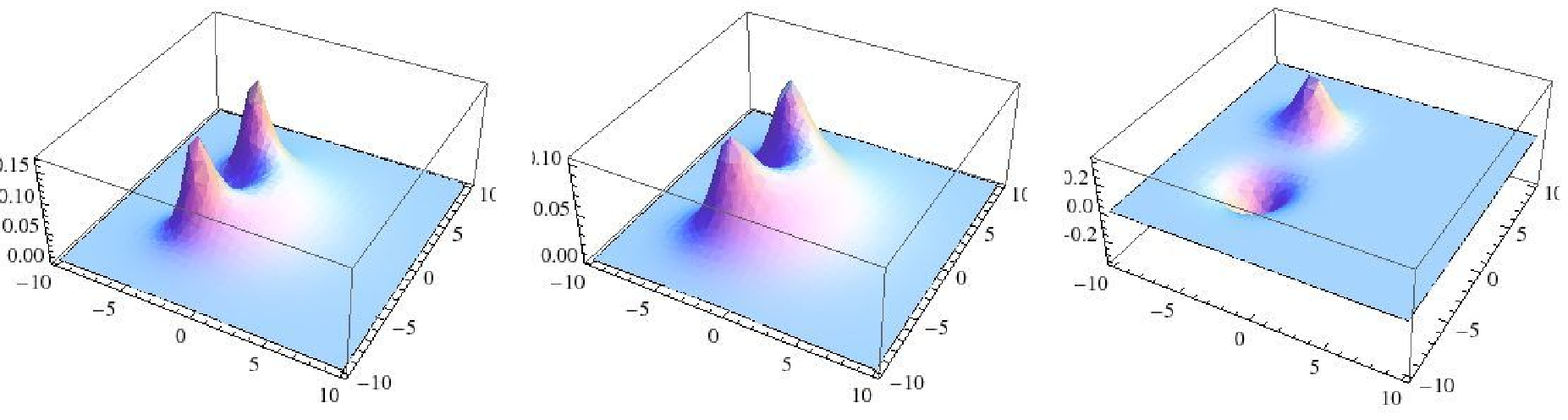} && \includegraphics[height=4cm]{u1u1_11_2_03a}\\
\includegraphics[width=10cm]{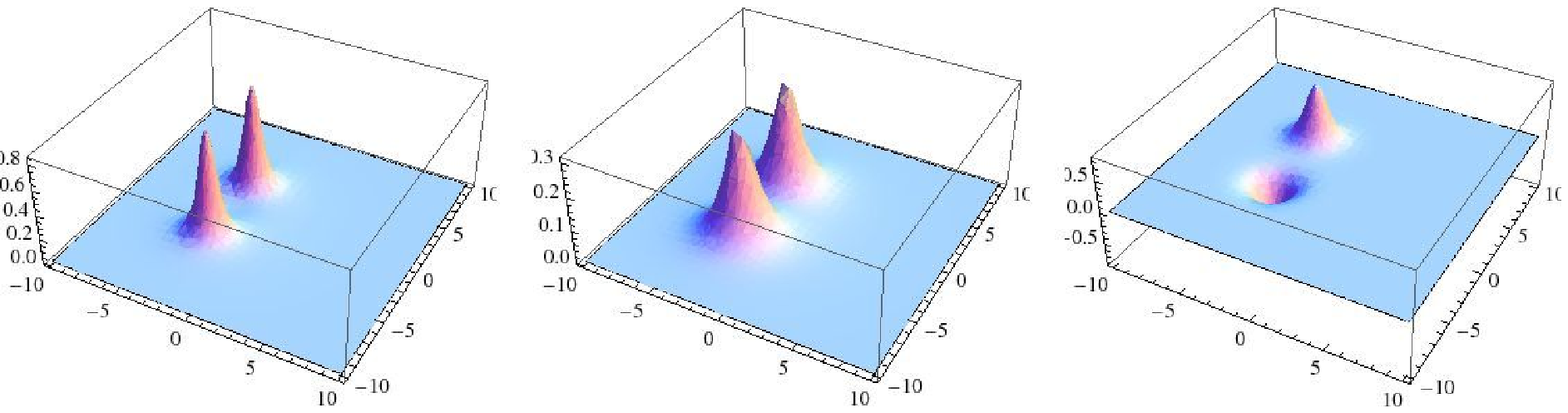} && \includegraphics[height=4cm]{u1u1_11_2_05a}
\end{tabular}
\caption{\footnotesize{
  The energy density (left-most) and the magnetic flux density
  $F_{12}^{(1)}$ (2nd from the left), $F_{12}^{(2)}$ (2nd from the
  right) and the boundary condition (right-most) for
  Eq.~(\ref{eq:1101}) with $\xi_1=1$ and $\xi_2 = 0$ and $e_1=1$,
  $e_2=2$. }}
\label{fig:frac1b2}
\end{center}
\end{figure}

Near the $\mathbb{Z}_2$ orbifold point 
we observe two peaks.    Although the energy density always looks the same, the
magnetic fluxes, especially of the second $U(1)_2$, depends on the
value of $\nu_2$. 
In Fig.~\ref{fig:frac1b1}, we show several numerical solutions for 
Eq.~(\ref{eq:1102}).
We also show a couple of solutions for Eq.~(\ref{eq:1101}) in
Fig.~\ref{fig:frac1b2}. 
In almost all regions, the configuration consists of one peak or two peaks
but sometimes we observe three peaks simultaneously.

On the other hand, at exactly a singular vacuum  $A_{\rm vev}=0$ (the
singular point on ${\cal M}$), the minimal vortex with tension
$T = \pi\xi_1$ is given by 
\begin{align} 
A_0 &= a\ ,\quad 
B_0 = B_{\rm vev} z + b\ ,\quad 
C_0 = C_{\rm vev}\ ,\quad (\nu_1,\nu_2) = (1/2,1/2)\ ,\\
A_0 &= a\ ,\quad 
B_0 = B_{\rm vev}\ ,\quad 
C_0 = C_{\rm vev}z+c\ ,\quad (\nu_1,\nu_2) = (1/2,-1/2)\ .
\end{align}
At $A=0$ ($\varphi = 0$) a $\mathbb{Z}_2$ symmetry remains unbroken
which is a typical orbifold singularity.  
As a result, the $U(1)_{1}$ fiber $F$ is the half ($\alpha =0 \to
\pi$) at the orbifold point as compared to that in other points of the
vacuum moduli, where $\alpha =0 \to  2\pi$. 
The global structure of the vortex-sigma model lumps in this model is
thus somewhat similar to the model of Section~\ref{sec:21model}. 
At the $\mathbb{Z}_2$  orbifold singularity  $\pi_{1}(F)$ and
$\pi_{2}({\cal M})$ make a jump, and this explains the appearance of
the double peaks.

As in the model in Section~\ref{sec:simplemodel}, we cannot take 
$B_{\rm vev} = C_{\rm vev} = 0$ as a boundary condition since the
second $U(1)_2$ is unbroken at infinity.

These aspects can be made more explicit in the strong gauge coupling
limit $e_1,e_2 \to \infty$, where the K\"ahler potential is simpler
(than in the model of Section~\ref{sec:simplemodel}) by construction and
the solutions can be analytically solved.  
Since the Coulomb phase leads to singular solutions, we here turn on 
the another FI parameter $\xi_2$ ($|\xi_2|<\xi_1$) for $U(1)_2$.
Working in a supersymmetric context, 
elimination of  the gauge superfields $V_1$ and $V_2$ from 
\beq K = |A|^2e^{-V_1} + |B|^2e^{-V_1-V_2} +
|C|^2e^{-V_1+V_2} +\xi_1 V_1 + \xi_2 V_2\ ,  \eeq
yields  the K\"ahler potential
\beq K = \xi_1 \log \left( 1 + \sqrt{\lambda^2 +
  (1-\lambda^2)|\tilde\varphi|^2}\right) 
-|\xi_2| \log \left( |\lambda| + \sqrt{ \lambda^2 + (1-\lambda^2) |
  \tilde\varphi|^2}\right) \ . \label{sweetpotato_kahlerquotient} \eeq 
where the inhomogeneous coordinate $\tilde\varphi$ and $\lambda$ are
defined by 
$\tilde\varphi \equiv \frac{2BC}{A^2}$ and 
$\lambda \equiv \frac{\xi_2}{\xi_1}$.
The BPS solutions are given by the holomorphic functions
\beq
\tilde \varphi(z) = \frac{2B_0(z)C_0(z)}{A_0^2(z)} \ , 
\eeq
and characterized by the quantized tension
\beq
T = 2\pi \sum_i\xi_i \nu_i \ ,\qquad
\nu_i \equiv \frac{1}{\pi} \int dx^2\, \p\bar\p \log |s_i|^2 \ ,
\eeq
where $|s_i|^2$ is given by
\beq
|s_1|^2 &=& 
\frac{|A_0|^2}{1-\lambda^2}
\left(1+\sqrt{\lambda^2 + (1-\lambda^2) |\tilde\varphi|^2}\right) \ ,\\
|s_2|^2 &=& \frac{|A_0|^2}{|C_0|^2} 
\frac{-\lambda + \sqrt{\lambda^2 + (1 -
    \lambda^2)|\tilde\varphi|^2}}{2(1+\lambda)}
=\frac{|B_0|^2}{|A_0|^2}
\frac{2(1-\lambda)}{\lambda + \sqrt{\lambda^2 + (1 -
    \lambda^2)|\tilde\varphi|^2}}  \ .
\eeq

\subsection{An alternative $U(1)\times SU(2)$ model}
\label{sec:sun_simplified}

The next example with the same base space is the $U(1)\times SU(2)$
theory with the same Lagrangian as in Section~\ref{sec:u1u2}, 
except for a different $U(1)$ charge assignment for the $A$ field:
\begin{center}
\begin{tabular}{c||cc}
& $U(1)$ & $SU(2)$ \\
\hline
\hline
$A$ & $1$ & $\mathbbm{1}$ \\
${\bf B}$ & $1$ & $\square$ 
\end{tabular}
\end{center}
The gauge group action on the fields is 
\beq
(A,{\bf B}) \to (e^{i\alpha}A,e^{i\alpha} g {\bf B})\ ,\quad
e^{i\alpha} \in U(1)\ ,\quad g \in SU(2)\ .
\eeq
Note that $(\alpha, g) = ( \pm \pi, -{\bf 1}_2)$ is not an identity operator, so
that the  gauge group is truly  $U(1) \times SU(2)$. 
The vacuum manifold is given by
\beq
M  = \{ (A,{\bf B}) \quad | \quad |A|^2{\bf 1}_2 + 2 {\bf B}{\bf
  B}^\dagger = \xi {\bf 1}_2 \} \ .
\eeq
By using $SU(2)$, we can bring ${\bf  B} = {\cal B} {\bf 1}_2$. Then
we see that the vacuum manifold is isomorphic to $S^3$ and the vacuum
moduli space is isomorphic to $S^3/U(1) \simeq \mathbb{C}P^1$. The
vacuum manifold is the lemon space as drawn in Fig.~\ref{fig:lemon}.  
There are two singularities: one (at $A=0$ )  is a  $\mathbb{Z}_2$ conical singularity
and the other is 
a Coulomb singularity where $SU(2)$ gauge symmetry is restored  (${\bf
  B}=0$). 

The construction of the BPS vortex in this model is the same as the
one in Section~\ref{sec:u1u2}. 
However a difference appears in Eq.~(\ref{eq:mm_u1xu2}) as
\beq
H = \left(A,\ {\bf B}) = (s^{-1} A_0(z),\ s^{-1} {S'}^{-1}{\bf
  B}_0(z)\right)\ . 
\eeq
This makes a difference for the choice of the $U(1)$ winding number
$\nu$ which should be chosen to be consistent with a given boundary
condition $(A,{\bf B}) \to (A_{\rm vev},{\cal B}_{\rm vev}{\bf 1}_2)$ 
satisfying $|A_{\rm vev}|^2 + 2|{\cal B}_{\rm vev}|^2 = \xi$. To see
this, we again make use of the holomorphic $SU(2)$ gauge invariant 
$I = \det {\bf B} = s^{-2}\det{\bf B}_0(z)$. 
For a generic point $A_{\rm vev} \neq 0$ or 
${\cal B}_{\rm vev} \neq 0$, the asymptotic behavior is of the form 
\beq
A_0 \to A_{\rm vev} z^\nu + \cdots \ , \quad
\det {\bf B}_0 \to {\cal B}_{\rm vev}^2 z^{2\nu} + \cdots \ .
\eeq
Holomorphy requires $\nu$ to be semi-positive integer.
Thus the minimal configuration with $\nu=1$ has the mass $T = 2\pi\xi$
and it is generated by the moduli matrix 
\beq
A_0 = A_{\rm vev} z + a \ ,\quad
\det {\bf B}_0 = {\cal B}_{\rm vev}^2 (z^2 + b_1 z + b_2) \ .
\eeq
These conditions do not uniquely determine the matrix 
${\bf B}_0(z)$. The moduli matrix is the same as one for $U(2)$ local
vortex, for instance
\beq
{\bf B}_0(z) = 
{\cal B}_{\rm vev}
\left(
\begin{array}{cc}
1 & c_1 z + c_2\\
0 & z^2 + b_1 z + b_2
\end{array}
\right) \ .
\eeq
Thus the moduli space for the single vortex is 
\beq
{\cal V}_{\rm gen} = \mathbb{C} \times {\cal V}^{U(2)}_{k=2}\ ,
\eeq
where ${\cal V}^{U(2)}_{k=2}$ stands for the moduli space of $k=2$
$U(2)$ local vortices and the first factor $\mathbb{C}$ corresponds to
the complex parameter $a$ in $A_0$. 
Note that the minimal configuration includes two non-Abelian vortices
since the vacuum moduli space has a
$\mathbb{Z}_2$ singularity. Once we choose a generic point as the
boundary condition, we cannot remove one of two 
non-Abelian vortices from the configuration. If we do that, the
configuration meets a singularity. 

Only when we choose the special point $A_{\rm vev}=0$, we can avoid
the $\mathbb{Z}_2$ conical singularity. 
The $U(1)$ winding number $\nu$ can be a half integer and the minimal
configuration is obtained by $\nu = 1/2$ 
\beq
A_0 = a \ ,\qquad
\det {\bf B}_0 = {\cal B}_{\rm vev}^2 (z + b)\ .
\eeq
This reflects the fact that at the orbifold point $A=0$ of ${\cal M}$,
the $U(1)$ fiber makes a jump (becomes a half of what is at regular
points). 
The {\it vortex  moduli space}  is the same as that of $k=1$ $U(2)$ local vortex \cite{HT,ABEKY}
and is given by
\beq
{\cal V}_{\rm sp} = {\cal V}^{U(2)}_{k=1}\ ,\qquad
{\cal V}^{U(2)}_{k=1} = \mathbb{C} \times \mathbb{C}P^1\ .
\eeq

As in Section~\ref{sec:u1u2} we cannot take the  singular point ${\cal B}_{\rm vev} = 0$ as a boundary 
condition for constructing vortex solutions. 
The lumps in the strong gauge coupling limit always hit 
the singularity.

\section {$U(1)\times SO(N)$ model} \label{sec5}

We now consider the fractional vortices occurring in a model
with gauge group $U(1)\times SO(N)$.
Vortices with orientational modes (non-Abelian vortices) in these
models, in a maximally color-flavor locked vacuum, have recently
been constructed and studied  \cite{FGK,General,grande}. 

For our purposes here, we shall consider only the even-dimensional
orthogonal groups, i.e.~$N=2M$. The matter content is $\NF=N$
flavors of squarks in the fundamental (vector) representation of the
$SO(N)$ group, all with the same unit charge with respect to the
$U(1)$ group:  
\begin{center}
\begin{tabular}{c||cc}
& $U(1)$ & $SO(N)$ \\
\hline
\hline
$H$ & $1$ & $\square$ 
\end{tabular}
\end{center}
As the ${\mathbb Z}_{2}$ element (i.e.~$- {\mathbf 1}$) of the $SO(N)$
group is also an element of $U(1)$, the gauge group is really
$U(1)\times SO(N)/{\mathbb Z}_{2}$. 

The vacuum moduli have been studied in Ref.~\cite{Eto:2008qw} and it
turns out that it has a rather rich structure. By color and flavor
transformation, the scalar VEV can be put in the canonical form
\beq \brc H\ckt = \diag\left(v_1, v_2, \cdots, v_{2M}\right)\ ,  \qquad 
\sum_{i=1}^{2M} v_{i}^{2}  = \xi \ , \qquad 
v_{i}\in \mathbb{R} \ . \label{eq:genericvac}\eeq
Note that, in contrast to the $U(N)$ models with $N_{f}=N$ flavors,
where vacuum conditions force the VEV of $H$ to be proportional to an
$N\times N$ unit matrix, the weaker condition here leaves the
possibility of having arbitrary values $v_{i}$ subject to 
the constraint, $\sum_{i=1}^{2M}  v_{i}^{2}  = \xi$.   A large vacuum
degeneracy is present here.

At a generic point in ${\cal M}$, where $v_{i} \ne 0, \,\, \forall i$,
and all distinct, the gauge and flavor groups  
\beq L = \frac{U(1)\times SO(2M)}{\mathbb{Z}_{2}} \ , \qquad  
G_{F} = SU(2M) \ , \eeq
are completely broken. The fiber $F$ is given by the $L$ orbits of
the points $H=\diag(v_{1}, v_{2}, \ldots, v_{2M})$.

On the points where some (at least two) of the  $v_{i}$'s vanish, the
unbroken gauge group $L_{0}$ is strictly smaller than $L$. The gauge
orbit $F$ is now generated by $L / L_{0}$ and has a smaller dimension
than in the case of a vortex constructed on a generic point of 
${\cal M}$. Thus even though in all cases 
\beq \pi_{1}(F) = {\mathbb Z} \ , \eeq
its actual (e.g.) minimal element goes through discontinuous changes
whenever we hit a singularity (or a singularity curve) on ${\cal M}$.
Also, in such a point, the global symmetry group $G_{F}$  is different
from that at surrounding points, and the consequent internal {\it
  vortex moduli}  also undergoes a discontinuous change.
As a singular surface (e.g.~with a given number of vanishing
$v_{i}'s$) contains a smaller subspace of singular points (some of the
remaining $v_{i}$'s vanishing there), etc., one ends up with a
rather rich structure of a (stratified) singular manifold ${\cal M}$,
and of the vortices and related sigma-model lumps as the fiber defined
over it. We shall leave the study of these varieties of phenomena for a
separate investigation: here we will take all vacuum moduli to be
non-vanishing.

Our fractional vortex solution is closely related to the ``fractional lump'' which 
was found by some of us recently \cite{Eto:2008qw}. We choose in the following the scalar VEV to
be of color-flavor diagonal form,  and moreover proportional to the  unit matrix form,  
\beq \langle H \rangle = \sqrt{\xi}\mathbf{1}_{2M} \  
\label{eq:colorflavorvac}\eeq
leaving a residual
global color-flavor symmetry $SO(N)_{\rm c+f}$  unbroken. 
The standard moduli-matrix Ansatz is 
\beq H = s^{-1}(z,\bar{z}){S'}^{-1}(z,\bar{z})H_0(z) \ , \eeq
where $s\in U(1)^{\mathbb{C}}, S'\in SO(N)^{\mathbb{C}}$.  The gauge field 
BPS equations lead to 
\begin{align}
\partial\bar{\partial}\log\omega &=
-\frac{e^2}{4N}\left(\frac{1}{\omega}\Tr\left(\Omega_0{\Omega'}^{-1}\right)
- v^2\right) \ , \label{master1} \\
\bar{\partial}\left(\Omega'\partial{\Omega'}^{-1}\right) &=
\frac{g^2}{8\omega}\left(\Omega_0{\Omega'}^{-1} -
J^\dag\left(\Omega_0{\Omega'}^{-1}\right)^{\rm T}J\right) \ , \label{master2}
\end{align}
where $\omega = ss^\dag,\ \Omega' = S'{S'}^\dag,\ \Omega_0 =
H_0H_0^\dag$.     $J$ is the invariant tensor of $SO(N=2M)$
\beq J = \begin{pmatrix} 0&\mathbbm{1}_{M}\\\mathbbm{1}_{M}&0 
\end{pmatrix} \ . \eeq
The tension of the vortex remains
\beq T = 2v^2\int_{\mathbb{C}} d^2x\ \bar{\partial}\partial\log\omega 
= \pi v^2 k \ . \eeq
Following the construction of \cite{General}, we have the constraint
\beq H_0^{\rm T}J H_0 = z^k J + \mathcal{O}\left(z^{k-1}\right) \ ,
\label{SO-holomorphic} \eeq
for vortex solution of winding number $k$. 

In order to study the minimal winding vortex configuration more
concretely,  we choose  
\beq H_0 = \begin{pmatrix} z\mathbbm{1}_M - \mathbf{Z}&\mathbf{C}\\
0&\mathbbm{1}_M \end{pmatrix} \ , \qquad \mathbf{Z} =
\diag(z_1,z_2,\ldots,z_M) \ , \qquad \mathbf{C} =
\diag(c_1,c_2,\ldots,c_M)\  . \eeq
To solve the master equations (\ref{master1}) and (\ref{master2}), we
set
\beq \Omega' = \diag\left(e^{\chi_1'},\ldots,e^{\chi_M'},
e^{-\chi_1'},\ldots,e^{-\chi_M'}\right) \ , \label{SO-ansatz} \eeq
where the determinant one is manifest. Taking $\omega = e^{\psi}$, we
obtain 
\begin{align}
\bar{\partial}\partial\psi &= -\frac{e^2}{8M}
\left[\sum_{i=1}^M\left\{\left(|z-z_i|^2+|c_i|^2\right)e^{-(\psi+\chi_i')}
+e^{-(\psi-\chi_i')}\right\} - v^2\right] \ , \\
\bar{\partial}\partial\chi_i' &= -\frac{g^2}{8}\left[\left(
|z-z_i|^2+|c_i|^2\right)e^{-(\psi+\chi_i')}-e^{-(\psi-\chi_i')}\right]
\ , \qquad \forall\, i\in[1,M] \ .
\end{align}
If we now take the infinite gauge coupling limit $e\to\infty,\ g\to\infty$, we
obtain the following lump solution
\begin{align}
e^{\chi_i'} &= \sqrt{|z-z_i|^2+|c_i|^2} \ , \\
e^\psi &= \frac{2}{v^2}\sum_{i=1}^M e^{\chi_i'} =
\frac{2}{v^2}\sum_{i=1}^M \sqrt{|z-z_i|^2+|c_i|^2} \ , 
\end{align}
which has the energy density
\beq \mathcal{E} = 2\xi\bar{\partial}\partial\log
\left\{\sum_{i=1}^M \sqrt{|z-z_i|^2+|c_i|^2}\right\} \ . \eeq
This is  the fractional lump solution found in  Ref.~\cite{Eto:2008qw}. 

\begin{figure}[h!tp]
\begin{center}
\begin{tabular}{cc}
\includegraphics[width=7cm]{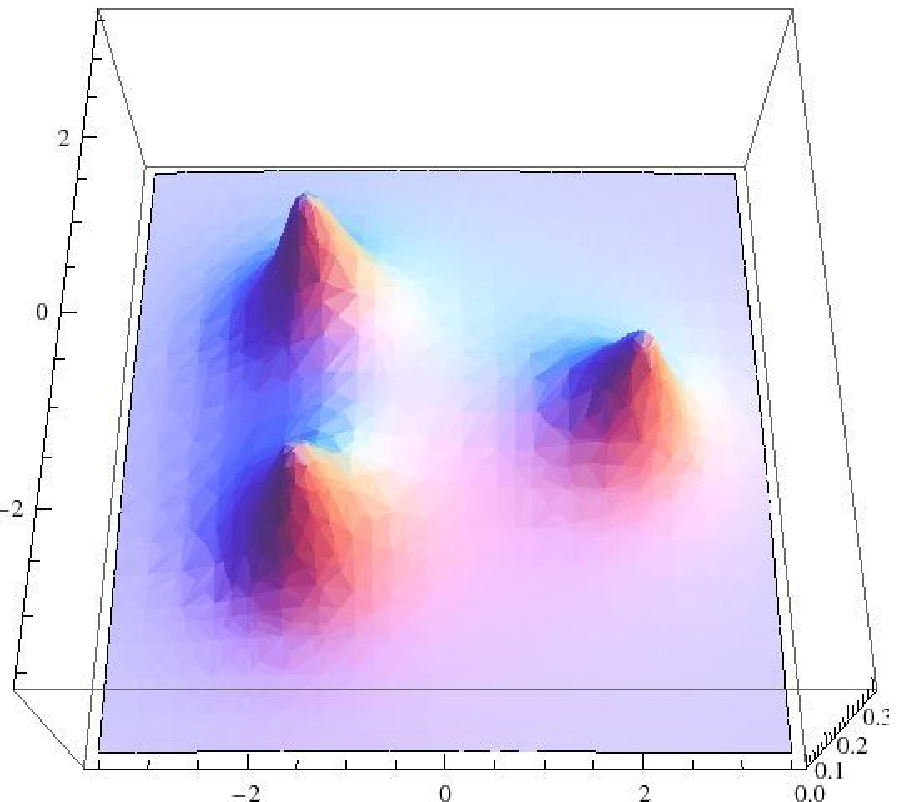} & \includegraphics[width=7cm]{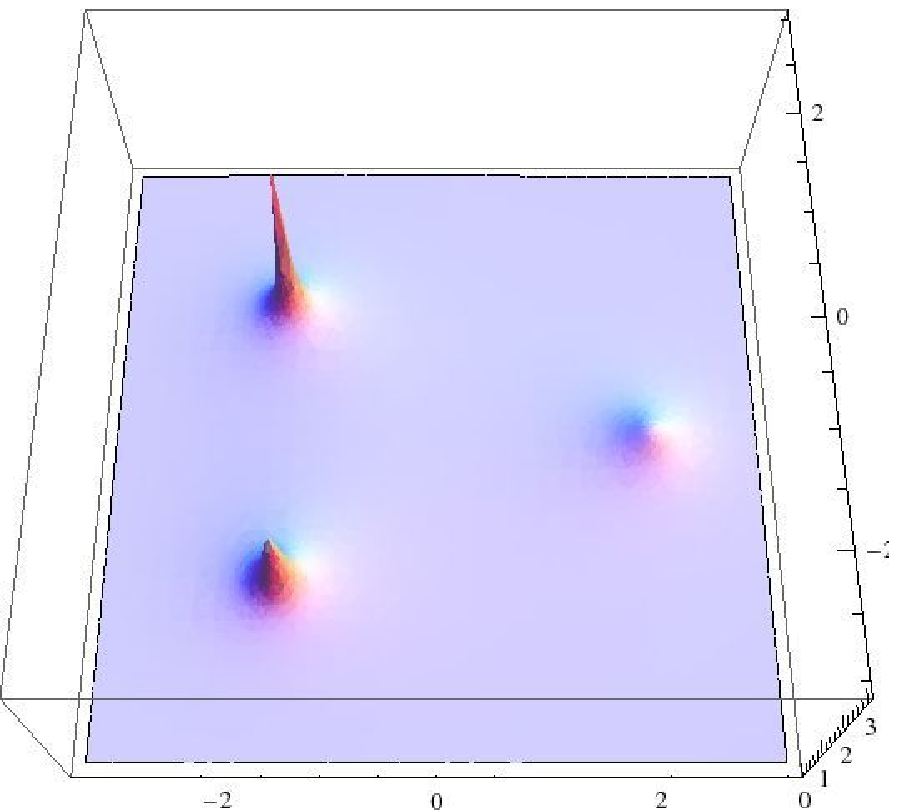}
\end{tabular}
\caption{\footnotesize{
  The energy density of three fractional vortices (lumps) in the
  $U(1)\times SO(6)$ model in the strong coupling approximation.  The positions are
  $z_1=-\sqrt{2}+i\sqrt{2}, z_2=-\sqrt{2}-i\sqrt{2},
  z_3=2$. \textit{Left panel}: the size parameters are chosen as $c_1
  = c_2 = c_3 = 1/2$. \textit{Right panel}: the size parameters are
  chosen as $c_1 = 0, c_2 = 0.1, c_3 = 0.3$. Notice that one peak is
  singular ($z_1$) and the other two are regularized by the finite
  (non-zero) parameters $c_{2,3}$. }}
\label{fig:fracson}
\end{center}
\end{figure}

The vortex energy profile in the strong-coupling approximation for  the $U(1)\times SO(6)$ model  is shown
in Figure \ref{fig:fracson}.
Three fractional peaks are clearly seen.  The positions of the peaks can be understood as follows. If $c_{i}=0$ one of the 
${\hat U}(1) \subset   U(1) \times SO(2M)$, constructed as the
diagonal combination of $U(1)$ and one of the $U(1)$ Cartan subalgebra
of $SO(2M)$,  is restored at the points  $z=z_{i}$  ($i=1,2,\ldots
M$).  The sharp peak in the right panel of Fig.~\ref{fig:fracson}  can
be thought locally (in $z$)  to be an ANO vortex.  If $c_{i} \ne 0$
the situation around a fractional peak at $z=z_{i}$ is more similar to
the power-behaved semi-local vortex of the EAH model.  
 The number of peaks reflects obviously the rank of the group
 considered (here  ${\rm rank} \{SO(6)\}=3$),  but the number of the
 possible fractional peaks depends on the point of the vacuum moduli
 (a particular VEV) considered. For instance, if two of $v_{i}$ are
 taken to be zero, the maximum number of the fractional peaks would be
 two, and so on.  

In the supersymmetric version of the models based on the 
$U(1)\times SO(N)$ gauge groups, the K\"ahler potential in terms of a
meson $M$ has been determined in Ref.~\cite{Eto:2008qw}, 
\beq K = \xi\log\Tr\sqrt{M M^\dag} \ . \label{SOKahler}\eeq
If we relax the vacuum moduli to be equal (\ref{eq:colorflavorvac}), 
thus having the possibility of distinct $\{v_i\}$'s in
Eq.~(\ref{eq:genericvac}),
it will prove convenient to work directly with the mesons of $SO(2M)$
\beq M = 
\begin{pmatrix}
e^u(z-a) & \pm ia \\
\pm ia & e^{-u}(z+a)
\end{pmatrix} \ , \eeq
with $a,u\in\mathbb{R}$. The meson VEV will be 
$\diag(v_1^2,v_2^2)=\diag(e^u,e^{-u})$. 
Using the K\"ahler potential (\ref{SOKahler})
we readily obtain the energy density
\beq \mathcal{E} = \xi\bar{\p}\p
\log\left(\left|z-a\tanh(u)\right|^2 + \frac{a^2}{\cosh^2(u)}\right)
\ . \eeq
Furthermore, we can construct a typical example of fractional
vortices, in a $U(1)\times SO(2N)$ model in the lump limit as follows
\beq \mathcal{E} = 2\xi\bar{\p}\p
\log\left(\sum_{i=1}^{N}m_i\sqrt{\left|z-a_i\tanh(u_i)\right|^2 +
  \frac{a_i^2}{\cosh^2(u_i)}}\right) \ , \eeq
with  $v_{2i-1}^2=m_i\,e^{u_i},v_{2i}^2=m_i\,e^{-u_i}$.
For each $SO(2)$ subgroup, we have in this construction a possibility
for amplification $m_i$, and $a_i, u_i$ which serve as a position- and
an effective size-parameters. One can observe that $m_i$ controls the
relative weight of the energy distributed to the $i$-th fractional
vortex.

\section{Conclusion}\label{sec6}
In this paper we have given a simple account of the fractional
vortices which have minimally quantized magnetic flux (winding) but
with non-trivial substructures in the energy distribution in the
transverse plane. 
They could often appear in various generalizations of the Abelian Higgs
model. 
The common characteristic features these models share are a
non-trivial vacuum degeneracy  and the BPS saturated nature of the
vortex solutions.  
We have generalized the moduli matrix formalism
\cite{Isozumi:2004vg,Eto:2005yh,Eto:2006pg} to clarify all possible
moduli parameters of the minimally quantized fractional vortices. 

The {\it  vacuum moduli} ${\cal M}$  in these
models turns out, in general, to be a singular manifold, i.e., a
manifold with singularities.  
Vortex solutions approach the vacuum configuration far from the
center, and trace various closed gauge orbits $F$.  

We have classified fractional vortices into two types; the first type
appears when ${\cal M}$ has a $\mathbb{Z}_n$ singularity where the
gauge symmetry is not restored while the second type occurs when
${\cal M}$ has a 2-cycle with a deformed geometry.
The existence of a singularity is not essential for the fractional
vortices of the second type.
Indeed, we have observed that smooth fractional lump solutions become
singular as the smooth manifold ${\cal M}$ is deformed into a singular
manifold (e.g.~when some FI parameters are turned off). 
Even when ${\cal M}$ has such singularities, we have found
smooth fractional {\it vortex} solutions.
The vortices share the same properties as those of the corresponding
lumps wrapping on ${\cal M}$ smoothened.

An interesting aspect of our analysis, especially relevant to the systems
with the first type of fractional vortices and lumps, is the fact that
the latter often represent {\it a generalized fiber bundle over the
  singular manifold ${\cal M}$.} At a singularity (or on a singular
surface) the fiber space $F$ undergoes a discontinuous 
change either in its dimension or in its nature, or both. 
The {\it vortex moduli} also make a jump at such a  point (points).
These observations seem to point towards  interesting physical
applications as well as some novel kind of mathematical structures.

\section*{Note added} 

After completing this paper, 
a new paper by D. Tong and  B. Collie which discusses precisely the second type of our fractional vortices -- lumps, though in a different context,  
was posted on the ArXiv  \cite{Collie:2009iz}.

\section*{Acknowledgements}

One of us (K.K.) thanks Daniele Dorigoni, Sergio Spagnolo, and Mario Salvetti for discussions. 
W.V. is supported by Della Riccia grant which supports Italian young researchers for working abroad. W.V. thanks the Department of Applied  Mathematics and Theoretical Physics (DAMTP) of Cambridge for the nice hospitality. W.V, S.B.G. and M.E. thank David Tong for the nice and useful discussions. 
M.E. and K.O. are supported by the Research Fellowships of the Japan Society 
for the Promotion of Science for Research Abroad. 
The work of M.N.~is supported in part by Grant-in-Aid for Scientific
Research (No.~20740141) from the Ministry
of Education, Culture, Sports, Science and Technology-Japan.
One of us (K.K.)  thanks  the organizers and the participants of the Workshop ``Crossing Boundaries'',  Minneapolis, 14 -17 May 2009, in honor of the 60th birthday  of M. Shifman, for providing him with an opportunity to discuss this work, and especially David Tong for a discussion on the unpublished versions 
of our (and their) work.

\appendix

\section{General BPS  vortex equations}

\subsection{General arguments}
Let $W_{\mu}^{I}$ denote the gauge fields for $G=U(1)^n \times
G^{\prime}$ gauge group, where $I=- (n-1),\cdots,-1, 0$ are for the
Abelian gauge fields $U(1)^n$, the rest referring to the arbitrary simple
group $G^{\prime}$.   
We write the complex scalar (Higgs) fields as 
\[   H^{\alpha}\ , \qquad \alpha =  (r_{1}, A_{1})\ , (r_{2}, A_{2})
\ , \ldots, \]
where  $r_{i}$ and $A_{i}$ stand for the color and flavor  
indices of the $i$-th matter in the representation ${\underline {R_{i}}}$. 
A gauge transformation ($G$) induces 
\[  \delta H^{\alpha}  = i \, (\Lambda  H)^{\alpha} =  i\, \sum_{I} \Lambda^{I} (T^{I})^{\alpha}_{\beta}\, 
H^{\alpha} = i\, \sum_{I \le 0}  \Lambda^{I}\,(T^I)^\alpha_\beta \, H^{\beta}  + i\, 
\sum_{I > 0} \Lambda^{I}  (T^{I})^{\alpha}_{\beta}\, H^{\beta} \ , 
\]
Here $T^I$ for $I \le 0$ is the charge matrix for $U(1)_I$, i.e,
$Q_{i}^I$ are the $U(1)_I$ charges of the matter fields present
\[  (T^I)^\alpha_\beta = Q^I_\beta\,\delta_{\alpha\beta} \ ,  \qquad
\text{for}\quad I \le 0 \ .\]
$T^{I}$    $(I\ge 1)$  denotes the generators of the gauge group
$G^{\prime}$ in the (in general, reducible) representation  
\[      T^{I}=   t_{I}^{(1)} \otimes {\mathbf 1}_{N_{f}^{(1)}}  
\oplus  t_{I}^{(2)} \otimes {\mathbf 1}_{N_{f}^{(2)}}  \oplus
\cdots\ . \]
Accordingly, the covariant derivative is given by
\[ ( {\cal D}_{\mu}  H )^{\alpha} = [ (\partial_{\mu} - i W_{\mu} ) \,
  H ]^{\alpha} =  \left( \delta^{\alpha}_{\beta} \partial_{\mu} - i
W_{\mu}^{I} \, (T^{I})^{\alpha}_{\beta}  \right)  H^{\beta}\ ;
\]
in our models the energy (tension) has the following form and is
semi-positive definite:  
\[    T = \int d^{2}x \, \left[  \sum_{I} \frac{1}{2 \, g_{I}^{2} } (F_{12}^{I})^{2} 
+ \sum_{\alpha} \left(  |({\cal D}_{1} H)^{\alpha}|^{2} +  |({\cal D}_{2} H)^{\alpha}|^{2} \right) 
+  \sum_{I} \frac{g_{I}^{2}}{2}  ( H^{\dagger} T^{I} H  - \xi_{I})^{2}
\right]\ ,
\]
where 
$\xi_{I}=0, ({}^{\forall} I > 0)$ 
and $\xi_I$ for $I\le 0$ are assumed such that a 
supersymmetric vacuum exists and the theory is in Higgs phase.

The vacuum manifold is the zero locus of the superpotential
\beq
M_{\rm v} = \left\{H\ |\ D^I = H^\dagger T^I H - \xi_I = 0\ ,\ \forall I\right\}
\eeq
and the vacuum moduli is 
\beq
{\cal M}_{\rm v} = M/G \simeq H /\!\!/ G^{\mathbb{C}}\ .
\label{eq:vac_moduli}
\eeq
These quotients may be ill-defined where the vacuum moduli space is
singular or when some gauge symmetry is restored.

When we choose a point in the vacuum manifold where some of $U(1)$
gauge symmetries are spontaneously broken as $G = U(1)^n \times G' \to
U(1)^{n-r} \times g'$ ($g' \in G'$), the topologically stable vortices
appear with support from the non-trivial first homotopy group 
\[
\pi_1\left(U(1)^r\right) = \bigoplus_{i=1}^r \mathbb{Z} \ .
\]
This is related to $\pi_2({\cal M})$ by the homotopy sequence
(\ref{homotseq}). We now introduce the complex coordinates to write
down the BPS equations 
\begin{eqnarray*}    z =  x + i y\ ,  \quad   {\bar z} =  x - i y\ ,  \quad   
\partial \equiv  \frac{1}{2} (\partial_{x}- i   \partial_{y})\ , \quad    
{\bar \partial}\equiv \frac{1}{2} (\partial_{x}  +   i
\partial_{y})\ ,\qquad\qquad\\
\bar{\mathcal{D}} \equiv
\frac{1}{2}  ( \mathcal{D}_1+i\mathcal{D}_2  )\ , \    {\mathcal{D}} \equiv
\frac{1}{2}  ( \mathcal{D}_1-  i\mathcal{D}_2  )\ ,\ 
W = \frac{1}{2}(W_1^I - i W_2^I)T^I\ ,\ 
\bar W = \frac{1}{2}(W_1^I + i W_2^I)T^I\ .
\end{eqnarray*}
The Bogomol'nyi completion reads
\[   T =  \int d^{2}x \, \left[  \frac{1}{2 \, g_{I}^{2} } [ F_{12}^{I} \mp   g_{I}^{2}   
(H^{\dagger} T^{I} H  - \xi_{I} ) ]^{2}  +    
\sum_{\alpha}    | ({\cal D}_{1} H  \pm  i  {\cal D}_{2} H )^{\alpha}|^{2}
\mp   \xi_I   F_{12}^{I} \right]\ .
\]
This shows that the minimum of the tension is given by either  
the ``negative chirality'' (or left-winding) solutions satisfying  the
Bogomol'nyi equations 
\beq  F_{12}^{I} =   g_{I}^{2}   (H^{\dagger} T^{I} H  - \xi_{I} )\ ,
\qquad     \bar{\mathcal{D}}  H=0\ , \qquad   
T_{\min}=  -  \xi_I \int  d^{2}x \, F_{12}^{I} >0 \ ; 
\label{left}\ee
or by the ``positive chirality'' (right-winding)  solutions such that  
\[   F_{12}^{I} =  -   g_{I}^{2}   (H^{\dagger} T^{I} H  - \xi_{I} )\ ,
\qquad    {\mathcal{D}}  H=0\ , \qquad 
T_{\min}=   \xi_I \int  d^{2}x \, F_{12}^{I} >0 \ ; \]
Therefore, the tension is expressed by
\beq
T_{\min} = \sum_I 2\pi \xi_I \nu_I \ ,\qquad
\nu_I = \mp \frac{1}{2\pi} \int dx^2\, F_{12}^I \ ,
\label{eq:tension}
\eeq
where $-$ is for the negative chirality and $+$ is for the positive
chirality. The rational number $\nu_I$ (for $I\le 0$) stands for how
many times the corresponding solution winds $U(1)_I \sim S^1$ when we
go around once the boundary circle $S^1$ on the $z$-plane.

From now on, we concentrate on the BPS states with the negative
chirality which are the solutions of Eq.~(\ref{left}). The matter part
of the above vortex equation can be solved by the moduli matrix  
$H_0(z)$ which is a color-flavor mixed matrix whose elements are
holomorphic (polynomials) in $z$, for  the left-handed vortex solution
Eq.~(\ref{left}) \cite{Isozumi:2004vg,Eto:2005yh,Eto:2006pg}
\beq  H = S^{-1}  \,  H_{0} (z) \ ,\quad
\bar W = - i S^{-1}\bar \p S\ ,
\label{Ansatze} \ee
where $S$ is an element of the complex extension of the gauge group
\beq  S(z,\bar z)=  e^{{  }i\, \sum_{I} {\Lambda^I(z,\bar z)}   T^{I}
} \in G^{\mathbb{C}} \ ,
\qquad  {{\Lambda}}^{I}(z,\bar z) \in {\mathbb C}\ . 
\label{complexif}\ee
For a given $H_0$, $S$ is determined by the first equation in
Eq.~(\ref{left}). We assume existence and uniqueness of the
solutions. Thus all the complex constants (coefficients of the
polynomial functions) appearing in $H_0(z)$ represent the vortex moduli
parameters. Finally, $S$ and $H_{0}(z)$ are defined up to a $V(z)$
transformation which does not change (\ref{Ansatze}) and keeps $S$
inside $G^{\mathbb C}$ 
\[     H_{0}(z) \to  V(z) \, H_{0}(z)\ , \qquad  S \to   V(z) \, S \ , 
\]
where $V(z)$ is a holomorphic matrix belonging to the complexified
gauge group $G^{\mathbb{C}}$. When the vortex moduli space is seen as
a complex manifold (whose local coordinates are the moduli parameters
appearing in $H_{0}(z)$), the $V(z)$ transformations act as the
transition functions in two overlapping patches. 
Thus the (vortex) moduli space of  the 1/2 BPS vortices is formally expressed by 
\[
{\cal M}_{\rm vor} = \left\{ H_0\ |\ H_0 \sim V H_0 \ ,\ \bar\p H_0 =
0 \ ,\ \ \bar\p V = 0,\ V \in G^{\mathbb{C}}\right\} \ .
\]

Generally speaking, one should choose a boundary condition when one
wishes to solve some partial differential equations. Our strategy of
solving the BPS equations (\ref{left}) is somehow upside-down to such
an ordinary way. In fact, we have not fixed any boundary conditions
yet. Of course, we are talking about the topological solitons which
are characterized by the boundary conditions. So the remaining task is
to figure out the condition for the moduli matrix which yields
solutions satisfying the correct boundary condition. Note that the
condition for $H_0(z)$ depends on the boundary condition which is
nothing but VEV $\left<H\right>$, a point on the vacuum manifold. We
have to be careful to specify the moduli matrix especially when we
choose a singular point as the boundary condition. Furthermore, the
configuration (energy distribution) may change as varying the VEV even
if the moduli matrix is fixed. 

The concrete conditions for the moduli matrix $H_0(z)$ depend on
details of the models, such as gauge groups, representations of the
matter fields and the $U(1)$ charges. The case of $G = U(1) \times G'$
with $G'$ being an arbitrary simple group have been studied in
Ref.~\cite{General}.
In what follow, we will explain two typical cases. i) $G=U(1)$ gauge
theory with matter fields whose $U(1)$-charges are distinct. ii) there
are more than one Abelian group such as 
$G=U(1)_1 \times U(1)_2$. These models, and especially, their moduli
spaces, have not been studied so far.

\subsection{$G=U(1)_1\times U(1)_2$}

Let us consider a $G = U(1)_1 \times U(1)_2$ gauge theory  with three
Higgs fields $H = (A,B,C)$ with the following $U(1)$-charges: 
$Q_1 = (m,1,1)$ under the first $U(1)_1$ and $Q_2 = (0,1,-1)$ under
$U(1)_2$. The vacuum manifold and the vacuum moduli are 
\beq
M &=& \{A,B,C\ |\ m|A|^2 + |B|^2 + |C|^2 = \xi_1 \ ,\ |B|^2 - |C|^2 =
\xi_2\} \simeq S^3 \ ,\\
{\cal M} &=& M/(U(1)_1 \times U(1)_2) \simeq
\mathbb{C}P^1/\mathbb{Z}_m \ .
\eeq
Here $\xi_{1,2}$ are the FI-terms of $U(1)_{1,2}$.
We choose them in the region  $-\xi_1 \le \xi_2 \le \xi_1$ $(\xi_1 \ge 0)$ to
get the system in a Higgs vacuum.  
In the main text of this paper we set
$\xi_2 = 0$ but here we consider a more general situation.
At generic point of $M$, both $U(1)_1$ and $U(1)_2$ are broken, hence
we have topologically stable vortex solutions with topological
characters $\nu_I$ 
\beq
&&\nu_I = \frac{1}{\pi} \int dx^2\, \p\bar\p \log |s_I|^2 \ ,\quad
\bar W^I = - i\bar\p\log s_I,\quad(I=1,2)\ ,\label{eq:2u1}\\
&&H = (A,B,C) = \left(s_1^{-m} A_0(z),\ s_1^{-1}s_2^{-1}
B_0(z),\ s_1^{-1}s_2 C_0(z)\right) \ ,
\eeq
The solutions are determined by 
$U(1)_1 \times U(1)_2$ winding numbers $(\nu_1,\nu_2)$.
Eq.~(\ref{eq:2u1}) determines the asymptotic behavior of $s_I$
\beq
|s_I|^2 \to |z|^{2\nu_I}\qquad \text{as} \quad |z| \to \infty \ .
\eeq
Then the Higgs fields asymptotically behaves as
\beq
(A_0,B_0,C_0)  \simeq  \left( 
|z|^{m \nu_1} A_{\rm vev},\ 
|z|^{\nu_1 + \nu_2} B_{\rm vev},\ 
|z|^{\nu_1 - \nu_2} C_{\rm vev}
\right) \ ,\quad \text{as}\quad |z| \to \infty \ ,
\eeq
$(A_{\rm vev},B_{\rm vev},C_{\rm vev})$ being a point in the vacuum
manifold.

\subsubsection{$m=1$}

Let us first consider a generic point,  i.e.,  
$(A_{\rm vev},B_{\rm vev},C_{\rm vev}) \neq (0,0,0)$.
Holomorphy forces us to choose 
$\nu_1 \in \mathbb{Z}_+$, $\nu_1 + \nu_2 \in \mathbb{Z}_+$ and
$\nu_1 - \nu_2 \in \mathbb{Z}_+$. Let us rewrite 
\beq
\nu_1 \equiv k_1 \in \mathbb{Z}_+ \ ,\quad
\nu_1 + \nu_2 \equiv k_2 \in \mathbb{Z}_+ \ ,
\qquad 2k_1 \ge k_2 \ge 0  \ .
\eeq
The tension can be expressed as
\beq
\frac{T}{2\pi} = \nu_1 \xi_1 + \nu_2 \xi_2 = (\xi_1-\xi_2) k_1 + \xi_2 k_2 \ge
\frac{\xi_1 + \xi_2}{2} k_2 \ge 0 \ .
\label{eq:tension_gene_m=1}
\eeq
Thus the moduli matrix for $(k_1,k_2)$ configuration is
\beq
A_0(z) = A_{\rm vev} z^{k_1} + \cdots \ ,\quad
B_0(z) = B_{\rm vev} z^{k_2} + \cdots \ ,\quad
C_0(z) = C_{\rm vev} z^{2k_1-k_2} + \cdots \ .
\eeq
The number of complex moduli parameters is 
${\rm dim}_{\mathbb{C}}{\cal M}_{(k_1,k_2)} = 3k_1$. 
The minimum configuration depends on $\xi_2$.
When $0 < \xi_2 < \xi_1$, the minimum configuration is 
$T_{(1,0)} = 2\pi (\xi_1 - \xi_2)$. When $-\xi_1 < \xi_2 < 0$, 
$T_{(1,2)} = 2\pi (\xi_1 + \xi_2)$ is minimum.
The next lightest configuration has  always $T_{(1,1)} = 2\pi \xi_1$.
If $\xi_2 = 0$, all these three are degenerate, namely
$T_{(1,0)} = T_{(1,1)} = T_{(1,2)}$.

Let us next choose the case of $A_{\rm vev} = 0$ with 
$B_{\rm vev},C_{\rm vev} \neq 0$. 
In this case, the condition changes as 
$\nu_1 + \nu_2 \equiv k_1 \in \mathbb{Z}_+$ and
$\nu_1 - \nu_2 \equiv k_2  \in \mathbb{Z}_+$. In this case, the
$U(1)$-charges $\nu_{1,2}$ are 
half quantized and the energy is
\beq
\frac{T}{2\pi} = \frac{\xi_1 + \xi_2}{2} k_1 + \frac{\xi_1 - \xi_2}{2}
k_2 \ge 0 \ .
\eeq
The moduli matrix for $(k_1,k_2)$ configuration should be chosen as
\beq
A_0(z) = a z^{\alpha} + \cdots\ , \quad
B_0(z) = B_{\rm vev} z^{k_1} + \cdots\ ,\quad
C_0(z) = C_{\rm vev} z^{k_2} + \cdots\ ,
\eeq
where $\alpha$ is an arbitrary positive integer less than 
$\nu_1 = (k_1 + k_2)/2$. 
The number of complex moduli parameters is 
${\rm dim}_{\mathbb{C}}{\cal M}_{(k_1,k_2)} = k_1 + k_2 + \alpha + 1$.
The minimum configuration is $T_{(1,0)} = \pi (\xi_1 + \xi_2)$ for 
$-\xi_1 < \xi_2 < 0$ while 
$T_{(0,1)} = \pi (\xi_1 - \xi_2)$ for $0 < \xi_2 < \xi_1$.
$T_{(1,1)} = 2 \pi \xi_1$ is the third lightest configuration.
Note that these tensions are one half of that of the vortices for 
$A_{\rm vev} \neq 0$.
If $\xi_2 = 0$, we again observe degeneracy 
$T_{(1,0)} = T_{(0,1)} < T_{(1,1)}$.

The vacuum $B_{\rm vev} = C_{\rm vev} = 0$ is possible when $\xi_2 = 0$.
We shall not consider this case since the second
$U(1)$ factor is now in a Coulomb phase.

\subsubsection{$m=2$}

Let us repeat the analysis  for case  $m=2$.
First we consider $(A_{\rm vev},B_{\rm vev},C_{\rm vev}) \neq (0,0,0)$.
We find the condition
$2\nu_1 \in \mathbb{Z}_+$, $\nu_1 + \nu_2 \in \mathbb{Z}_+$ and
$\nu_1 - \nu_2 \in \mathbb{Z}_+$. Rewrite
\beq
\nu_1 + \nu_2 \equiv k_1 \in \mathbb{Z}_+ \ ,\quad
\nu_1 - \nu_2 \equiv k_2 \in \mathbb{Z}_+ \ .
\eeq
Note that $2\nu_1 = k_1 + k_2 \in \mathbb{Z}_+$ is automatically satisfied.
The energy is given by
\beq
\frac{T}{2\pi} = 
\xi_1 \frac{k_1+k_2}{2} + \xi_2 \frac{k_1 - k_2}{2}
= \frac{\xi_1 + \xi_2}{2}k_1 + \frac{\xi_1-\xi_2}{2}k_2
\ge 0 \ . 
\label{eq:tenxion_2U(1)_1}
\eeq
Since the $U(1)$ charges are half quantized, this tension formula is slightly 
 different from Eq.~(\ref{eq:tension_gene_m=1}) with $m=1$. 
The moduli matrix for $(k_1,k_2)$ configuration is
\beq
A_0(z) = A_{\rm vev} z^{k_1+k_2} + \cdots\ ,\quad
B_0(z) = B_{\rm vev} z^{k_1} + \cdots\ ,\quad
C_0(z) = C_{\rm vev} z^{k_2} + \cdots\ .
\eeq
Dimension of the moduli space is 
${\rm dim}_{\mathbb{C}}{\cal M}_{(k_1,k_2)} = 2k_1+2k_2$.
The minimal energy configuration is
$T_{(1,0)} = \pi (\xi_1 - \xi_2)$ for $\xi_2 > 0$ or
$T_{(1,1)} = \pi (\xi_1 + \xi_2)$ for $\xi_2 <0$.
If $\xi_2 = 0$, these two are degenerate.

Configurations with $A_{\rm vev}=0$, or 
$B_{\rm vev}=0$ (or $C_{\rm vev}=0$) as the boundary condition are not
special but belong to the above category.
$B_{\rm vev} = C_{\rm vev} = 0$ is only possible when $\xi_2 = 0$. Again, 
in this case the $U(1)_2$ gauge symmetry  is restored (Coulomb phase).

\subsubsection{$m\ge3$}

As we have seen, the difference coming from the choice of $m$ appears
only when $(A_{\rm vev},B_{\rm vev},C_{\rm vev}) \neq (0,0,0)$:
\beq
m \nu_1 \equiv k_1 \ ,\quad \nu_1 + \nu_2 = k_2\ ,\quad\Rightarrow\quad
\nu_1 - \nu_2 = 2\nu_1 - k_2 = \frac{2k_1}{m} - k_2 \in \mathbb{Z}_+\ .
\eeq
When $m$ is odd, we should choose $k_1 = m_{\rm odd} k_1'$ with 
$k_1' \in \mathbb{Z}_+$.
Then  $k_2$ can be taken in the region 
$2k_1' \ge k_2 \ge 0$.  The corresponding tension is 
\beq
\frac{T_{\rm odd}}{2\pi} = \frac{\xi_1 - \xi_2}{m_{\rm odd}} k_1 + \xi_2 k_2
= (\xi_1 - \xi_2)k_1' + \xi_2 k_2 \ .
\eeq
The moduli matrix is
\beq
A_0(z) = A_{\rm vev} z^{m_{\rm odd} k_1'} + \cdots \ ,\quad
B_0(z) = B_{\rm vev} z^{k_2} + \cdots \ ,\quad
C_0(z) = C_{\rm vev} z^{2 k_1' - k_2} + \cdots \ .
\eeq
The dimension of the moduli space is 
${\rm dim}_{\mathbb{C}}{\cal M}_{(k_1,k_2)}^{\rm odd} = (m_{\rm odd}+2)k_1'$.

When $m$ is even, we must choose 
$k_1 = \frac{m_{\rm even}}{2}k_1'$ with $k_1' \in \mathbb{Z}_+$.
Thus the $U(1)$ charges are half quantized. 
Holomorphy forces $k_1' \ge k_2 \ge 0$.
The tension is given by
\beq
\frac{T_{\rm even}}{2\pi} = \frac{\xi_1 - \xi_2}{m_{\rm even}} k_1 + \xi_2 k_2
= \frac{\xi_1 - \xi_2}{2}k_1' + \xi_2 k_2\ .
\eeq
The moduli matrix is given by
\beq
A_0(z) = A_{\rm vev} z^{\frac{m_{\rm even}}{2} k_1'} + \cdots \ ,\quad
B_0(z) = B_{\rm vev} z^{k_2} + \cdots\ ,\quad
C_0(z) = C_{\rm vev} z^{k_1' - k_2} + \cdots\ .
\eeq
The dimension of the moduli space is 
${\rm dim}_{\mathbb{C}}{\cal M}_{(k_1,k_2)}^{\rm even} 
= \left(\frac{m_{\rm even}}{2}+1\right)k_1'$.

\subsection{$G=U(1) \times SU(N)$}

Let us consider $G=U(1) \times G'$ and $G'=SU(N)$ with Higgs fields 
$H = (A,{\bf B})$ where $A$ is a singlet scalar field and ${\bf B}$ is
a collection of $N$  fields   in the fundamental representation  ${\underline N}$   of
$SU(N)$, written as an $N\times N$ matrix.
We assign the $U(1)$  charges $(m,1)$   to the fields $(A,{\bf B})$,  
respectively. 
The vacuum manifold is given by the $D$-flatness conditions and the
vacuum moduli is obtained by dividing it by $G$ as
\beq
M = \left\{(A,{\bf B})\ |\ m |A|^2 {\bf 1}_N + N {\bf B}{\bf
  B}^\dagger = \xi {\bf 1}_N \right\} \ ,\quad
{\cal M} = M/G \simeq \mathbb{C}P^1/\mathbb{Z}_m \ . .
\eeq
The vacuum condition is simplified by using 
${\bf B} = B_{\rm vev} {\bf 1}_N$ as
\beq
m|A_{\rm vev}|^2 + N|B_{\rm vev}|^2 = \xi \ .
\eeq
The moduli matrix formalism is summarized as
\beq
(A,{\bf B}) = s^{-1} \left(A_0(z),\ S^{-1} {\bf B}_0(z) \right) \ ,\quad
\bar W_{U(1)} = - i \bar\p\log s  \ ,\quad
\bar W_{SU(N)} = - i S^{-1}\bar\p S \ ,
\eeq
with $s \in \mathbb{C}^*$ and $S \in SL(N,\mathbb{C})$.
$U(1)$ winding number is given by
\beq
\nu = \frac{1}{\pi} \int dx^2\, \p\bar\p \log |s|^2 \ , 
\quad\Rightarrow\quad
|s|^2 \sim |z|^{2\nu}\quad {\rm as}\quad |z| \to \infty \ .
\eeq
When we have some non-Abelian gauge group $G'$, we can consider the
holomorphic $G'$ invariant 
\beq
I \equiv \det {\bf B} = s^{-N} \det {\bf B}_0(z) 
\to |z|^{-N\nu} \det {\bf B}_0(z) \equiv B_{\rm vev}^N
\quad{\rm as}\quad |z| \to \infty \ .
\eeq
Since all the elements in ${\bf B}_0(z)$ is holomorphic in $z$, 
$\det{\bf B}_0(z)$ is also a polynomial function of $z$.
Combining this and the behavior of $A$
\beq
A = s^{-1} A_0(z) \to |z|^{-m\nu} A_0(z) \equiv A_{\rm vev} \ ,
\quad{\rm as}\quad |z| \to \infty \ ,
\eeq
we can consistently determine $\nu$.

When we choose $(A_{\rm vev},B_{\rm vev}) \neq (0,0)$, the moduli
matrix asymptotically behave as
\beq
A_0(z) \to |z|^{m\nu} A_{\rm vev} \ ,\quad
\det{\bf B}_0(z) \to |z|^{\nu N}B_{\rm vev}^N\ ,\quad
{\rm as}\quad |z| \to \infty \ .
\eeq
Thus we should choose the $U(1)$ winding number $\nu$ in such a way
that 
\beq
m \nu \equiv k \in \mathbb{Z}_+ \ ,\quad
N \nu =\frac{N}{m} \,k = \frac{N'}{m'}\,k \in \mathbb{Z}_+ \ ,
\eeq
where we have assumed $N = n_0 N'$ and $m = n_0 m'$ with 
$n_0 \in \mathbb{Z}_+$ ($n_0 = {\rm g.c.d}(N,m)$).
Thus $k$ must be $k = m' k'$ with $k' \in \mathbb{Z}_+$, so that the
$U(1)$ winding number is fractionally quantized 
$\nu = \frac{k'}{n_0}$ if $n_0 \ge 2$. 
The corresponding tension is $T = 2\pi \xi \frac{k'}{n_0}$.
The moduli matrices for the configuration with $k'$ may be chosen as
\beq
A_0(z) = A_{\rm vev} z^{\frac{m}{n_0} k'} + \cdots \ ,\quad
\det B_0(z) = B_{\rm vev}^N z^{\frac{N}{n_0}k'} + \cdots\ .
\eeq
Note that we have specified only $\det{\bf B}_0(z)$, so there are many
possibilities for ${\bf B}_0(z)$ per se, it is nothing but the
orientational moduli. 
The moduli space for $k'=1$ is 
${\cal M}_{k'=1} \simeq \mathbb{C}^{\frac{N+m}{n_0}} \times \mathbb{C}P^{N-1}$.

When we choose $A_{\rm vev} = 0$ and $B_{\rm vev} \neq 0$ 
as the boundary condition, we should satisfy
\beq
A \to |z|^{-m\nu} A_0(z) = 0 \ ,\quad
I \to |z|^{-N\nu} \det {\bf B}_0(z) = B_{\rm vev}^N
\quad{\rm as}\quad |z| \to \infty \ .
\eeq
Thus we can choose $N\nu \equiv k \in \mathbb{Z}_+$, such that the 
$U(1)$ winding is $1/N$ quantized and corresponding tension is 
$T = 2\pi \xi \frac{k}{N}$. The moduli matrices are
\beq
A_0(z) = a z^\alpha + \cdots \ ,\quad
\det {\bf B}_0 = B_{\rm vev}^N z^k + \cdots \ ,
\eeq
where $\alpha$ is a semi-positive definite integer 
$\alpha < m\nu = m \frac{k}{N}$.


\begin{thebibliography}{99}


\bibitem{Abrikosov}
A.~A.~Abrikosov, 
 Sov.\ Phys.\ JETP {\bf 5}
(1957) 1174 [Zh.\ Eksp.\ Teor.\ Fiz.\ {\bf 32} (1957) 1442].

\bibitem{NielsenOlesen}
H.~B.~Nielsen and P.~Olesen,
Nucl.\ Phys.\ B {\bf 61} (1973) 45.


\bibitem{Babaev}
  E.~Babaev,
  Phys.\ Rev.\ Lett.\  {\bf 89}, 067001 (2002)
  [arXiv:cond-mat/0111192].
  
\bibitem{Torons}
  G.~'t Hooft,  
Commun. Math. Phys. 81 (1981),
267.



\bibitem{Caloron} 

B.~J.~ Harrington and  H.~K.~Shepard, Phys. Rev. {\bf D17} (1978) 2122; Phys. Rev. {\bf D18}  (1978) 2990;  D.~J.~Gross, R.~D.~Pisarski and L.~G.~Yaffe,  Rev. Mod. Phys. {\bf 53} (1981) 43.

\bibitem{Tong:2002hi}
  D.~Tong,
  Phys.\ Rev.\  D {\bf 66}, 025013 (2002)
  [arXiv:hep-th/0202012].


\bibitem{Eto:2004rz}
  M.~Eto, Y.~Isozumi, M.~Nitta, K.~Ohashi and N.~Sakai,
  Phys.\ Rev.\  D {\bf 72}, 025011 (2005)
  [arXiv:hep-th/0412048];
  M.~Eto, T.~Fujimori, Y.~Isozumi, M.~Nitta, K.~Ohashi, K.~Ohta and N.~Sakai,
  Phys.\ Rev.\  D {\bf 73}, 085008 (2006)
  [arXiv:hep-th/0601181]; 
  M.~Eto, T.~Fujimori, M.~Nitta, K.~Ohashi, K.~Ohta and N.~Sakai,
  Nucl.\ Phys.\  B {\bf 788}, 120 (2008)
  [arXiv:hep-th/0703197].

\bibitem{Bruckmann:2007zh}
  F.~Bruckmann,
  Phys.\ Rev.\ Lett.\  {\bf 100}, 051602 (2008)
  [arXiv:0707.0775 [hep-th]];
  D.~Harland,
  arXiv:0902.2303 [hep-th];
  W.~Brendel, F.~Bruckmann, L.~Janssen, A.~Wipf and C.~Wozar,
  arXiv:0902.2328 [hep-th].

\bibitem{Collie:2009iz}
  B.~Collie and D.~Tong,
  arXiv:0905.2267 [hep-th].



 \bibitem{EAH}   
  T.~Vachaspati and A.~Achucarro,
  Phys.\ Rev.\  D {\bf 44}, 3067 (1991);
  A.~Achucarro and T.~Vachaspati,
  Phys.\ Rept.\  {\bf 327} (2000) 347
  [arXiv:hep-ph/9904229].

\bibitem{Achucarro:1992hs}
  A.~Achucarro, K.~Kuijken, L.~Perivolaropoulos and T.~Vachaspati,
  Nucl.\ Phys.\  B {\bf 388}, 435 (1992).


\bibitem{Hindmarsh}  M.~ Hindmarsh, 
Nucl.Phys. \ {\bf B392} (1993) 461. 
e-Print: hep-ph/9206229;  
M.~ Hindmarsh, R.~ Holman, T. W.~ Kephart and T.~ Vachaspati,
 Nucl.Phys.  \ {\bf B404} (1993) 794. 
e-Print: hep-th/9209088

  
 

\bibitem{HT}
  A.~Hanany and D.~Tong,
  JHEP {\bf 0307} (2003) 037
  [arXiv:hep-th/0306150].


\bibitem{ABEKY}
R.~Auzzi, S.~Bolognesi, J.~Evslin, K.~Konishi and A.~Yung,
Nucl.\ Phys.\ B {\bf 673} (2003) 187
[arXiv:hep-th/0307287].


 \bibitem{ABEK}
  R.~Auzzi, S.~Bolognesi, J.~Evslin and K.~Konishi,
  Nucl.\ Phys.\  B {\bf 686} (2004) 119
  [arXiv:hep-th/0312233].

\bibitem{tmon}
  D.~Tong,
  Phys.\ Rev.\  D {\bf 69} (2004) 065003
  [arXiv:hep-th/0307302].


 \bibitem{SY}  
   M.~Shifman and A.~Yung,
  Phys.\ Rev.\  D {\bf 70} (2004) 045004
  [arXiv:hep-th/0403149].
     
  \bibitem{HT2}
  A.~Hanany and D.~Tong,
  JHEP {\bf 0404} (2004) 066
  [arXiv:hep-th/0403158].


\bibitem{GSY}
  A.~Gorsky, M.~Shifman and A.~Yung,
  Phys.\ Rev.\  D {\bf 71} (2005) 045010
  [arXiv:hep-th/0412082].


\bibitem{Isozumi:2004vg}
  Y.~Isozumi, M.~Nitta, K.~Ohashi and N.~Sakai,
  Phys.\ Rev.\  D {\bf 71}, 065018 (2005)
  [arXiv:hep-th/0405129];
  Phys.\ Rev.\ Lett.\  {\bf 93}, 161601 (2004)
  [arXiv:hep-th/0404198];
  Phys.\ Rev.\  D {\bf 70}, 125014 (2004)
  [arXiv:hep-th/0405194].

\bibitem{Eto:2005yh}
  M.~Eto, Y.~Isozumi, M.~Nitta, K.~Ohashi and N.~Sakai,
  Phys.\ Rev.\ Lett.\  {\bf 96}, 161601 (2006)
  [arXiv:hep-th/0511088].

\bibitem{Eto:2006pg}
M.~Eto, Y.~Isozumi, M.~Nitta, K.~Ohashi and N.~Sakai,
J.\ Phys.\ A {\bf 39} (2006) R315
[arXiv:hep-th/0602170];
  ``Solitons in supersymmetric gauge theories: Moduli matrix approach,''
  arXiv:hep-th/0607225.


 \bibitem{Duality}
 M. Eto, L. Ferretti,  K. Konishi, G. Marmorini, M. Nitta, K. Ohashi, W. Vinci,  N. Yokoi,
Nucl.Phys. {\bf B780} 161-187, 2007
   [arXiv: hep-th/0611313].

 

\bibitem{Shifman:2007ce}
  M.~Shifman and A.~Yung,
  Rev.\ Mod.\ Phys.\  {\bf 79} (2007) 1139
  [arXiv:hep-th/0703267].

\bibitem{Tong:2005un}
  D.~Tong,
  ``TASI lectures on solitons,''
  arXiv:hep-th/0509216.

\bibitem{Tong:2008qd}
  D.~Tong,
  ``Quantum Vortex Strings: A Review,''
  arXiv:0809.5060 [hep-th].

\bibitem{HashiTong}  
  K.~Hashimoto and D.~Tong,
  JCAP {\bf 0509} (2005) 004
  [arXiv:hep-th/0506022].

\bibitem{ASY}   R.~Auzzi, M.~Shifman and A.~Yung,
  Phys.\ Rev.\  D {\bf 73} (2006) 105012
  [Erratum-ibid.\  D {\bf 76} (2007) 109901]
  [arXiv:hep-th/0511150].

\bibitem{sevenHW}
  M.~Eto, K.~Konishi, G.~Marmorini, M.~Nitta, K.~Ohashi, W.~Vinci and N.~Yokoi,
  Phys.\ Rev.\  D {\bf 74} (2006) 065021
  [arXiv:hep-th/0607070].

\bibitem{SYSemi}
  M.~Shifman and A.~Yung,
  Phys.\ Rev.\  D {\bf 73} (2006) 125012
  [arXiv:hep-th/0603134].
 
\bibitem{SemiL}  
    M. Eto, J. Evslin, K. Konishi, G. Marmorini, M. Nitta, K. Ohashi, W. Vinci,  N. Yokoi,   
  Phys.\ Rev.\  D {\bf 76} (2007) 105002
  [arXiv:0704.2218 [hep-th]].

   
 \bibitem{AEV1} 
  R.~Auzzi, M.~Eto and W.~Vinci,
  JHEP {\bf 0711} (2007) 090
  [arXiv:0709.1910 [hep-th]].


\bibitem{AEV2} 
  R.~Auzzi, M.~Eto and W.~Vinci,
  JHEP {\bf 0802} (2008) 100
  [arXiv:0711.0116 [hep-th]].
  M.~Eto,
  arXiv:0810.4895 [hep-th].


\bibitem{KF} 
  L.~Ferretti and K.~Konishi,
  ``Duality and confinement in SO(N) gauge theories,''
  in ``''Sense of Beauty in Physics: a volume in honour of Adriano Di Giacomo'',
  Ed. by M. D'Elia, et. al.,  Edizioni PLUS (Univ.  Pisa. Press), 2000, 
  arXiv:hep-th/0602252.

\bibitem{heterotic}
  M.~Edalati and D.~Tong,
  JHEP {\bf 0705} (2007) 005
  [arXiv:hep-th/0703045];
  D.~Tong,
  JHEP {\bf 0709} (2007) 022
  [arXiv:hep-th/0703235];
  M.~Shifman and A.~Yung,
  Phys.\ Rev.\  D {\bf 77} (2008) 125016
  [arXiv:0803.0158 [hep-th]];
  M.~Shifman and A.~Yung,
  Phys.\ Rev.\  D {\bf 77} (2008) 125017
  [arXiv:0803.0698 [hep-th]]. 
  
  
  \bibitem{DKO}
  D.~Dorigoni, K.~Konishi and K.~Ohashi,
  Phys. Rev. {\bf D 79}  (2009) 045011.   
  

  
  \bibitem{FGK}
  L.~Ferretti, S.~B.~Gudnason and K.~Konishi,
  Nucl.\ Phys.\  B {\bf 789}, 84 (2008)
  [arXiv:0706.3854 [hep-th]].

  \bibitem{General}
  M.~Eto, T.~Fujimori, S.~B.~Gudnason, K.~Konishi, M.~Nitta, K.~Ohashi and W.~Vinci,
  Phys.\ Lett.\  B {\bf 669}, 98 (2008)
  [arXiv:0802.1020 [hep-th]].


\bibitem{grande}
  M.~Eto {\it et al.},
  JHEP (in press) [arXiv:0903.4471 [hep-th]].

\bibitem{Higashi} K.~Higashijima, M.~Nitta,
Prog. Theor. Phys. {\bf 103} (2000) 635. 
e-Print: hep-th/9911139;  
Prog. Theor. Phys. {\bf 103} (2000) 833. 
e-Print: hep-th/9911225. 

 
\bibitem{Eto:2008qw}
  M.~Eto, T.~Fujimori, S.~B.~Gudnason, M.~Nitta and K.~Ohashi,
  Nucl.\ Phys.\  B {\bf 815}, 495 (2009)
  [arXiv:0809.2014 [hep-th]].
 
   
  

  
  \end{thebibliography}
  \end{document}